\documentclass[12pt]{article}

\pdfoutput=1

\usepackage[utf8]{inputenc}
\usepackage[T1]{fontenc}
\usepackage{lmodern}
\usepackage{physics,mathtools,simplewick}
\newcommand*{\sk}{-3pt}
\usepackage{array,multirow,url}
\usepackage{graphicx}
\graphicspath{{./images}}
\usepackage{float}
\usepackage{listings}
\usepackage{cite}
\usepackage{xcolor}
\usepackage{microtype}

\newcommand\Scale[2][1]{\scalebox{#1}{\mbox{\ensuremath{\displaystyle #2}}}}

\usepackage{pifont}
\makeatletter
\newcommand{\pushright}[1]{\ifmeasuring@#1\else\omit\hfill$\displaystyle#1$\fi\ignorespaces}
\newcommand{\pushleft}[1]{\ifmeasuring@#1\else\omit$\displaystyle#1$\hfill\fi\ignorespaces}
\makeatother
\newcommand{\dalambert}{\mathop{{}\mathding{113}}\nolimits}
\makeatletter
\newcommand{\mathding}[1]{\mathpalette\math@ding{#1}}
\newcommand{\math@ding}[2]{%
  \begingroup
  \mbox{\normalfont\fontsize{%
    \ifx#1\displaystyle\f@size\else
    \ifx#1\textstyle\f@size\else
    \ifx#1\scriptstyle\sf@size\else
    \ssf@size\fi\fi\fi}{0}\selectfont
    \ding{#2}%
  }%
  \endgroup
}
\makeatother

\newcommand*\rot{\rotatebox{90}}
\newcommand{\cA}{\contraction{}{\phi}{{}^{1}_{1}\phi_{1}^{2}\phi_{1}^{3}\phi_{1}^{4}}{\phi}}
\newcommand{\cB}{\contraction{}{\phi}{{}^{1}_{1}\phi_{1}^{2}\phi_{1}^{3}\phi_{1}^{4} \phi_{1}^{z}}{\phi}}
\newcommand{\cC}{\contraction{}{\phi}{{}^{1}_{1}\phi_{1}^{2}\phi_{1}^{3}\phi_{1}^{4} \phi_{1}^{z}\phi_{1}^{z}}{\phi}}
\newcommand{\cD}{\contraction{}{\phi}{{}^{1}_{1}\phi_{1}^{2}\phi_{1}^{3}\phi_{1}^{4} \phi_{1}^{z}\phi_{1}^{z}\phi_{1}^{z}}{\phi}}
\newcommand{\dA}{\contraction[2ex]{\phi_{1}^{1}}{\phi}{{}^{2}_{1}\phi_{1}^{3}\phi_{1}^{4}}{\phi}}
\newcommand{\dB}{\contraction[2ex]{\phi_{1}^{1}}{\phi}{{}^{2}_{1}\phi_{1}^{3}\phi_{1}^{4} \phi_{1}^{z}}{\phi}}
\newcommand{\dC}{\contraction[2ex]{\phi_{1}^{1}}{\phi}{{}^{2}_{1}\phi_{1}^{3}\phi_{1}^{4} \phi_{1}^{z}\phi_{1}^{z}}{\phi}}
\newcommand{\dD}{\contraction[2ex]{\phi_{1}^{1}}{\phi}{{}^{2}_{1}\phi_{1}^{3}\phi_{1}^{4} \phi_{1}^{z}\phi_{1}^{z}\phi_{1}^{z}}{\phi}}
\newcommand{\eA}{\contraction[3ex]{\phi_{1}^{1}\phi_{1}^{2}}{\phi}{{}^{3}_{1}\phi_{1}^{4}}{\phi}}
\newcommand{\eB}{\contraction[3ex]{\phi_{1}^{1}\phi_{1}^{2}}{\phi}{{}^{3}_{1}\phi_{1}^{4} \phi_{1}^{z}}{\phi}}
\newcommand{\eC}{\contraction[3ex]{\phi_{1}^{1}\phi_{1}^{2}}{\phi}{{}^{3}_{1}\phi_{1}^{4} \phi_{1}^{z}\phi_{1}^{z}}{\phi}}
\newcommand{\eD}{\contraction[3ex]{\phi_{1}^{1}\phi_{1}^{2}}{\phi}{{}^{3}_{1}\phi_{1}^{4} \phi_{1}^{z}\phi_{1}^{z}\phi_{1}^{z}}{\phi}}
\newcommand{\fA}{\contraction[4ex]{\phi_{1}^{1}\phi_{1}^{2}\phi_{1}^{3}}{\phi}{{}_{1}^{4}\phi}{\phi}}
\newcommand{\fB}{\contraction[4ex]{\phi^{1}\phi_{1}^{2}\phi_{1}^{3}}{\phi}{{}_{1}^{4} \phi_{1}^{z}}{\phi}}
\newcommand{\fC}{\contraction[4ex]{\phi_{1}^{1}\phi_{1}^{2}\phi_{1}^{3}}{\phi}{{}_{1}^{4}\phi_{1}^{z}\phi_{1}^{z}}{\phi}}
\newcommand{\fD}{\contraction[4ex]{\phi_{1}^{1}\phi_{1}^{2}\phi_{1}^{3}}{\phi}{{}_{1}^{4}\phi_{1}^{z}\phi_{1}^{z}\phi_{1}^{z}}{\phi}}
\newcommand{\grE}{\phi_{1}^{1}\phi_{1}^{2}\phi_{1}^{3}\phi_{1}^{4}\phi_{1}^{z}\phi_{1}^{z}\phi_{1}^{z}\phi_{1}^{z}}
\newcommand{\hA}{\contraction{}{\phi}{{}^{1}_{1}\phi_{1}^{2}\phi_{2}^{3}\phi_{2}^{4}}{\phi}}
\newcommand{\hB}{\contraction{}{\phi}{{}^{1}_{1}\phi_{1}^{2}\phi_{2}^{3}\phi_{2}^{4}\phi_{1}^{z}}{\phi}}
\newcommand{\iA}{\contraction[2ex]{\phi^{1}_{1}}{\phi}{{}_{1}^{2}\phi_{2}^{3}\phi_{2}^{4}}{\phi}}
\newcommand{\iB}{\contraction[2ex]{\phi^{1}_{1}}{\phi}{{}_{1}^{2}\phi_{2}^{3}\phi_{2}^{4}\phi_{1}^{z}}{\phi}}
\newcommand{\jC}{\contraction[3ex]{\phi^{1}_{1}\phi_{1}^{2}}{\phi}{{}_{2}^{3}\phi_{2}^{4}\phi_{1}^{z}\phi_{1}^{z}}{\phi}}
\newcommand{\jD}{\contraction[3ex]{\phi^{1}_{1}\phi_{1}^{2}}{\phi}{{}_{2}^{3}\phi_{2}^{4}\phi_{1}^{z}\phi_{1}^{z}\phi_{2}^{z}}{\phi}}
\newcommand{\kC}{\contraction[4ex]{\phi^{1}_{1}\phi_{1}^{2}\phi_{2}^{3}}{\phi}{{}_{2}^{4}\phi_{1}^{z}\phi_{1}^{z}}{\phi}}
\newcommand{\kD}{\contraction[4ex]{\phi^{1}_{1}\phi_{1}^{2}\phi_{2}^{3}}{\phi}{{}_{2}^{4}\phi_{1}^{z}\phi_{1}^{z}\phi_{2}^{z}}{\phi}}
\newcommand{\grF}{\phi_{1}^{1}\phi_{1}^{2}\phi_{2}^{3}\phi_{2}^{4}\phi_{1}^{z}\phi_{1}^{z}\phi_{2}^{z}\phi_{2}^{z}}
\newcommand{\laA}{\contraction{}{\phi}{{}_{1}^{1}\phi_{1}^{2}}{\phi}}
\newcommand{\laB}{\contraction{}{\phi}{{}_{1}^{1}\phi_{1}^{2}\phi_{1}^{z}}{\phi}}
\newcommand{\laC}{\contraction{}{\phi}{{}_{1}^{1}\phi_{1}^{2}\phi_{1}^{z}\phi_{1}^{z}}{\phi}}
\newcommand{\laD}{\contraction{}{\phi}{{}_{1}^{1}\phi_{1}^{2}\phi_{1}^{z}\phi_{1}^{z}\phi_{1}^{z}}{\phi}}

\newcommand{\lbA}{\contraction[2ex]{\phi_{1}^{1}}{\phi}{{}_{1}^{2}}{\phi}}
\newcommand{\lbB}{\contraction[2ex]{\phi_{1}^{1}}{\phi}{{}_{1}^{2}\phi_{1}^{z}}{\phi}}
\newcommand{\lbC}{\contraction[2ex]{\phi_{1}^{1}}{\phi}{{}_{1}^{2}\phi_{1}^{z}\phi_{1}^{z}}{\phi}}
\newcommand{\lbD}{\contraction[2ex]{\phi_{1}^{1}}{\phi}{{}_{1}^{2}\phi_{1}^{z}\phi_{1}^{z}\phi_{1}^{z}}{\phi}}

\newcommand{\lcA}{\contraction[3ex]{\phi_{1}^{1}\phi_{1}^{2}}{\phi}{{}_{1}^{z}}{\phi}}
\newcommand{\lcB}{\contraction[3ex]{\phi_{1}^{1}\phi_{1}^{2}}{\phi}{{}_{1}^{z}\phi_{1}^{z}}{\phi}}
\newcommand{\lcC}{\contraction[3ex]{\phi_{1}^{1}\phi_{1}^{2}}{\phi}{{}_{1}^{z}\phi_{1}^{z}\phi_{1}^{z}}{\phi}}

\newcommand{\ldA}{\contraction[3ex]{\phi_{1}^{1}\phi_{1}^{2}\phi_{1}^{z}}{\phi}{{}_{1}^{z}}{\phi}}
\newcommand{\ldB}{\contraction[3ex]{\phi_{1}^{1}\phi_{1}^{2}\phi_{1}^{z}}{\phi}{{}_{1}^{z}\phi_{1}^{z}}{\phi}}

\newcommand{\leA}{\contraction[3ex]{\phi_{1}^{1}\phi_{1}^{2}\phi_{1}^{z}\phi_{1}^{z}}{\phi}{{}_{1}^{z}}{\phi}}

\newcommand{\grG}{\phi_{1}^{1}\phi_{1}^{2}\phi_{1}^{z}\phi_{1}^{z}\phi_{1}^{z}\phi_{1}^{z}}
\title{Two interacting scalar fields:\\ practical renormalization}
\date{}
\author{S. R. Ju\'{a}rez Wysozka\\
\small{Departamento de F\'{\i}sica, Escuela Superior de F\'{\i}sica y Matem\'{a}ticas}\\[\sk]
\small{Instituto Polit\'{e}cnico Nacional. U.P Adolfo L\'{o}pez Mateos}\\[\sk]
\small{C.P. 07738. Ciudad de M\'{e}xico, M\'{e}xico}\\[3pt]
Piotr Kielanowski, Edgar Uribe Longoria\\
\small{Departamento de F\'{\i}sica, Centro de Investigaci\'{o}n y Estudios
Avanzados}\\[\sk]
\small{C.P. 07000, Ciudad de M\'{e}xico, M\'{e}xico}\\[3pt]
Liliana Vázquez Mercado\\[\sk]
\small{Departamento de F{\'\i}sica, Centro Universitario de Ciencias Exactas e Ingenier{\'\i}as}\\[\sk]
\small{Universidad de Guadalajara. Av. Revoluci\'on 1500, Colonia Ol{\'\i}mpica, C.P. 44430}\\[\sk] \small{Guadalajara, Jalisco, México}\\
\small{\textit{E-mail:} \texttt{rebecajw@gmail.com, piotr.kielanowski@cinvestav.mx}}\\[\sk] \small{\texttt{edgar.uribe@cinvestav.mx, liliana.vmercado@academicos.udg.mx}}}

\begin{document}
\maketitle
\begin{abstract}
  The main theme of the paper is the detailed discussion of the
  renormalization of the quantum field theory comprising two
  interacting scalar fields. The potential of the model is the
  fourth-order homogeneous polynomial of the fields, symmetric with
  respect to the transformation $\phi_{i}\rightarrow{-\phi_{i}}$. We
  determine the Feynman rules for the model and then we present a
  detailed discussion of the renormalization of the theory at one
  loop. Next, we derive the one loop renormalization group equations
  for the running masses and coupling constants. At the level of two
  loops, we use the \textit{FeynArts} package of \textit{Mathematica}
  to generate the two loops Feynman diagrams and calculate in detail
  the \textit{setting sun} diagram.
\end{abstract}

\section{Introduction}\label{sec:intro}
Renormalization in quantum field theory is an indispensable tool to
obtain precise results at higher orders of perturbation theory. The
full process of renormalization is a complex task, and it is best
learned with examples. We will consider here a theory of two real,
interacting scalar fields described by the following Lagrangian
density
\begin{equation}
  \label{eq:1}
  \begin{gathered} \mathcal{L}(\phi_{1},\phi_{2})=\frac{1}{2}\partial_{\mu}\phi_{1}\partial^{\mu}\phi_{1} -\frac{1}{2}m_{1}^{2}\phi_{1}^{2} + \frac{1}{2}\partial_{\mu}\phi_{2}\partial^{\mu}\phi_{2} -\frac{1}{2}m_{2}^{2}\phi_{2}^{2}\\ -\frac{\lambda_{1}}{4!}\phi_{1}^{4} -\frac{\lambda_{2}}{4!}\phi_{2}^{4} -\frac{\lambda_{3}}{4}\phi_{1}^{2}\phi_{2}^{2}.
  \end{gathered}
\end{equation}
The interaction potential density in~\eqref{eq:1}
\begin{equation}
  \mathcal{V}(\phi_{1},\phi_{2})=\frac{\lambda_{1}}{4!}\phi_{1}^{4} +\frac{\lambda_{2}}{4!}\phi_{2}^{4} +\frac{\lambda_{3}}{4}\phi_{1}^{2}\phi_{2}^{2}\label{eq:3}
\end{equation}
should be positive definite, so we assume that the coupling
constans $\lambda_{1}$, $\lambda_{2}$ and $\lambda_{3}$ fulfill the
following conditions
\begin{equation}
  \label{eq:4}
  \lambda_{1}>0,\quad\lambda_{2}>0,\quad\lambda_{1}\lambda_{2}-9\lambda_{3}^{2}>0.
\end{equation}
From the Lagrangian density~\eqref{eq:1} one obtains the classical
equations of motion for the fields $\phi_{1}$ and $\phi_{2}$
\begin{subequations}\label{eq:2}
  \begin{align}
    (\dalambert+m_{1}^{2})\phi_{1}+\frac{\lambda_{1}}{3!} \phi_{1}^{3} + \frac{\lambda_{3}}{2}\phi_{1}\phi_{2}^{2}=0,\label{eq:2a}\\
    (\dalambert+m_{2}^{2})\phi_{2}+\frac{\lambda_{2}}{3!} \phi_{2}^{3} + \frac{\lambda_{3}}{2}\phi_{1}^{2}\phi_{2}=0.\label{eq:2b}
  \end{align}
\end{subequations}

In our paper, we concentrate on the renormalization of the model
described by the Lagrangian density~\eqref{eq:1}. The contents of the
paper are the following: In Section~\ref{sec:feynman-rules}, we recall
the notion of the Green's function and we establish the Feynman rules
of the considered model. In Section~\ref{sec:two-four-point}, we
analyze the divergent diagrams for the two and four-point Green's
functions. We draw the corresponding Feynman diagrams and derive the
analytical expressions corresponding to these diagrams. In Section
\ref{sec:calc-loop-integr}, we discuss the mathematical methods that
are used in the calculation of the Feynman diagrams, containing the
loop integrals. We discuss the Feynman prescription, Wick rotation and
two methods of regularization of the divergent integrals:
Pauli-Villars regularization and dimensional
regularization. Section~\ref{sec:renormalization} is devoted to the
discussion of the one loop renormalization of our model. First, we
discuss the regularization of the one loop divergent diagrams and
then, we explain how the divergencies are removed in the one loop
diagrams, through the procedure of renormalization. In
Section~\ref{sec:renorm-group-at}, we derive the one loop
renormalization group equations for the masses $m_{1}$, $m_{2}$ and
the coupling constants $\lambda_{1}$, $\lambda_{2}$ and~$\lambda_{3}$
in our theory. Section~\ref{sec:renorm-at-two} is devoted to the
discussion of the two loop renormalization of the model. After
Conclusions, we include several appendices, where some additional
details of the calculations are presented. We also give the listings
of the \textit{Mathematica} programs, that were used in the discussion
of the two loop renormalization.

\section{Feynman rules}
\label{sec:feynman-rules}

\begin{figure}[!hb]\centering
  \includegraphics[width=0.9\linewidth]{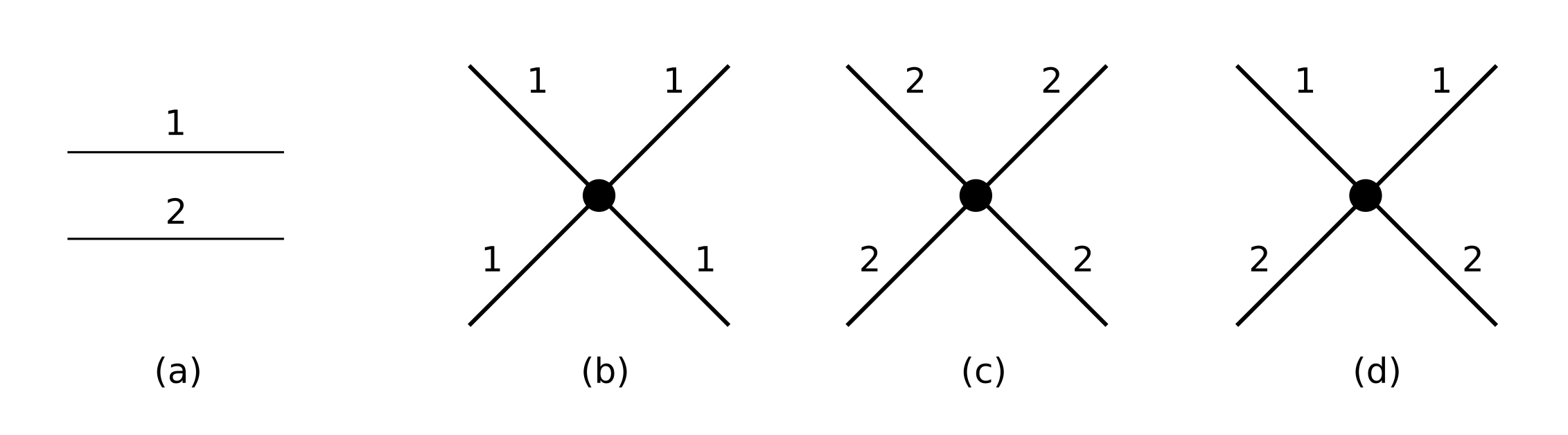}
  \caption{\label{fig:1} Elements of the Feynman diagrams, (a)
    propagators of particles~1 and~2, (b) vertex with the coupling
    constant $\lambda_{1}$, (c) vertex with the coupling constant
    $\lambda_{2}$, (d) vertex with the coupling constant
    $\lambda_{3}$.}
\end{figure}

Feynman rules for the Feynman diagrams in the momentum space are
determined from the Lagrangian density and serve to derive the
analytical expressions for the amplitudes of the given processes. The
Lagrangian density~\eqref{eq:1} contains two fields $\phi_{1}$ and
$\phi_{2}$ and three interaction vertices, so the Feynman diagrams
will contain the propagators for the fields $\phi_{1}$ and $\phi_{2}$
and the vertices corresponding to the coupling constants
$\lambda_{1}$, $\lambda_{2}$ and $\lambda_{3}$. The graphical
representations of the basic elements of the Feynman diagrams stemming
from the Lagrangian density are given in Fig.~\ref{fig:1}.

The Green's functions $G^{(n)}(x_{1},x_{2},\ldots,x_{n})$ or $n$~point
functions are the quantities from which the $S$-matrix elements are
calculated. They are the vacuum expectation values of the time ordered
product of fields and are equal to
\begin{equation}
  \label{eq:5}
  G^{(n)}_{i_{1}\cdots i_{n}}(x_{1},x_{2},\ldots,x_{n})=\mel{0}{T(\phi_{i_{1}}(x_{1}),\phi_{i_{2}}(x_{2}), \ldots,\phi_{i_{n}}(x_{n}))}{0}.
\end{equation}
Here $T$ denotes the time ordered product of the fields.

The Green's function in the momentum space
$G^{(n)}_{i_{1}\cdots i_{n}}(p_{1},p_{2},\ldots,p_{n})$ is the Fourier
transform of the Green's function
$G^{(n)}_{i_{1}\cdots i_{n}}(x_{1},x_{2},\ldots,x_{n})$
\begin{multline}
  \label{eq:10}
  (2\pi)^{4}\delta(p_{1}+\cdots + p_{n})
  G^{(n)}_{i_{1}\cdots i_{n}}(p_{1},p_{2},\ldots,p_{n})\\
  =\int\prod_{k=1}^{n}\dd[4]x_{k}x^{-ip_{k}x_{k}}G^{(n)}_{i_{1}\cdots
    i_{n}}(x_{1},x_{2},\ldots,x_{n}).
\end{multline}
The amputated Green's function is obtained from the momentum space
Green's function by removing the propagators $\Delta_{k}(p_{k})$ of
external fields
\begin{equation}
  \label{eq:16}
  ^{\text{amp}}G^{(n)}_{i_{1}\cdots i_{n}}(p_{1},p_{2},\ldots,p_{n}) =
  \prod_{k=1}^{n}\Bigg[\frac{1}{i\Delta_{k}(p_{k})}\Bigg] G^{(n)}_{i_{1}\cdots i_{n}}(p_{1},p_{2},\ldots,p_{n}).
\end{equation}

The $S$-matrix elements are obtained from the
Lehmann-Symanzik-Zi\-mmer\-mann reduction formula
\begin{multline}
  \label{eq:9}
  \mel{q_{1},\ldots,q_{n}}{S}{p_{1},p_{2}}=
  \Big[i\int\dd[4]x_{1}e^{-ip_{1}x_{1}}(\dalambert+m_{1}^{2})\Big]\Big[i\int\dd[4]x_{2}e^{-ip_{2}x_{2}}(\dalambert+m_{2}^{2})\Big]\\
  \times\Big[i\int\dd[4]z_{1}e^{iq_{1}z_{1}}(\dalambert+M_{i_{1}}^{2})\Big]
  \cdots\Big[i\int\dd[4]z_{n}e^{iq_{n}z_{n}}(\dalambert+M_{i_{n}}^{2})\Big]\\
  \times\mel{0}
  {T(\phi_{1}(x_{1}),\phi_{2}(x_{2}),\phi_{i_{1}}(z_{1}),\ldots,\phi_{i_{n}}(z_{n}))}{0}
\end{multline}
and for the Green's function the following perturbative expansion holds
\begin{multline}\label{eq:6}
  G^{(n)}_{i_{1}\cdots i_{n}} (x_{1},x_{2},\ldots,x_{n})=\sum_{l=0}^{\infty}\frac{(-i)^{l}}{l!} \int_{-\infty}^{\infty} \dd[4]{z_{1}}\cdots\dd[4]{z_{l}}\\
  \times
  \mel{0}{T(\phi_{i_{1}}^{\text{I}}(x_{1}),\ldots,\phi_{i_{n}}^{\text{I}}(x_{n}),
    \mathcal{V}(\phi_{1}^{\text{I}}(z_{1}),\phi_{2}^{\text{I}}(z_{1})),\ldots,
    \mathcal{V}(\phi_{1}^{\text{I}}(z_{l}),\phi_{2}^{\text{I}}(z_{l}))}{0}_{c}.
\end{multline}
The superscript~I means that the fields are taken in the interaction
picture and the subscript $c$ denotes the connected part.

The Green's functions contain the vacuum expectation value of time
ordered product of the fields. In order to calculate this vacuum
expectation value the time ordered product is expanded in terms of
the normal products of the fields. The Feynman diagrams are the
graphical representation of terms of this expansion.

The propagators are calculated from the $G_{i_{1}i_{2}}^{(2)}$ Green's
functions
\begin{equation*}
  \mel{0}{T(\phi_{i_{1}}^{\text{I}}(x)\phi_{i_{2}}^{\text{I}}(y))}{0}
  =i\Delta_{i_{1}}(x-y)\delta_{i_{1}i_{2}}=\int\frac{\dd[4]{p}}{(2\pi)^{4}}e^{ip(x-y)}\frac{i}{p^{2}-m_{i_{1}}^{2}+i\epsilon}\delta_{i_{1}i_{2}}.
\end{equation*}
The Green's function $G^{(2)}_{12}$ vanishes, because the Lagrangian
density~\eqref{eq:1} contains only even powers of fields and the
symmetry $\phi_{i} \rightarrow-\phi_{i}$ holds.

The vertices are obtained from the $G^{(4)}$ Green's functions and the
non\linebreak  vanishing functions are
$G^{(4)}_{1111}(x_{1},x_{2},x_{3},x_{4})$,
$G^{(4)}_{2222}(x_{1},x_{2},x_{3},x_{4})$ and \linebreak
$G^{(4)}_{1122}(x_{1},x_{2},x_{3},x_{4})$. The vertices of the Feynman
diagrams obtained from these functions with the analytic
correspondence are given in Table~\ref{tab:1}.

The Green's functions $G^{(4)}_{1111}(x_{1},x_{2},x_{3},x_{4})$ and
$G^{(4)}_{1122}(x_{1},x_{2},x_{3},x_{4})$ at the lowest order
including interactions are obtained from the equations
\begin{align}
     G^{(4)}_{1111}(x_{1},x_{2},&x_{3},x_{4})\nonumber\\
    \label{eq:9a}
   =\int_{-\infty}^{\infty}\dd[4]{z}&\mel{0}
    {T(\phi_{1}^{\text{I}}(x_{1}),\phi_{1}^{\text{I}}(x_{2}),
      \phi_{1}^{\text{I}}(x_{3}),\phi_{1}^{\text{I}}(x_{4}),
      \mathcal{V}(\phi_{1}^{\text{I}}(z),\phi_{2}^{\text{I}}(z))) }
    {0}_{c},\\
    G^{(4)}_{1122}(x_{1},x_{2},&x_{3},x_{4})\nonumber\\
    \label{eq:9b}
    =\int_{-\infty}^{\infty}\dd[4]{z}&\mel{0}
    {T(\phi_{1}^{\text{I}}(x_{1}),\phi_{1}^{\text{I}}(x_{2}),
      \phi_{2}^{\text{I}}(x_{3}),\phi_{2}^{\text{I}}(x_{4}),
      \mathcal{V}(\phi_{1}^{\text{I}}(z),\phi_{2}^{\text{I}}(z))) }
    {0}_{c}.
\end{align}
and $G^{(4)}_{2222}(x_{1},x_{2},x_{3},x_{4})$ is obtained from
$G^{(4)}_{1111}(x_{1},x_{2},x_{3},x_{4})$ by the change of indices
$1\leftrightarrow2$.

The integrand of Eq.~\eqref{eq:9a} contains the term
$\mathcal{V}(\phi_{1}^{\text{I}}(z),\phi_{2}^{\text{I}}(z)))$ given in
Eq.~\eqref{eq:3}, which contains three terms. The connected diagrams
are obtained only from the term $(\lambda_{1}\phi_{1}^{4})/4!$, so
only this term has to be included in the calculation. A similar
situation occurs in Eq.~\eqref{eq:9b} which corresponds to the term
$(\lambda_{3}\phi_{1}^{2}\phi_{2}^{2})/4$ in Eq.~\eqref{eq:3}.

The terms in the Lagrangian density~\eqref{eq:1} corresponding to the
vertices contain the factors $1/4!$ and $1/4$. In
Appendix~\ref{sec:wicks-theorem}, we present Wick's theorem, which is
used in Appendices~\ref{sec:calculation-g4_1111}
and~\ref{sec:calculation-g4_1122} to show that these factors are
canceled in Feynman diagrams and the Feynman rules are such as those
given in Table~\ref{tab:1}.~\footnote{In case of diagrams with loops
  the situation becomes more complicated and it is necessary to
  introduce the symmetry factors.}

\begin{table}[ht]\centering
  \caption{\label{tab:1}Analytical expressions corresponding to the
    elements of the Feynman diagrams.}\vspace*{10pt}
  \begin{tabular}{c|rcl}
    \multirow{2}{*}{\rot{\parbox[c]{2.4cm}{\centering Propagators}}}&
                                                                      \begin{minipage}[c]{0.2\linewidth}
                                                                        \includegraphics[width=\linewidth,keepaspectratio]{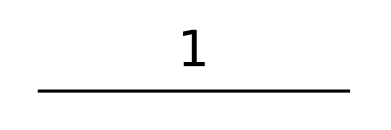}
                                                                      \end{minipage}&
                                                                                      $\longrightarrow$&$\dfrac{i}{p^{2}-m_{1}^{2}+i\epsilon}$\\
                                                                    &\begin{minipage}[c][20mm][c]{0.2\linewidth}
                                                                       \includegraphics[width=\linewidth,keepaspectratio]{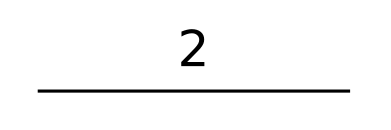}
                                                                     \end{minipage}&
                                                                                     $\longrightarrow$&$\dfrac{i}{p^{2}-m_{2}^{2}+i\epsilon}$\\[6pt]
    \hline \multirow{3}{*}{\rot{\parbox[c]{6cm}{\centering Vertices}}}
                                                                    &\begin{minipage}[c][30mm][c]{0.2\linewidth}
                                                                       \includegraphics[width=\linewidth,keepaspectratio]{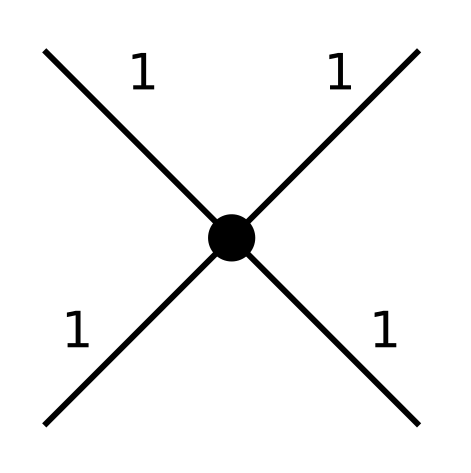}
                                                                     \end{minipage}&
                                                                                     $\longrightarrow$&$-i\lambda_{1}$\\
                                                                    &\begin{minipage}[c]{0.2\linewidth}
                                                                       \includegraphics[width=\linewidth,keepaspectratio]{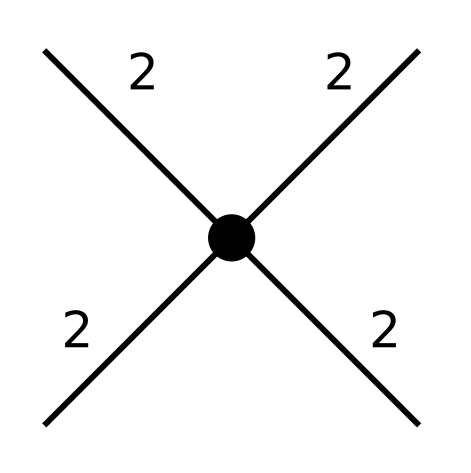}
                                                                     \end{minipage}&
     $\longrightarrow$&$-i\lambda_{2}$,\\
&\begin{minipage}[c]{0.2\linewidth}
\includegraphics[width=\linewidth,keepaspectratio]{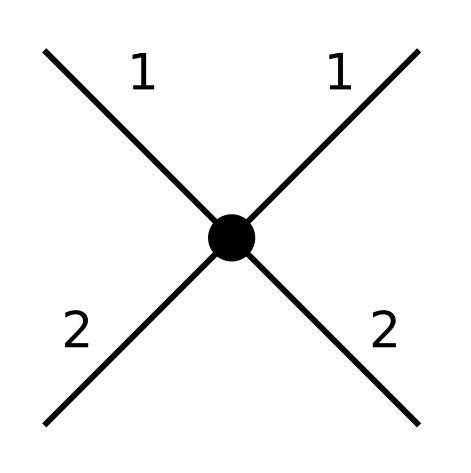}
\end{minipage}&
$\longrightarrow$&$-i\lambda_{3}$
  \end{tabular}
\end{table}

To conclude this section let us list the rules for the construction
of the analytical expression of the transition amplitude in the
momentum representation from the Feynman diagram:
\begin{itemize}
\item The amplitude contains an overall factor of $i$.
\item Each vertex produces the factor $-i\lambda_{j}$: $j=1$ for four
  particles of type~1, $j=2$ for four particles of type~2, $j=3$ for
  two particles of type~1 and two particles of type~2.
\item For each internal line of a particle of the type~$j$ with
  four-momentum $k$ there corresponds the propagator
  $\displaystyle\frac{i}{k^{2}-m_{j}^{2}+i\epsilon}$.
\item Four momentum is conserved at each vertex.
\item Integrate over each internal four momentum within a loop (not
  fixed by the four momentum conservation) by applying the integral
  $\int\frac{\dd^{4}k}{(2\pi)^{4}}$.
\end{itemize}

\section{Two and four point divergent diagrams}
\label{sec:two-four-point}

In this section we will discuss the lowest order divergent diagrams and
the analytical expressions corresponding to them.

\subsection{Diagrams for the propagators}
\label{sec:diagrams-propagators}

\begin{figure}[!hb]
  \includegraphics[width=0.58\linewidth]{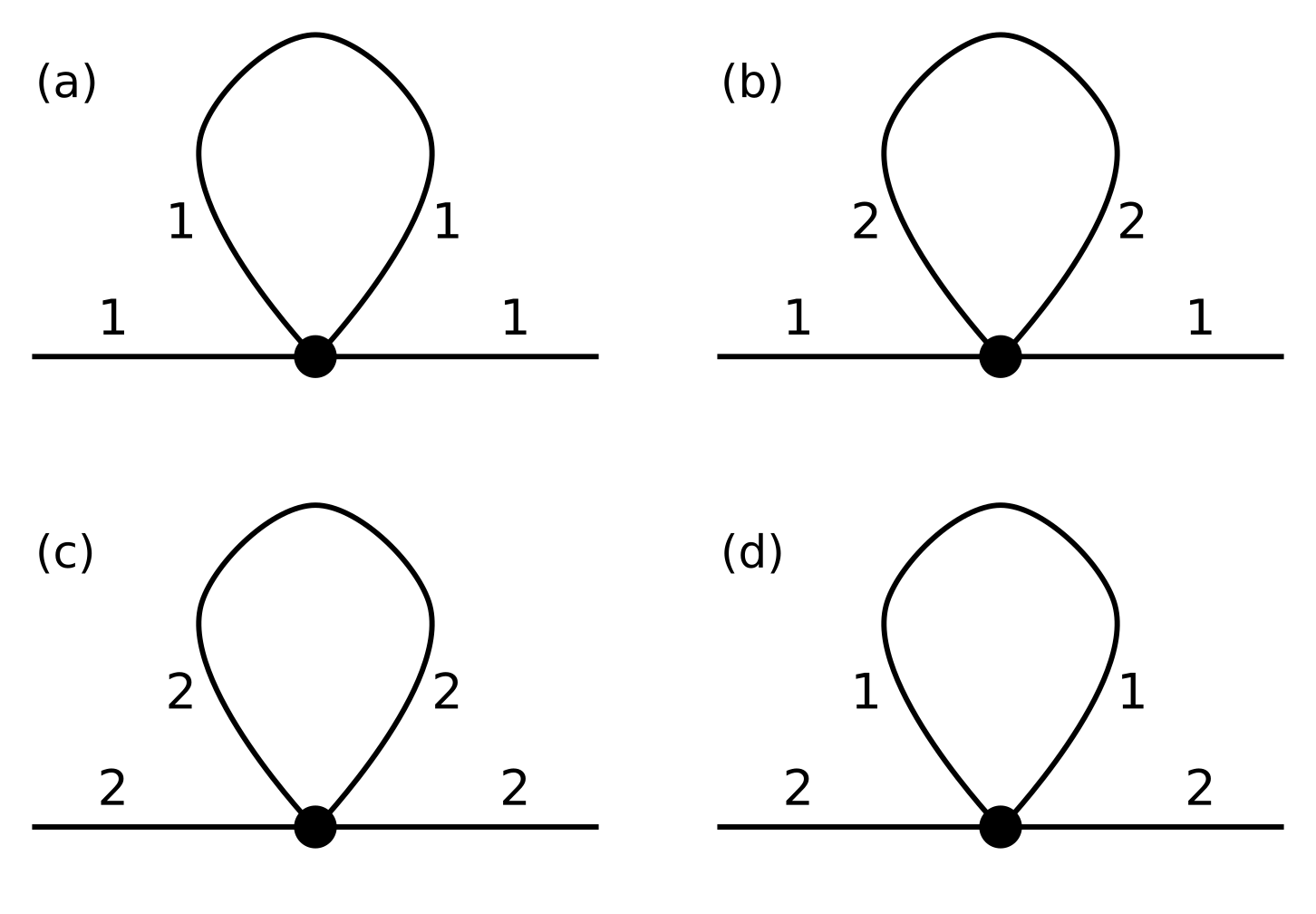}\centering
  \caption{\label{fig:2} Divergent two point Feynman diagrams; (a)
    propagator of particle~1 with a loop containing particle~1, (b)
    propagator of particle~1 with a loop containing particle~2, (c)
    propagator of particle~2 with a loop containing particle~2, (d)
    propagator of particle~2 with a loop containing particle~1. The full
    propagator requires the summation of different diagrams.}
\end{figure}

In Fig.~\ref{fig:2} there are the first order corrections to the
propagators of particles~1 and~2. We will consider in detail only the
diagrams for the propagator of particle~1, because the results for
particle~2 are obtained from those of particle~1 by a simple change of
indices.

The analytical expression in the coordinate space obtained from
Eq.~\eqref{eq:6} that generates the diagrams in Fig.~\ref{fig:2}~(a)
and~(b) is
\begin{equation}
  \label{eq:17}
  (-i)\int_{-\infty}^{\infty}\dd[4]z_{1}\mel{0}{T(\phi_{1}^{\text{I}}(x_{1}), \phi_{1}^{\text{I}}(x_{2}), \mathcal{V}(\phi_{1}^{\text{I}}(z_{1}), \phi_{2}^{\text{I}}(z_{1})))}{0}_{c}.
\end{equation}
After inserting the explicit form of the potential
$\mathcal{V}(\phi_{1}^{\text{I}}(z_{1}), \phi_{2}^{\text{I}}(z_{1}))$
from Eq.~\eqref{eq:3} we obtain separately for diagrams in
Fig.~\ref{fig:2}~(a) and~(b)
\begin{subequations}\label{eq:18}
  \begin{align}
    \label{eq:18a} &(-i)\frac{\lambda_{1}}{4!}\int_{-\infty}^{\infty}\dd[4]z_{1}\mel{0} {T(\phi_{1}^{\text{I}}(x_{1}), \phi_{1}^{\text{I}}(x_{2}), {\big(\phi_{1}^{\text{I}}(z_{1})\big)}^{4})}{0}_{c},\\
    \label{eq:18b}
&(-i)\frac{\lambda_{3}}{4}\int_{-\infty}^{\infty}\dd[4]z_{1}\mel{0} {T(\phi_{1}^{\text{I}}(x_{1}), \phi_{1}^{\text{I}}(x_{2}), {\big(\phi_{1}^{\text{I}}(z_{1})\big)}^{2},
{\big(\phi_{2}^{\text{I}}(z_{1})\big)}^{2}) }{0}_{c}.
  \end{align}
\end{subequations}
From these equations we obtain in Appendix~\ref{sec:symm-fact-diagr}
that the \textit{symmetry factors}\footnote{Also known as the
  \textit{combinatorial factors}.} for the diagrams in
Fig.~\ref{fig:2} are equal $1/2 $.

In the momentum space we obtain the following expressions for the
diagrams in Fig.~\ref{fig:2}
\begin{subequations}\label{eq:19}
  \begin{align}
    \label{eq:19a}
    &(i)(-i\lambda_{1})\frac{1}{2}\int\frac{\dd[4]k}{(2\pi)^{4}} \frac{i}{k^{2}-m_{1}^{2}+i\epsilon} =\frac{\lambda_{1}}{2} \int\frac{\dd[4]k}{(2\pi)^{4}} \frac{i}{k^{2}-m_{1}^{2}+i\epsilon},\\
    \label{eq:19b}
    &(i)(-i\lambda_{3})\frac{1}{2}\int\frac{\dd[4]k}{(2\pi)^{4}} \frac{i}{k^{2}-m_{2}^{2}+i\epsilon} =\frac{\lambda_{3}}{2} \int\frac{\dd[4]k}{(2\pi)^{4}} \frac{i}{k^{2}-m_{2}^{2}+i\epsilon},\\
    \label{eq:19c}
    &(i)(-i\lambda_{2})\frac{1}{2}\int\frac{\dd[4]k}{(2\pi)^{4}} \frac{i}{k^{2}-m_{2}^{2}+i\epsilon} =\frac{\lambda_{2}}{2} \int\frac{\dd[4]k}{(2\pi)^{4}} \frac{i}{k^{2}-m_{2}^{2}+i\epsilon},\\
    \label{eq:19d}
    &(i)(-i\lambda_{3})\frac{1}{2}\int\frac{\dd[4]k}{(2\pi)^{4}} \frac{i}{k^{2}-m_{1}^{2}+i\epsilon} =\frac{\lambda_{3}}{2} \int\frac{\dd[4]k}{(2\pi)^{4}} \frac{i}{k^{2}-m_{1}^{2}+i\epsilon}.
  \end{align}
\end{subequations}
The $1/2$ in Eqs.~\eqref{eq:19} is the symmetry factor and the
integrals are quadratically divergent.

\subsection{Diagrams for the vertices}
\label{sec:diagrams-vertices}

\begin{figure}[!hb]
  \includegraphics[width=0.95\linewidth]{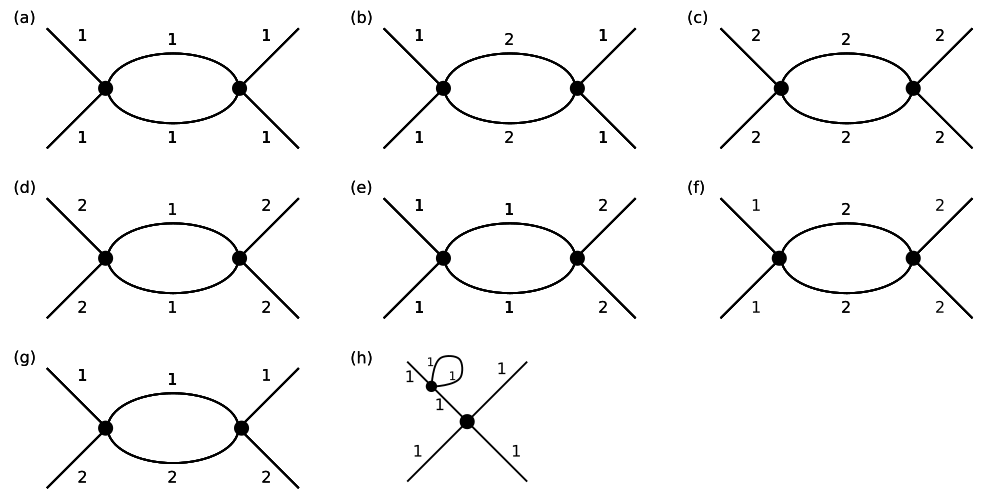}\centering
  \caption{\label{fig:3} Patterns for the divergent four point Feynman
    diagrams; (a), (b) correction to the vertex with the coupling
    constant $\lambda_{1}$; (c), (d) correction to the vertex with the
    coupling constant $\lambda_{2}$; (e), (f), (g) correction to the
    vertex with the coupling constant $\lambda_{3}$; (h) this type of
    diagram is not included in the propagator correction because it is
    not one particle irreducible diagram (it can be divided into two
    separate diagrams by cutting only one internal line). To each
    pattern corresponds more than one Feynman diagram. See
    Appendix~\ref{sec:symm-fact-diagr-1}.}
\end{figure}

The analytical expressions in the coordinate space obtained from
Eq.~\eqref{eq:6} that generate the diagrams in Fig.~\ref{fig:3}
are
\renewcommand*{\sk}{0.9}
\begin{subequations}\label{eq:22}
  \begin{align} &\Scale[\sk]{\frac{(-i)^{2}}{2!}\int_{-\infty}^{\infty}\dd[4]z_{1}\dd[4]z_{2}}\nonumber\\ &\Scale[\sk]{\times\mel{0}{T(\phi_{1}^{\text{I}}(x_{1}), \phi_{1}^{\text{I}}(x_{2}), \phi_{1}^{\text{I}}(x_{3}) \phi_{1}^{\text{I}}(x_{4}),  \mathcal{V}(\phi_{1}^{\text{I}}(z_{1}), \phi_{2}^{\text{I}}(z_{1})), \mathcal{V}(\phi_{1}^{\text{I}}(z_{2}), \phi_{2}^{\text{I}}(z_{2})))}{0}_{c},} \label{eq:22a}\\ &\Scale[\sk]{\frac{(-i)^{2}}{2!}\int_{-\infty}^{\infty}\dd[4]z_{1}\dd[4]z_{2}}\nonumber\\ &\Scale[\sk]{\times\mel{0}{T(\phi_{2}^{\text{I}}(x_{1}), \phi_{2}^{\text{I}}(x_{2}), \phi_{2}^{\text{I}}(x_{3}) \phi_{2}^{\text{I}}(x_{4}),  \mathcal{V}(\phi_{1}^{\text{I}}(z_{1}), \phi_{2}^{\text{I}}(z_{1})), \mathcal{V}(\phi_{1}^{\text{I}}(z_{2}), \phi_{2}^{\text{I}}(z_{2})))}{0}_{c},} \label{eq:22b}\\ &\Scale[\sk]{\frac{(-i)^{2}}{2!}\int_{-\infty}^{\infty}\dd[4]z_{1}\dd[4]z_{2}}\nonumber\\ &\Scale[\sk]{\times\mel{0}{T(\phi_{1}^{\text{I}}(x_{1}), \phi_{1}^{\text{I}}(x_{2}), \phi_{2}^{\text{I}}(x_{3}) \phi_{2}^{\text{I}}(x_{4}),  \mathcal{V}(\phi_{1}^{\text{I}}(z_{1}), \phi_{2}^{\text{I}}(z_{1})), \mathcal{V}(\phi_{1}^{\text{I}}(z_{2}), \phi_{2}^{\text{I}}(z_{2})))}{0}_{c}.} \label{eq:22c}
  \end{align}
\end{subequations}\vspace*{-10pt}
\\
After inserting the explicit form of the potential
$\mathcal{V}(\phi_{1}^{\text{I}}(z_{1}), \phi_{2}^{\text{I}}(z_{1}))$
we obtain separately for the diagrams in Fig.~\ref{fig:3} (a)--(g)
\begin{subequations}\label{eq:23}
  \begin{align}
        &(-i)^{2}\left(\frac{\lambda_{1}}{4!}\right)^{2}\frac{1}{2!} \int_{-\infty}^{\infty} \dd[4]z_{1}\dd[4]z_{2}\nonumber\\
    & \pushright{\times\mel{0}{T(\phi_{1}^{\text{I}}(x_{1}), \phi_{1}^{\text{I}}(x_{2}),\phi_{1}^{\text{I}}(x_{3}), \phi_{1}^{\text{I}}(x_{4}), {\big(\phi_{1}^{\text{I}}(z_{1})\big)}^{4}, {\big(\phi_{1}^{\text{I}}(z_{2})\big)}^{4})}{0}_{c},}    \label{eq:23a}\\
    &(-i)^{2}\left(\frac{\lambda_{3}}{4}\right)^{2} \frac{1}{2!}\int_{-\infty}^{\infty} \dd[4]z_{1}\dd[4]z_{2}\nonumber\\
        & \phantom{AAAAA}\times\langle0\vert T(\phi_{1}^{\text{I}}(x_{1}), \phi_{1}^{\text{I}}(x_{2}),\phi_{1}^{\text{I}}(x_{3}), \phi_{1}^{\text{I}}(x_{4}), \nonumber\\ &\pushright{\big(\phi_{1}^{\text{I}}(z_{1})\big)^{2},\big(\phi_{2}^{\text{I}}(z_{1})\big)^{2},
\big(\phi_{1}^{\text{I}}(z_{2})\big)^{2}\big(\phi_{2}^{\text{I}}(z_{2})\big)^{2}) \vert0\rangle_{c},}    \label{eq:23b}\\
    &(-i)^{2}\left(\frac{\lambda_{2}}{4!}\right)^{2} \frac{1}{2!}\int_{-\infty}^{\infty} \dd[4]z_{1}\dd[4]z_{2}\nonumber\\
    & \pushright{\times\mel{0}{T(\phi_{2}^{\text{I}}(x_{1}), \phi_{2}^{\text{I}}(x_{2}),\phi_{2}^{\text{I}}(x_{3}), \phi_{2}^{\text{I}}(x_{4}), {\big(\phi_{2}^{\text{I}}(z_{1})\big)}^{4}, {\big(\phi_{2}^{\text{I}}(z_{2})\big)}^{4})}{0}_{c},}    \label{eq:23c}\\
    &(-i)^{2}\left(\frac{\lambda_{3}}{4}\right)^{2} \frac{1}{2!}\int_{-\infty}^{\infty} \dd[4]z_{1}\dd[4]z_{2}\nonumber\\
    & \phantom{AAAAA}\times\langle0\vert T(\phi_{2}^{\text{I}}(x_{1}), \phi_{2}^{\text{I}}(x_{2}),\phi_{2}^{\text{I}}(x_{3}), \phi_{2}^{\text{I}}(x_{4}), \nonumber\\ &\pushright{\big(\phi_{1}^{\text{I}}(z_{1})\big)^{2},\big(\phi_{2}^{\text{I}}(z_{1})\big)^{2},
\big(\phi_{1}^{\text{I}}(z_{2})\big)^{2}\big(\phi_{2}^{\text{I}}(z_{2})\big)^{2}) \vert0\rangle_{c},}    \label{eq:23d}\\
   &(-i)^{2}\frac{\lambda_{1}}{4!}\frac{\lambda_{3}}{4} \frac{2}{2!}\int_{-\infty}^{\infty} \dd[4]z_{1}\dd[4]z_{2}\nonumber\\
        & \phantom{AAAAA}\times\langle0\vert T(\phi_{1}^{\text{I}}(x_{1}), \phi_{1}^{\text{I}}(x_{2}),\phi_{2}^{\text{I}}(x_{3}), \phi_{2}^{\text{I}}(x_{4}),\nonumber\\
&\pushright{\big(\phi_{1}^{\text{I}}(z_{1})\big)^{4},\big(\phi_{1}^{\text{I}}(z_{2})\big)^{2},
      \big(\phi_{2}^{\text{I}}(z_{2})\big)^{2}) \vert0\rangle_{c},} \label{eq:23e}\\
   &(-i)^{2}\frac{\lambda_{2}}{4!}\frac{\lambda_{3}}{4} \frac{2}{2!}\int_{-\infty}^{\infty} \dd[4]z_{1}\dd[4]z_{2}\nonumber\\
    & \phantom{AAAAA}\times\langle0\vert T(\phi_{2}^{\text{I}}(x_{1}), \phi_{2}^{\text{I}}(x_{2}),\phi_{1}^{\text{I}}(x_{3}), \phi_{1}^{\text{I}}(x_{4}),\nonumber\\
&\pushright{\big(\phi_{2}^{\text{I}}(z_{1})\big)^{4},\big(\phi_{1}^{\text{I}}(z_{2})\big)^{2}, \big(\phi_{2}^{\text{I}}(z_{2})\big)^{2}) \vert0\rangle_{c},} \label{eq:23f}\\
    &(-i)^{2}\left(\frac{\lambda_{3}}{4}\right)^{2} \frac{1}{2!}\int_{-\infty}^{\infty} \dd[4]z_{1}\dd[4]z_{2}\nonumber\\
    & \phantom{AAAAA}\times\langle0\vert T(\phi_{1}^{\text{I}}(x_{1}), \phi_{2}^{\text{I}}(x_{2}),\phi_{1}^{\text{I}}(x_{3}), \phi_{2}^{\text{I}}(x_{4}), \nonumber\\ &\pushright{\big(\phi_{1}^{\text{I}}(z_{1})\big)^{2},\big(\phi_{2}^{\text{I}}(z_{1})\big)^{2}, \big(\phi_{1}^{\text{I}}(z_{2})\big)^{2}\big(\phi_{2}^{\text{I}}(z_{2})\big)^{2}) \vert0\rangle_{c}.}    \label{eq:23g}
  \end{align}
\end{subequations}

The three Feynman diagrams in the momentum space corresponding to
Eq.~\eqref{eq:23b} are shown in Fig.~\ref{fig:4} and the analytic
expressions for these diagrams are
\begin{subequations}\label{eq:25}
  \begin{align}
    \label{eq:25a}
    -\frac{i\lambda_{1}^{2}}{2}\int\frac{\dd[4]k}{(2\pi)^{4}} \frac{i}{k^{2}-m_{1}^{2}+i\epsilon}\cdot \frac{i}{(p_{1}+p_{2}-k)^{2}-m_{1}^{2}+i\epsilon},\\
    \label{eq:25b}
    -\frac{i\lambda_{1}^{2}}{2}\int\frac{\dd[4]k}{(2\pi)^{4}} \frac{i}{k^{2}-m_{1}^{2}+i\epsilon}\cdot \frac{i}{(p_{1}-p_{3}-k)^{2}-m_{1}^{2}+i\epsilon},\\
    \label{eq:25c}
    -\frac{i\lambda_{1}^{2}}{2}\int\frac{\dd[4]k}{(2\pi)^{4}} \frac{i}{k^{2}-m_{1}^{2}+i\epsilon} \cdot \frac{i}{(p_{1}-p_{4}-k)^{2}-m_{1}^{2}+i\epsilon}
  \end{align}
\end{subequations}
and similarly one can obtain the momentum space diagrams for the
remaining equations~\eqref{eq:23}. The integrals in Eqs.~\eqref{eq:25}
are logarithmically divergent.

\section{Calculation of the loop integrals}
\label{sec:calc-loop-integr}

The loop integrals contain the integration over momentum. Those
integrals contain the products of the particle propagators and
integration of such a product is complicated. The Feynman
prescription, which we will discuss first, facilitates the
integration.

Another convenient scheme that facilitates the integration is the Wick
rotation which allows the conversion of the integral from the Minkowski
space to the Euclidean space.

As we have seen in Section~\ref{sec:two-four-point} the loop diagrams
are divergent.  None of the physical quantities are infinite, so the
loop diagrams require a special treatment in the field
theory. Ultimately these infinities are removed by the procedure of
renormalization, but first these diagrams are made finite by
introducing a cut-off into their analytical expressions. The idea of
the cut-off is to make the integral convergent by adding to the
integral a modification, which depends on an additional parameter with
a property that at a certain limit of the parameter, the integral
tends to the original divergent integral. In such a way, we can
identify a type of divergence, which is then removed. A~simple cut-off
placed as an upper limit of integration is physically inconvenient,
because it breaks the translation invariance of the theory. We will
discuss here two methods of removing the divergence of the loop
integrals: the \textit{Pauli-Villars regularization} and the
\textit{dimensional regularization.} These methods of regularization
have desirable properties from a physical point of view.

\subsection{Feynman prescription}
\label{sec:feynman-prescription}

The Feynman prescription converts the product of two fractions into an
integral of one fraction and adds additional integrals to the
expression. This facilitates the integration of the product of the
propagators
\begin{equation}
  \label{eq:26}
  \frac{1}{AB}=\int_{0}^{1}\frac{\dd{\xi}}{(A\xi+(1-\xi)B)^{2}} =\int_{0}^{1}\int_{0}^{1}\frac{\dd{\xi_{1}}\dd{\xi_{2}}\delta(1-\xi_{1}-\xi_{2})} {(A\xi_{1}+B\xi_{2})^{2}}.
\end{equation}
Eq.~\eqref{eq:26} can be generalized to the product of $n$ terms in
the denominator
\begin{equation}
  \label{eq:27}
  \frac{1}{A_{1}\cdots A_{n}}= (n-1)!\int_{0}^{1}\cdots\int_{0}^{1} \frac{\dd{\xi_{1}} \cdots\dd{\xi_{n}}\delta(1-\xi_{1}-\cdots-\xi_{n})}{(A_{1}\xi_{1}+ \cdots +A_{n}\xi_{n})^{n}}.
\end{equation}
Another generalization is obtained by the differentiation of
Eq.~\eqref{eq:26} with respect to $A$
\begin{equation}
  \label{eq:28}
  \frac{1}{A^{n}B}=\int_{0}^{1}\frac{n!\dd{\xi}}{(A\xi+(1-\xi)B)^{n+1}} =\int_{0}^{1}\int_{0}^{1}\frac{n!\dd{\xi_{1}}\dd{\xi_{2}}\delta(1-\xi_{1}-\xi_{2})} {(A\xi_{1}+B\xi_{2})^{n+1}}.  
\end{equation}

\subsection{Wick rotation}
\label{sec:wick-rotation}

\begin{figure}[!ht]\centering
  \includegraphics[width=0.5\linewidth]{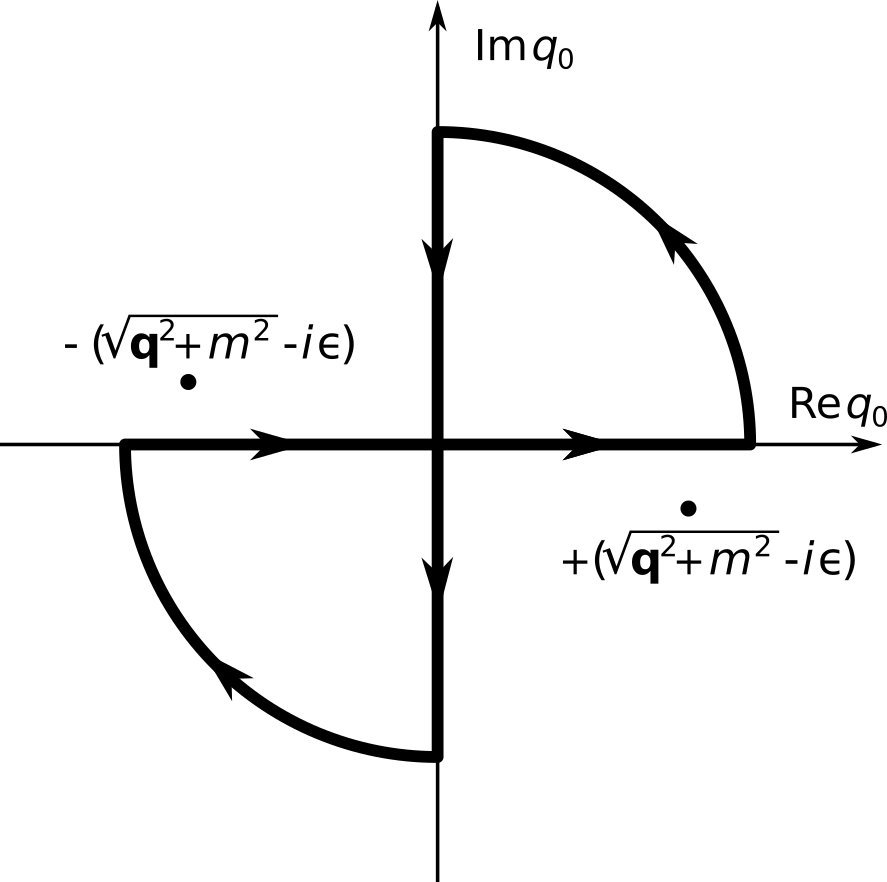}
  \caption{\label{fig:5} Integration path in the $q_{0}$ and the poles
    for Eq.~\eqref{eq:30}}
\end{figure}

The momentum integrals are performed in the Minkowski space, where the
length of a vector $q=(q_{0},q_{1},q_{2},q_{3})$ is equal
$q^{2}= q_{0}^{2}-q_{1}^{2}-q_{2}^{2}-q_{3}^{2}$. Let us consider the
integral
\begin{equation}
  \label{eq:30}
  \int\frac{\dd[4]{q}}{(2\pi)^{4}}\frac{1}{(q^{2}-m^{2}+i\epsilon)^{n}}
\end{equation}
in the Minkowski space. The integrated function as the function of
complex $q_{0}$ has poles at
$q_{0}= \pm(\sqrt{\vb{q}^{2}+m^{2}}-i\epsilon)$. This means that the
contour integral in the complex $q_{0}$ plane over the path shown in
Fig.~\ref{fig:5} does not include the poles of the integrand, so it
vanishes.
When the integration radius goes to infinity the integrals over the
circular segments tend to zero and it means that the following
relation holds
\begin{equation}
  \label{eq:31}
  \int_{-\infty}^{+\infty}\dd{q_{0}}\int\frac{\dd[3]{\vb{q}}}{(2\pi)^{4}} \frac{1}{(q^{2}-m^{2}+i\epsilon)^{n}}=
  \int_{-i\infty}^{+i\infty}\dd{q_{0}}\int\frac{\dd[3]{\vb{q}}}{(2\pi)^{4}} \frac{1}{(q^{2}-m^{2}+i\epsilon)^{n}}.
\end{equation}
The variable $q_{0}$ on the right hand side of Eq.~\eqref{eq:31} is
pure imaginary. To convert the integration over real numbers we make
the change of variables
\begin{align*}
  q_{0}\rightarrow iq_{4},\quad
  \dd{q_{0}}\rightarrow i\dd{q_{4}},\quad
  \dd[4]{q}=\dd{q_{0}}\dd[3]{\vb{q}} \rightarrow i\dd{q_{4}}\dd[3]{\vb{q}}=i\dd[4] {q_{\text{E}}}\\
  q^{2}=q_{0}^{2}-q_{1}^{2}-q_{2}^{2}-q_{3}^{2}\rightarrow -q_{1}^{2}-q_{2}^{2}-q_{3}^{2}-q_{4}^{2}=-q_{\text{E}}^{2}.
\end{align*}
Then the Minkowski space integral~\eqref{eq:30} is converted into the
Euclidean space integral and
\begin{equation}
  \label{eq:32}
  \int\frac{\dd[4]{q}}{(2\pi)^{4}}\frac{1}{(q^{2}-m^{2}+i\epsilon)^{n}} =
  i(-1)^{n} \int\frac{\dd[4]{q_{\text{E}}}}{(2\pi)^{4}}. \frac{1}{(q_{\text{E}}^{2}+m^{2})^{n}}.
\end{equation}
The $i\epsilon$ term in the denominator on the right hand side is not
necessary, because the function $(q_{\text{E}}^{2}+m^{2})$ is not
singular.

It should be noted that the Wick rotation is valid for any function
that has no singularities in the first and third quadrant of the
complex $q_{0}$ plane.

\subsection{Pauli-Villars regularization}
\label{sec:pauli-vill-regul}

The loop diagram contains an integral over the intermediate momentum
$\dd[4]{q}$ of a product of the propagators
$\sim\frac{1}{q^{2}-m^{2}}$. The integral with one propagator is
quadratically divergent and the integral with two propagators is
logarithmically divergent. The Pauli-Villars
regularization~\cite{RevModPhys.21.434} introduces a cut-off~$\Lambda$
by adding to the propagator terms, which vanish at a limit
$\Lambda\rightarrow\infty$. In such a way the asymptotic behavior of
the propagator is modified and the loop integral becomes finite, but
dependent on the cut-off. The prototype of the Pauli-Villars
regularization is illustrated by the equation
\begin{multline}
  \label{eq:29}
  \frac{i}{q^{2}-m^{2}+i\epsilon}\rightarrow
  \left[\frac{i}{q^{2}-m^{2}+i\epsilon}\right]^{\text{PV}}_{\Lambda}=
  \frac{i}{q^{2}-m^{2}+i\epsilon} -
  \frac{i}{q^{2}-\Lambda+i\epsilon}\\ = \frac{i(m^{2}-\Lambda)}
  {(q^{2}-m^{2}+i\epsilon) (q^{2}-\Lambda+i\epsilon)}.
\end{multline}
Here $\Lambda$ is the cut-off parameter with the dimension of mass
squared. One can see that the modified propagator behaves for large
$q^{2}$ like $~1/q^{4}$, while the original propagator behaved like
$1/q^{2}$. This can improve the convergence of the loop integrals. The
general form of the Pauli-Villars regularization of the propagator is
\begin{multline}
  \label{eq:29a}
  \frac{i}{q^{2}-m^{2}+i\epsilon}\rightarrow
  \left[\frac{i}{q^{2}-m^{2}+i\epsilon}\right]^{\text{PV}}_{\Lambda}=
  \frac{i}{q^{2}-m^{2}+i\epsilon} -
  \sum_{j=1}^{n}\frac{iC_{j}}{q^{2}-\Lambda_{j}+i\epsilon}\\ =
  \frac{iC} {(q^{2}-m^{2}+i\epsilon)
    \prod_{j=1}^{n}(q^{2}-\Lambda_{j}+i\epsilon)}.
\end{multline}
Pauli-Villars regularization is physically equivalent to the
introduction of scalar particles obeying the Fermi-Dirac statistics
(ghosts) into the theory.

\subsection{Dimensional regularization}
\label{sec:dimens-regul}

Dimensional regularization~\cite{Bollini1972, tHooft_Veltman} is
another method of analyzing the divergent diagrams in quantum field
theory. The method consists in calculating the integral in $d$
dimensional spacetime. Suppose we need to calculate the integral
\begin{equation}
  \label{eq:33}
  \int\frac{\dd[4]{q_{\text{E}}}}{(2\pi)^{4}}\frac{1}{(q_{\text{E}}^{2}+m^{2})^{2}}.
\end{equation}
This integral is $4$-dimensional and is logarithmically divergent,
because for large $q_{\text{E}}$ the numerator behaves, like
$\sim q_{\text{E}}^{3}$ and the denominator like
$\sim q_{\text{E}}^{4}$. In the 3-dimensional space this integral
is finite. This suggests to calculate this integral in the space with
lower dimensionality $d$ and then, analytically continue the result to
$d\rightarrow 4$. This will isolate the singularity of the integral.

In the $d$-dimensional space, after separating the angular part, the
integral~\eqref{eq:33} becomes
\begin{equation}
  \label{eq:34}
  \int\frac{\dd[d]{q_{\text{E}}}}{(2\pi)^{d}}\frac{1}{(q_{\text{E}}^{2}+m^{2})^{2}}
  =\frac{1}{(2\pi)^{d}}\int\dd{\Omega_{d}} \int_{0}^{\infty}\frac{q_{\text{E}}^{d-1}\dd{q_{\text{E}}}}{(q_{\text{E}}^{2}+m^{2})^{2}}.
\end{equation}
For the angular part we have
\begin{equation}\label{eq:35}
  \frac{1}{(2\pi)^{d}}\int\dd{\Omega_{d}}= \frac{2}{(2\pi)^{d}} \frac{\pi^{\frac{d}{2}}}{\Gamma(\frac{d}{2})}
\end{equation}
and it is not singular at $d=4$. Here $\Gamma(z)$ is the Euler gamma
function. The remaining integral
\begin{equation}\label{eq:36}
  \int_{0}^{\infty}\frac{q_{\text{E}}^{d-1}\dd{q_{\text{E}}}} {(q_{\text{E}}^{2}+m^{2})^{2}} = \frac{\Gamma(\frac{d}{2})\Gamma(2-\frac{d}{2})} {2\Gamma(2)(m^{2})^{2-\frac{d}{2}}} = \frac{(d-2)\pi}{4\sin(\frac{d\pi}{2})m^{4-d}}
\end{equation}
is finite for $d<4$. The right hand side of Eq.~\eqref{eq:36} can be
calculated for continuous values of the space dimension $d$.
Expanding the right hand side of Eq.~\eqref{eq:36} around the point
$d=4$ one obtains
\begin{equation}
  \label{eq:37}
  \frac{\Gamma(\frac{d}{2})\Gamma(2-\frac{d}{2})}{2\Gamma(2)m^{2(2-\frac{d}{2})}} \approx -\frac{1}{d-4}-\frac{1}{2}(1+\ln m^{2})+\cdots
\end{equation}
and for the integral~\eqref{eq:34} one gets
\begin{equation}
  \label{eq:38}
  \int\frac{\dd[d]{q_{\text{E}}}}{(2\pi)^{d}}\frac{1}{(q_{\text{E}}^{2}+m^{2})^{2}} =
  -\frac{1}{8\pi^{2}(d-4)} -\frac{1}{16\pi^{2}}\big(\gamma+\ln(\frac{m^{2}}{4\pi})\big)+\cdots,
\end{equation}
where $\gamma=0.577216\ldots$ is the Euler's constant. From
Eq.~\eqref{eq:38} we see that the integral has a pole at $d=4$ and the
divergent part is equal $-1/(8\pi^{2}(d-4))$.

Eqs.~\eqref{eq:41} and~\eqref{eq:42} below are general formulas, which
are helpful in the practical calculation of integrals in the
dimensional regularization
\begin{equation}
  \label{eq:41}
  \int\frac{\dd[d]{q_{\text{E}}}}{(2\pi)^{d}}\frac{1}{(q_{\text{E}}^{2}+m^{2})^{r}} = \frac{\Gamma(r-\frac{d}{2})}{(4\pi)^{\frac{d}{2}}\Gamma(r)}\cdot \frac{1}{(m^{2})^{r-\frac{d}{2}}}
\end{equation}
and
\begin{equation}
  \label{eq:42}
  \Gamma(-n+z)= \frac{(-1)^{n}}{n!}\Big(\frac{1}{z}+\psi(n+1) +\frac{1}{2}z \Big( \frac{\pi^{2}}{3} +\psi^{2}(n+1)-\psi'(n+1)\Big) \Big)+\cdots
\end{equation}
Here
\begin{equation}
  \label{eq:43}
  \psi(z) = \frac{\dd \ln\Gamma(z)}{\dd z},\quad
  \psi(n+1)= \sum_{l=1}^{n}\frac{1}{l} -\gamma, \quad
  \psi'(n+1)= \frac{\pi^{2}}{6} -\sum_{l=1}^{n}\frac{1}{l^{2}}
\end{equation}
and again $\gamma$ is the Euler constant.

From Eq.~\eqref{eq:41}, by a change of variables, one obtains a more
general result
\begin{equation}
  \label{eq:44}
  \int\frac{\dd[d]{q_{\text{E}}}}{(2\pi)^{d}}\frac{1} {(q_{\text{E}}^{2}+m^{2}+2q_{\text{E}}\cdot p)^{r}} = \frac{\Gamma(r-\frac{d}{2})}{(4\pi)^{\frac{d}{2}}\Gamma(r)} \frac{1}{(m^{2}-p^{2})^{r-\frac{d}{2}}}
\end{equation}
and by differentiation of Eq.~\eqref{eq:44} with respect of $p_{\mu}$
one can obtain more useful formulas.

\paragraph{Dimension of the coupling constans}\phantom{AA}\\
The action $S=\int\mathcal{L}\,\dd[4]x$ is dimensionless\footnote{In
  natural units the dimension of action is
  $\text{position}\times\text{momentum}$ which has the same dimension
  as Planck's constant $\hbar$.} and it has to be dimensionless also
in the $d$~dimensional space. The dimension of the fields depends of
the dimension of the space-time and can be determined from the
Lagrangian density or from the fields commutation relations. The
dimension of the scalar field~$\phi$ in the 4-dimensional space is
$L^{-1}$ (L is length). In the $d$-dimensional space the dimension of
the scalar field is $L^{1-\frac{d}{2}}$. In order that the part of the
action in $d$ dimensions, obtained from the $\lambda\phi^{4}$, is
dimensionless the coupling constant $\lambda$ must have the dimension
$L^{4-d}$. To maintain the coupling constant dimensionless it is
rescaled by a parameter $\mu$ with the dimension of mass, which plays
the role of the cut-off. The coupling constant $\lambda$ then becomes
\begin{equation}
  \label{eq:40}
  \lambda\rightarrow\lambda\mu^{4-d}.
\end{equation}
In the considered model the three coupling constants $\lambda_{1}$,
$\lambda_{2}$ and $\lambda_{3}$ are all rescaled in such a way.

\section{Renormalization}
\label{sec:renormalization}

The Lagrangian density in quantum field theory is a function of the fields
and of the masses and coupling constants. Those masses and coupling
constants \textit{do not} correspond to observable quantities, e.g.,
the observable masses are obtained from the poles of the propagators
and are functions of the Lagrangian masses and coupling constants. At
the lowest order of perturbation theory and for the free fields
the Lagrangian masses do correspond to the observable masses, but it
is not very interesting for most of practical cases. The fact that
parameters of the Lagrangian have to be adjusted to physical
observables is called \textit{renormalization}. This means that the
renormalization has to be performed in any field theory, independent
of the presence of infinities.

The infinite expressions first appeared in quantum
electrodynamics. This caused an important difficulty, because it ruled
out the calculations at higher orders of perturbation theory. It
turned out that in quantum electrodynamics the mathematical structure
of the infinite expressions was identical to the mathematical structure of
the Lagrangian density. This means that the infinite expressions were
\textit{renormalizing} the parameters of the Lagrangian and were not
modifying the theory in any other way. Such a situation does not occur
in all quantum field theories, but only in a small class of
renormalizable theories, like quantum electrodynamics, theories with
non-abelian gauge symmetry and theories of scalar fields with self
interaction of type $\phi^{3}$ or $\phi^{4}$. The theory described
by the Lagrangian density in Eq.~\eqref{eq:1} is renormalizable.

Most of the practical calculations in quantum field theory are done
perturbatively with respect to the power of the coupling constants. The
exception to this rule is the renormalization, which is discussed with
respect to the number of loops present in the Feynman diagrams. We
will discuss the renormalization of our model up to two loops.

\subsection{One loop regularization}
\label{sec:one-loop-renorm}

\begin{figure}[!ht]\centering
  \includegraphics[width=0.8\linewidth]{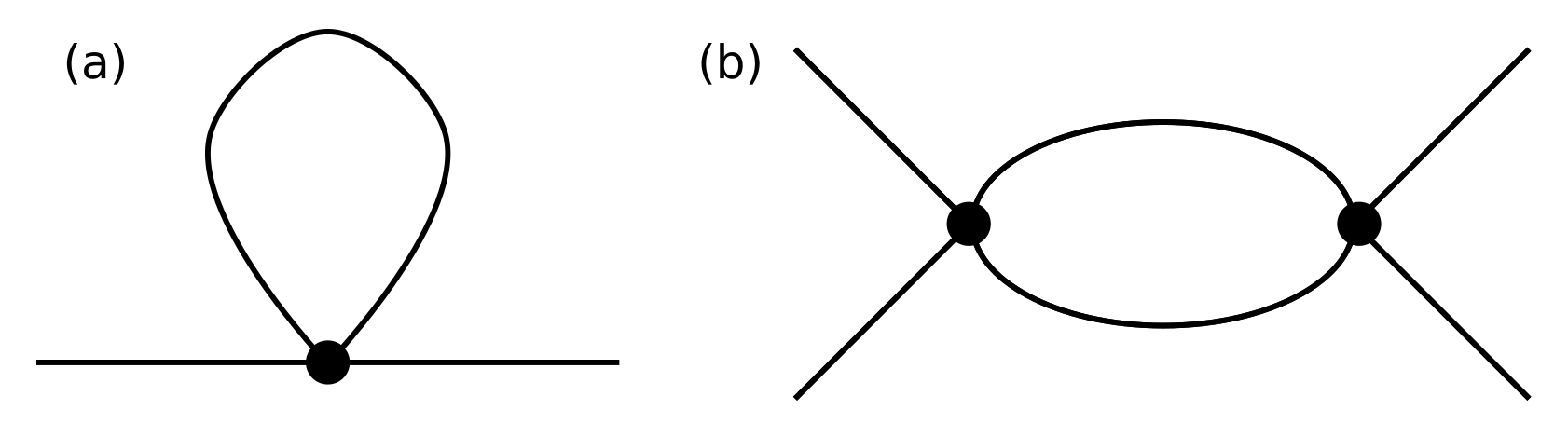}
  \caption{\label{fig:6} The topological structure of the one loop
    divergent diagrams for: (a) two point function, and (b) for the
    four point functions.}
\end{figure}

The process of renormalization starts with the identification of the
divergent diagrams. The one loop diagrams can appear at any order of
the perturbative calculations, but the structure of these diagrams is
such that one can extract a simpler one loop divergent sub-diagram and
the rest of the diagram is finite. This is the reason that one has to
concentrate on the renormalization of those simple
sub-diagrams\footnote{It can be shown that there is only a finite
  number of such diagrams.}. In our model those simple diagrams at one
loop correspond to the two and four point Green's functions. The
topological structure of those diagrams is shown in Fig.~\ref{fig:6}

\subsubsection{Regularization of the two point Green's
  functions} \label{sec:two-point-greens}

All the diagrams that correspond to the structure in
Fig.~\ref{fig:6}~(a) are shown in Fig.~\ref{fig:2} and the mathematical
expressions corresponding to those diagrams are given in
Eqs.~\eqref{eq:19}. For the diagram in Fig.~\ref{fig:2}~(a) the
mathematical formula in $d$~dimensions obtained from
Eq.~\eqref{eq:19a} becomes
\begin{multline}\label{eq:39}
  \frac{\lambda_{1}}{2} \int\frac{\dd[4]k}{(2\pi)^{4}}
  \frac{i}{k^{2}-m_{1}^{2}+i\epsilon}\rightarrow
  \frac{\lambda_{1}\mu^{4-d}}{2} \int\frac{\dd[d]k}{(2\pi)^{d}} \frac{i}{k^{2}-m_{1}^{2}+i\epsilon}\\
  \rightarrow \frac{\lambda_{1}\mu^{4-d}}{2}
  \int\frac{\dd[d]k_{\text{E}}} {(2\pi)^{d}}
  \frac{1}{k_{\text{E}}^{2}+m_{1}^{2}} =
  \frac{\lambda_{1}\mu^{4-d}}{2} \frac{\Gamma(1-\frac{d}{2})}{(4\pi)^{\frac{d}{2}}\Gamma(1)} \frac{1}{(m_{1}^{2})^{1-\frac{d}{2}}}\\
  \underset{d=4}{\rightarrow}
  \frac{\lambda_{1}m_{1}^{2}}{16\pi^{2}}\Big( \frac{1}{d-4}+
  \frac{1}{2} \big(\gamma-1 +\ln(\frac{m_{1}^{2}}{4\pi\mu^{2}}) \big)
  \Big)+\order{d-4}.
\end{multline}
In the same way we obtain the following result for the diagram in
Fig.~\ref{fig:2}~(b)
\begin{multline}
  \label{eq:46}
  \frac{\lambda_{3}}{2} \int\frac{\dd[4]k}{(2\pi)^{4}}
  \frac{i}{k^{2}-m_{2}^{2}+i\epsilon}\\
  \underset{d=4}{\rightarrow}
  \frac{\lambda_{3}m_{2}^{2}}{16\pi^{2}}\Big( \frac{1}{d-4}+
  \frac{1}{2} \big(\gamma-1 +\ln(\frac{m_{2}^{2}}{4\pi\mu^{2}}) \big)
  \Big)+\order{d-4}.
\end{multline}
The space-time dimension~$d$ is equal to~4 and the integrals in
Eqs.~\eqref{eq:39} and~\eqref{eq:46} have poles for this value of~$d$
and are divergent. The important property of Eqs.~\eqref{eq:39}
and~\eqref{eq:46} is that the singular part of the integral has been
separated from the finite part. The finite part depends on the cut-off
parameter $\mu^{2}$ and the infinite part does not. The $\mu^{2}$ is a
new free parameter of the theory\footnote{When we will discuss the
  renormalization group we will see that the physical predictions of
  the theory \textit{do not} depend on $\mu^{2}$.}, so the dependence
of the finite part on it, means that the finite parts of the integrals
are not uniquely determined.

\begin{figure}[!ht]\centering
  \includegraphics[width=0.7\linewidth]{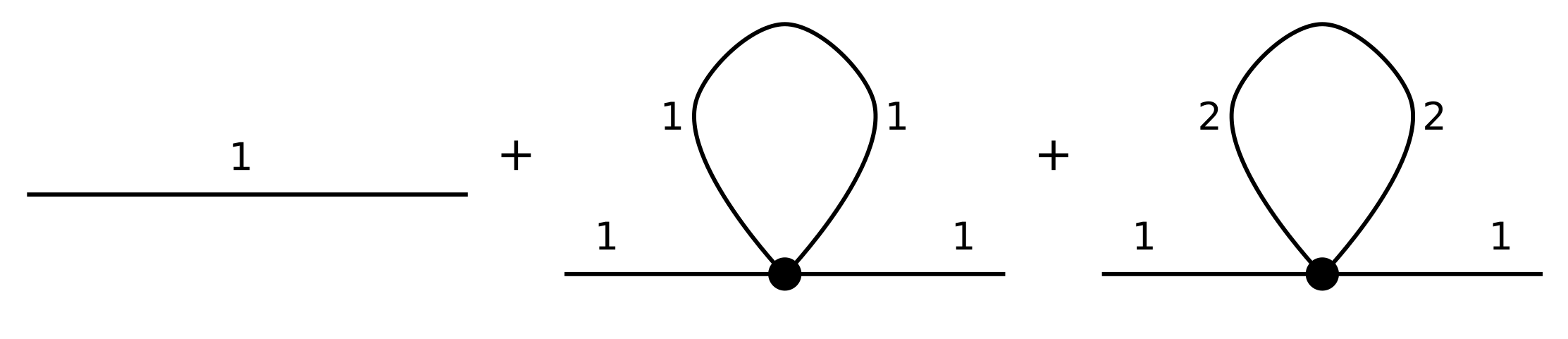}
  \caption{The propagator of particle~1 with one loop
    corrections.\label{fig:7}}
\end{figure}

Eqs.~\eqref{eq:39} and~\eqref{eq:46} are the one loop corrections to
the propagator of particle~1, so the full propagator at one loop order
will be the sum of the free propagator and the integrals~\eqref{eq:39}
and~\eqref{eq:46}. Graphically this sum is shown in Fig.~\ref{fig:7}.

\begin{figure}[!hb]\centering
  \includegraphics[width=0.8\linewidth]{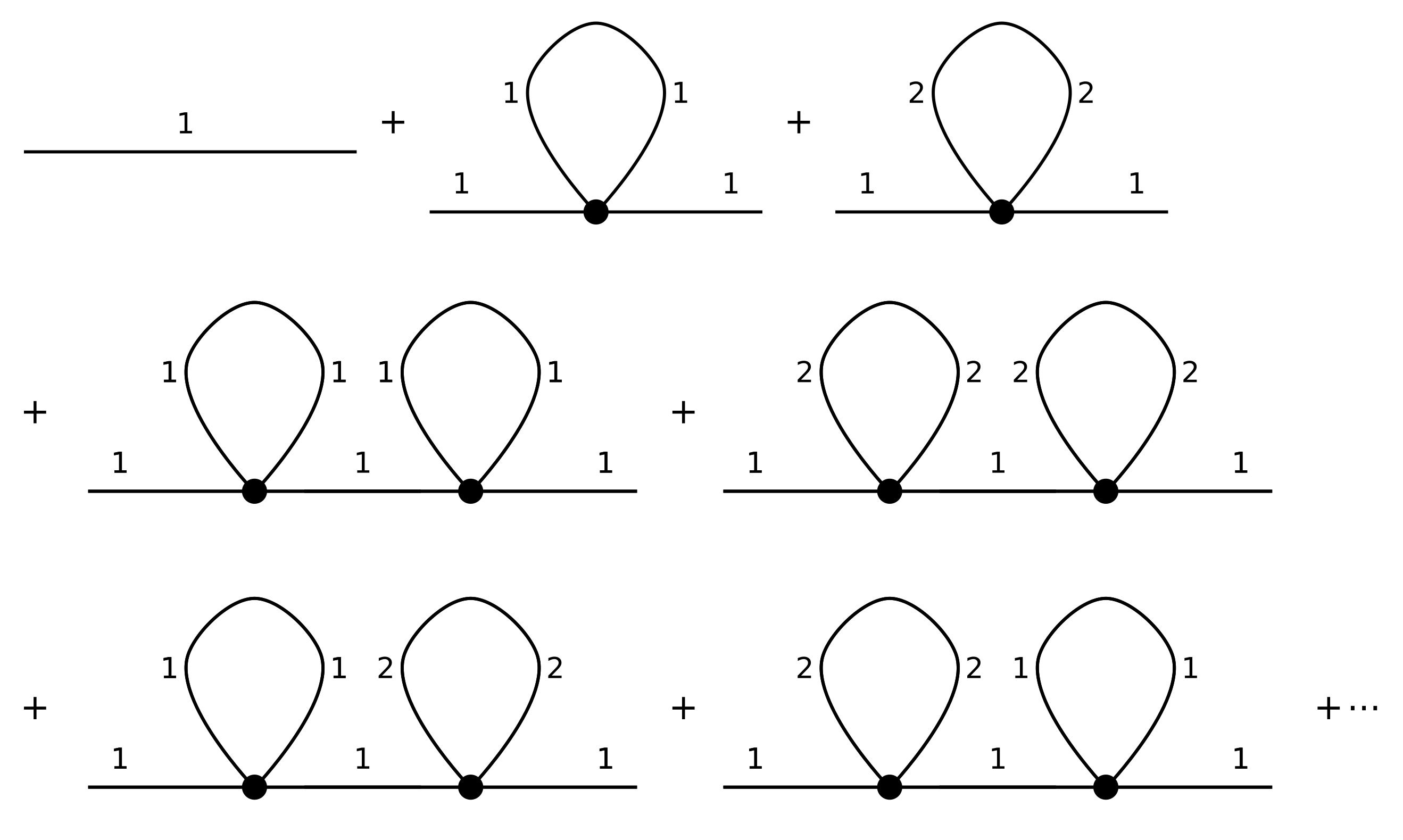}
  \caption{The next iteration for the propagator of particle~1 with
    one loop corrections.\label{fig:8}}
\end{figure}

The one loop corrections can be iterated and the next iteration is
shown in Fig.~\ref{fig:8}.
The iterations can be continued and as a result one obtains the
geometric series for the propagator of particle~1
\begin{multline}
  \label{eq:45}
  \frac{i}{p^{2}-m_{1}^{2}+i\epsilon}\\
  +\frac{i}{p^{2}-m_{1}^{2}+i\epsilon}(-i\Sigma_{1})
  \frac{i}{p^{2}-m_{1}^{2}+i\epsilon}
  +\frac{i}{p^{2}-m_{1}^{2}+i\epsilon}(-i\Sigma_{3}^{1}) \frac{i}{p^{2}-m_{1}^{2}+i\epsilon}\\
  +\frac{i}{p^{2}-m_{1}^{2}+i\epsilon}(-i\Sigma_{1}) \frac{i}{p^{2}-m_{1}^{2}+i\epsilon} (-i\Sigma_{1}) \frac{i}{p^{2}-m_{1}^{2}+i\epsilon}\\
  +\frac{i}{p^{2}-m_{1}^{2}+i\epsilon}(-i\Sigma_{3}^{1}) \frac{i}{p^{2}-m_{1}^{2}+i\epsilon} (-i\Sigma_{3}^{1}) \frac{i}{p^{2}-m_{1}^{2}+i\epsilon}\\
  +\frac{i}{p^{2}-m_{1}^{2}+i\epsilon}(-i\Sigma_{1}) \frac{i}{p^{2}-m_{1}^{2}+i\epsilon} (-i\Sigma_{3}^{1}) \frac{i}{p^{2}-m_{1}^{2}+i\epsilon}\\
  +\frac{i}{p^{2}-m_{1}^{2}+i\epsilon}(-i\Sigma_{3}^{1}) \frac{i}{p^{2}-m_{1}^{2}+i\epsilon} (-i\Sigma_{1}^{1}) \frac{i}{p^{2}-m_{1}^{2}+i\epsilon}+\cdots\\
  =\frac{i}{p^{2}-m_{1}^{2}+i\epsilon}\cdot
  \frac{1}{1-\frac{\displaystyle(-i(\Sigma_{1}+\Sigma_{3}^{1}))i}
    {\displaystyle p^{2}-m_{1}^{2}+i\epsilon}} =
  \frac{i}{p^{2}-m_{1}^{2}-\Sigma_{1}-\Sigma_{3}^{1}+i\epsilon}.
\end{multline}
In Eq.~\eqref{eq:45} we used the shorthand notation for the results of
Eqs.~\eqref{eq:39} and~\eqref{eq:46}
\begin{subequations}\label{eq:50}
  \begin{align}
    \label{eq:50a}
    &\Sigma_{1} =\frac{\lambda_{1}m_{1}^{2}}{16\pi^{2}}\Big(
      \frac{1}{d-4}+ \frac{1}{2} \big(\gamma-1 +\ln(\frac{m_{1}^{2}}{4\pi\mu^{2}})
      \big) \Big),\\
    \label{eq:50b}
    &\Sigma_{2} =\frac{\lambda_{2}m_{2}^{2}}{16\pi^{2}}\Big(
      \frac{1}{d-4}+ \frac{1}{2} \big(\gamma-1 +\ln(\frac{m_{2}^{2}}{4\pi\mu^{2}})
      \big) \Big),\\
    \label{eq:50c}
    &\Sigma_{3}^{1} =\frac{\lambda_{3}m_{2}^{2}}{16\pi^{2}}\Big(
      \frac{1}{d-4}+ \frac{1}{2} \big(\gamma-1 +\ln(\frac{m_{2}^{2}}{4\pi\mu^{2}})
      \big) \Big),\\
    \label{eq:50d}
    &\Sigma_{3}^{2} =\frac{\lambda_{3}m_{1}^{2}}{16\pi^{2}}\Big(
      \frac{1}{d-4}+ \frac{1}{2} \big(\gamma-1 +\ln(\frac{m_{1}^{2}}{4\pi\mu^{2}})
      \big) \Big).
  \end{align}
\end{subequations}
$\Sigma_{2}$ and $\Sigma_{3}^{2}$ are necessary for the propagator of
particle~2 and they are obtained from the $\Sigma_{1}$ and
$\Sigma_{3}^{1}$ by changing the indices~$1\leftrightarrow2$.

From Eq.~\eqref{eq:45} we see that after the one loop corrections for
the propagator the value of the mass $m_{1}^{2}$ was shifted
\begin{equation}
  \label{eq:47}
  m_{1}^{2}\rightarrow m_{1}^{2}+\Sigma_{1}+\Sigma_{3}^{1}.
\end{equation}
It means that the mass $m_{1}$ becomes \textit{renormalized} as the
result of the interactions.

The shift of the mass $m_{2}^{2}$ is
\begin{equation}
  \label{eq:48}
  m_{2}^{2}\rightarrow m_{2}^{2}+\Sigma_{2}+\Sigma_{3}^{2}.  
\end{equation}

\subsubsection{Regularization of the four point Green's
  functions}\label{sec:regul-two-point}

All the diagrams that correspond to the structure in
Fig.~\ref{fig:6}~(b), with all external particles of type~1, are shown
in Fig.~\ref{fig:4} and the mathematical expressions corresponding to
those diagrams are given in Eqs.~\eqref{eq:25}. For the diagram in
Fig.~\ref{fig:4}~(a) the mathematical formula in $d$~dimensions
obtained from Eq.~\eqref{eq:25a} is obtained by the following steps
(here we use the substitution: $p_{1}+p_{2}=p$)
\begin{subequations}
  \label{eq:49}
  \begin{align}
    &\label{eq:49a} -\frac{i\lambda_{1}^{2}}{2}\int\frac{\dd[4]k}{(2\pi)^{4}}
      \frac{i}{k^{2}-m_{1}^{2}+i\epsilon}\cdot
      \frac{i}{(p-k)^{2}-m_{1}^{2}+i\epsilon}\\
    &\label{eq:49b}\,\, \Scale[0.96]{
      =\frac{i\lambda_{1}^{2}}{2}\int_{0}^{1}\dd{z}\int\frac{\dd[4]k}{(2\pi)^{4}}
      \frac{1}{(z(k^{2}-m_{1}^{2}+i\epsilon)+(1-z)((p-k)^{2}-m_{1}^{2}+i\epsilon))^{2}}}
    \\
    &\label{eq:49c}=\frac{i\lambda_{1}^{2}}{2} \int_{0}^{1}\dd{z}\int\frac{\dd[4]k}{(2\pi)^{4}}
      \frac{1}{(zk^{2}+(1-z)(p^{2}+k^{2}-2pk)-m_{1}^{2} +i\epsilon)^{2}}\\
    &\label{eq:49d}=\frac{i\lambda_{1}^{2}}{2}\int_{0}^{1}\dd{z}\int\frac{\dd[4]k}{(2\pi)^{4}}
      \frac{1}{((k-(1-z)p)^{2}+z(1-z)p^{2}-m_{1}^{2} +i\epsilon)^{2}}\\
    &\label{eq:49e}\rightarrow
      \frac{i\lambda_{1}^{2}}{2}\int_{0}^{1}\dd{z}\int\frac{\dd[4]k}{(2\pi)^{4}}
      \frac{1}{(k^{2}+z(1-z)p^{2}-m_{1}^{2} +i\epsilon)^{2}}\\
    &\label{eq:49f}\rightarrow
      -\frac{\lambda_{1}^{2}}{2}\int_{0}^{1}\dd{z}\int\frac{\dd[4]k_{\text{E}}}{(2\pi)^{4}}
      \frac{1}{(k_{\text{E}}^{2}-z(1-z)p^{2}+m_{1}^{2})^{2}}\\
    &\label{eq:49g}\rightarrow
      -\frac{\lambda_{1}^{2}\mu^{2(4-d)}}{2}\int_{0}^{1}\dd{z}\int\frac{\dd[d]k_{\text{E}}}{(2\pi)^{d}}
      \frac{1}{(k_{\text{E}}^{2}-z(1-z)p^{2}+m_{1}^{2})^{2}}\\
    &\label{eq:49h}=-\frac{\lambda_{1}^{2}\mu^{2(4-d)}}{2}\int_{0}^{1}\dd{z} \frac{\Gamma(2-\frac{d}{2})}{(4\pi)^{\frac{d}{2}} \Gamma(2) (m_{1}^{2}-z(1-z)p^{2})^{2-\frac{d}{2}}}\\
    &\label{eq:49i}\underset{d=4}{\rightarrow} \frac{\lambda_{1}^{2}\mu^{4-d}}{16\pi^{2}}\Big(\frac{1}{d-4} +\frac{1}{2}(\gamma+ \int_{0}^{1}\dd{z}\ln(\frac{\vert z(1-z)p^{2}-m_{1}^{2}\vert}{4\pi\mu^{2}}))\Big).
  \end{align}
\end{subequations}
Let us explain each step in Eqs.~\eqref{eq:49}
\begin{description}
\item[Eq.~\eqref{eq:49a}$\rightarrow$~\eqref{eq:49b}] Feynman
  prescription.
\item[Eq.~\eqref{eq:49b}$\rightarrow$~\eqref{eq:49c}] Transformation
  of the denominator with propagators.
\item[Eq.~\eqref{eq:49c}$\rightarrow$~\eqref{eq:49d}] Transformation
  of the denominator with propagators.
\item[Eq.~\eqref{eq:49d}$\rightarrow$~\eqref{eq:49e}] Change of the
  integration variable $k\rightarrow (k-(1-z)p)$.
\item[Eq.~\eqref{eq:49e}$\rightarrow$~\eqref{eq:49f}] Wick rotation.
\item[Eq.~\eqref{eq:49f}$\rightarrow$~\eqref{eq:49g}] Transition to
  the $d$-dimensional integration.
\item[Eq.~\eqref{eq:49g}$\rightarrow$~\eqref{eq:49h}] Calculation of
  the $d$-dimensional integral, using Eq.~\eqref{eq:41}.
\item[Eq.~\eqref{eq:49h}$\rightarrow$~\eqref{eq:49i}] Calculation of
  the limit $d\rightarrow4$. Note, that the factor $\mu^{4-d}$ has not
  been included into the integral, because the dimension of the
  diagram is $\mu^{4-d}$.
\end{description}
The last integral in Eq.~\eqref{eq:49i} is elementary and can be
calculated, but an explicit integration does not introduce important
information.

To continue, we define the functions $F(r_{1},r_{2},r_{3})$ and
$F_{\Sigma}(s,t,u,r_{1},r_{2})$
\begin{align}
  \label{eq:51a}
  &F(r_{1},r_{2},r_{3})=\int_{0}^{1}\dd{z}\ln( \frac{\vert z(1-z)r_{1}-r_{2}\vert}{4\pi r_{3}}),\\
  &F_{\Sigma}(s,t,u,r_{1},r_{2}) = F(s,r_{1},r_{2}) +F(t,r_{1},r_{2}) +F(u,r_{1},r_{2}).
\end{align}
And with the help of the function~\eqref{eq:51a}, using
Eq.~\eqref{eq:49i}, we can write the analytical expressions
corresponding to the diagrams in Fig.~\ref{fig:4}
\begin{subequations}\label{eq:52}
  \begin{align}
    \label{eq:52a}
    \frac{\lambda_{1}^{2}\mu^{4-d}}{16\pi^{2}}\Big(\frac{1}{d-4} +\frac{1}{2}(\gamma+ F(s,m_{1}^{2},\mu^{2}))\Big), \\
    \label{eq:52b}    \frac{\lambda_{1}^{2}\mu^{4-d}}{16\pi^{2}}\Big(\frac{1}{d-4} +\frac{1}{2}(\gamma+ F(t,m_{1}^{2},\mu^{2}))\Big),\\
    \label{eq:52c}     \frac{\lambda_{1}^{2}\mu^{4-d}}{16\pi^{2}} \Big(\frac{1}{d-4} +\frac{1}{2}(\gamma+ F(u,m_{1}^{2},\mu^{2}))\Big).
  \end{align}
\end{subequations}
Here, $s$, $t$, and $u$ denote the Mandelstam variables
\begin{equation*}
  s=(p_{1}+p_{2})^{2},\quad t=(p_{1}-p_{3})^{2},\quad u=(p_{1}-p_{4})^{2}.
\end{equation*}

\begin{figure}[!ht]
  \includegraphics[width=0.75\linewidth]{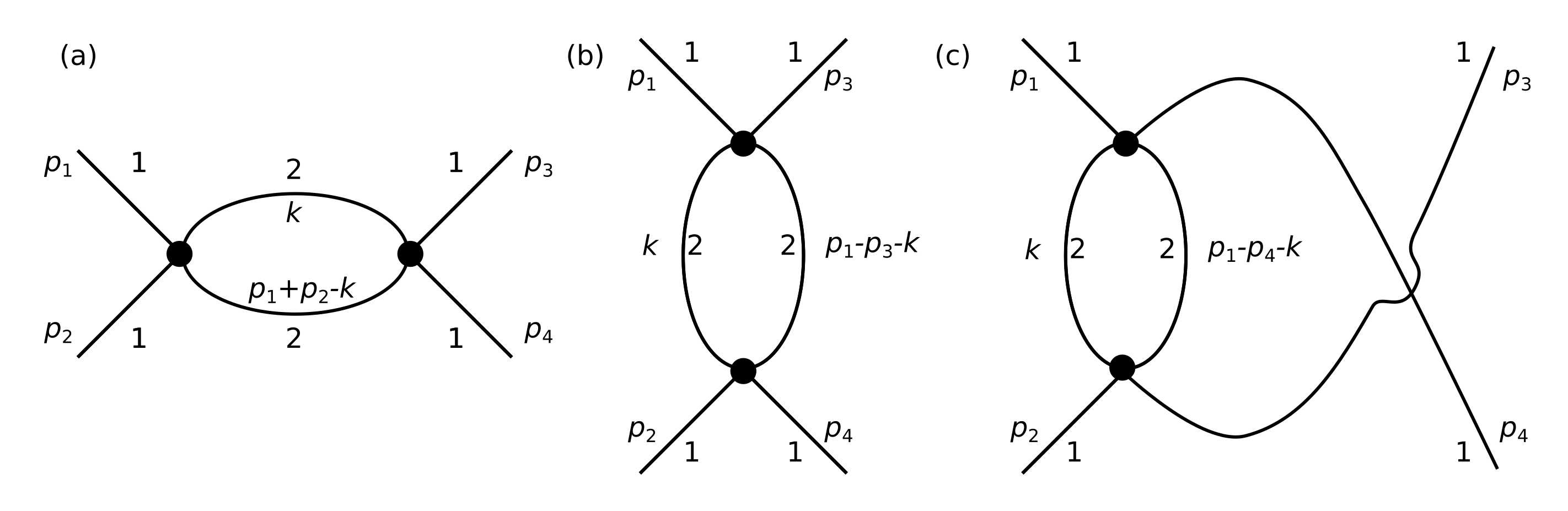}\centering
  \caption{\label{fig:9}Three Feynman diagrams with four external
    particles of the type~1 and the vertices of the type~3. }
\end{figure}

Next we will calculate the diagrams of the type shown in
Fig.~\ref{fig:3}~(b). There are three diagrams of this type which are
shown in Fig.~\ref{fig:9}.
The procedure of the derivation of the analytic expressions
corresponding to these diagrams is the same as in the previous case
and the result is
\begin{subequations}\label{eq:55}
  \begin{align}
    \label{eq:55a}
    \frac{\lambda_{3}^{2}\mu^{4-d}}{16\pi^{2}}\Big(\frac{1}{d-4} +\frac{1}{2}(\gamma+ F(s,m_{2}^{2},\mu^{2}))\Big), \\
    \label{eq:55b}
    \frac{\lambda_{3}^{2}\mu^{4-d}}{16\pi^{2}}\Big(\frac{1}{d-4} +\frac{1}{2}(\gamma+ F(t,m_{2}^{2},\mu^{2}))\Big), \\
    \label{eq:55c}  
    \frac{\lambda_{3}^{2}\mu^{4-d}}{16\pi^{2}}\Big(\frac{1}{d-4} +\frac{1}{2}(\gamma+ F(u,m_{2}^{2},\mu^{2}))\Big).
  \end{align}
\end{subequations}

The four body vertex with one loop contribution for four external
particles of the type~1 is thus equal
\begin{multline}
  \label{eq:53}
  \displaystyle\lambda_{1}\mu^{4-d} + \frac{\lambda_{1}^{2}\mu^{4-d}}{16\pi^{2}}\Big(\frac{3}{d-4} +\frac{3}{2}\gamma+ \frac{1}{2}\big(F_{\Sigma}(s,t,u,m_{1}^{2},\mu^{2})\big)\Big)\\
  +\frac{\lambda_{3}^{2}\mu^{4-d}}{16\pi^{2}}\Big(\frac{3}{d-4}
  +\frac{3}{2}\gamma+
  \frac{1}{2}\big(F_{\Sigma}(s,t,u,m_{2}^{2},\mu^{2})\big)\Big).
\end{multline}
By changing the indices $1\leftrightarrow 2$ we obtain from
Eq.~\eqref{eq:53} the four body vertex with one loop contribution for
four external particles of the type~2
\begin{multline}
  \label{eq:54}
  \displaystyle\lambda_{2}\mu^{4-d} + \frac{\lambda_{2}^{2}\mu^{4-d}}{16\pi^{2}}\Big(\frac{3}{d-4} +\frac{3}{2}\gamma+ \frac{1}{2}\big(F_{\Sigma}(s,t,u,m_{2}^{2},\mu^{2})\big)\Big)\\
  +\frac{\lambda_{3}^{2}\mu^{4-d}}{16\pi^{2}}\Big(\frac{3}{d-4}
  +\frac{3}{2}\gamma+
  \frac{1}{2}\big(F_{\Sigma}(s,t,u,m_{1}^{2},\mu^{2})\big)\Big).
\end{multline}

\begin{figure}[!ht]\centering
  \includegraphics[width=0.7\linewidth]{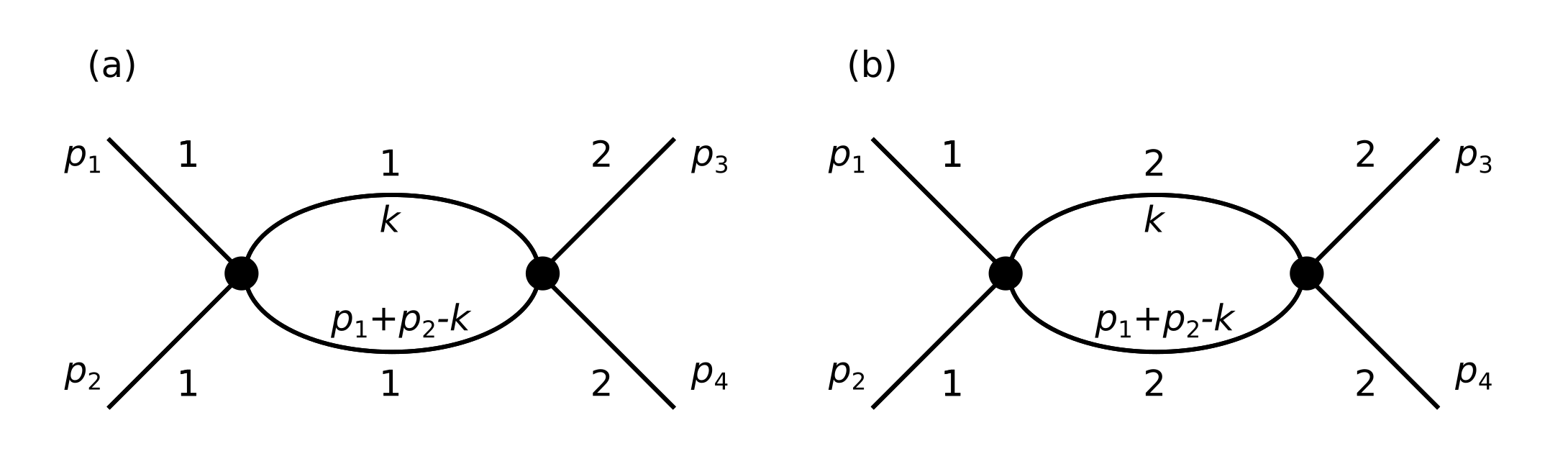}
  \caption{\label{fig:10} Feynman diagrams corresponding to the
    patterns in Fig.~\ref{fig:3}~(e) and~(f).}
\end{figure}

For the patterns in Fig.~\ref{fig:3}~(e) and~(f) there is only one
Feynman diagram, each in the momentum space shown in Fig.~\ref{fig:10}
and the analytical expressions are:\\
For the pattern in Fig.~\ref{fig:3}~(e)
\begin{align}
  &-\frac{i\lambda_{1}\lambda_{3}}{2}\int\frac{\dd[4]k}{(2\pi)^{4}}
    \frac{i}{k^{2}-m_{1}^{2}+i\epsilon}\cdot
    \frac{i}{(p_{1}+p_{2}-k)^{2}-m_{1}^{2}+i\epsilon}\nonumber\\
  &\label{eq:52aa}\phantom{AAAAAAAAAAAAAA}\underset{d=4}{\rightarrow} \frac{\lambda_{1}\lambda_{3} \mu^{4-d}}{16\pi^{2}}\Big(\frac{1}{d-4} +\frac{1}{2}(\gamma+ F(s,m_{1}^{2},\mu^{2}))\Big)\\ 
  \shortintertext{and for the pattern in Fig.~\ref{fig:3}~(f)}
  &-\frac{i\lambda_{2}\lambda_{3}}{2}\int\frac{\dd[4]k}{(2\pi)^{4}}
    \frac{i}{k^{2}-m_{2}^{2}+i\epsilon}\cdot
    \frac{i}{(p_{1}+p_{2}-k)^{2}-m_{2}^{2}+i\epsilon}\nonumber\\
  &\label{eq:52bb}\phantom{AAAAAAAAAAAAAA}\underset{d=4}{\rightarrow} \frac{\lambda_{2}\lambda_{3} \mu^{4-d}}{16\pi^{2}}\Big(\frac{1}{d-4} +\frac{1}{2}(\gamma+ F(s,m_{2}^{2},\mu^{2}))\Big).
\end{align}

\begin{figure}[!hb]
  \includegraphics[width=0.75\linewidth]{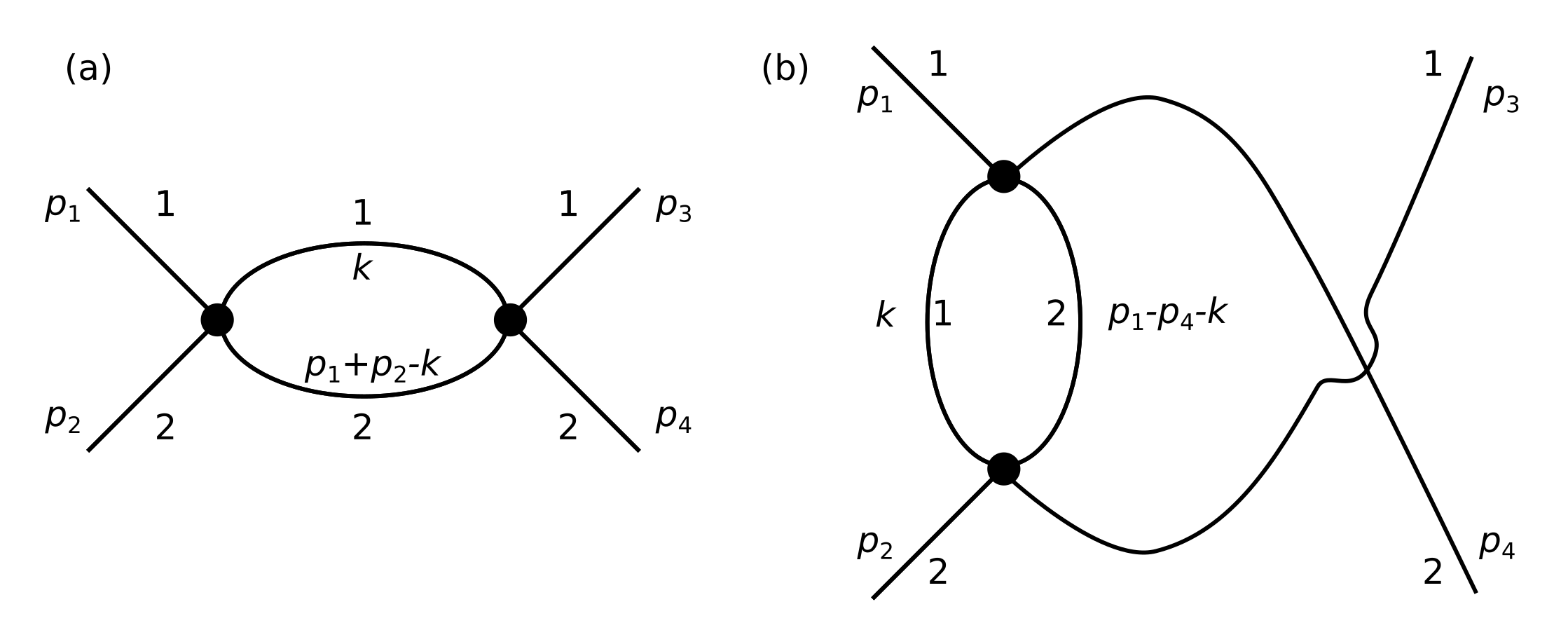}\centering
  \caption{\label{fig:11}Two Feynman diagrams corresponding to the
    pattern in Fig.~\ref{fig:3}~(g). }
\end{figure}

The last diagram that we consider is with pattern in
Fig.~\ref{fig:3}~(g) for which we have two diagrams in momentum space
shown in Fig.~\ref{fig:11}
and the analytic expressions corresponding to these diagrams are
\begin{align}
  &\Scale[0.97]{-i\lambda_{3}^{2}\int\frac{\dd[4]k}{(2\pi)^{4}}
    \frac{i}{k^{2}-m_{1}^{2}+i\epsilon}\cdot
    \frac{i}{(p_{1}+p_{2}-k)^{2}-m_{2}^{2}+i\epsilon}}\nonumber\\
  &\Scale[0.97]{\underset{d=4}{\rightarrow} \frac{\lambda_{3}^{2}\mu^{4-d}}{8\pi^{2}}\Big(\frac{1}{d-4} +\frac{1}{2}(\gamma+ \int_{0}^{1}\dd{z}\ln(\frac{\vert z(1-z)s-zm_{1}^{2}-(1-z) m_{2}^{2}  \vert}{4\pi\mu^{2}}))\Big)} \label{eq:56}\\
  \shortintertext{and}
  &\Scale[0.97]{-i\lambda_{3}^{2}\int\frac{\dd[4]k}{(2\pi)^{4}}
    \frac{i}{k^{2}-m_{1}^{2}+i\epsilon}\cdot
    \frac{i}{(p_{1}-p_{4}-k)^{2}-m_{2}^{2}+i\epsilon}}\nonumber\\
  &\Scale[0.97]{\underset{d=4}{\rightarrow} \frac{\lambda_{3}^{2}\mu^{4-d}}{8\pi^{2}}\Big(\frac{1}{d-4} +\frac{1}{2}(\gamma+ \int_{0}^{1}\dd{z}\ln(\frac{\vert z(1-z)u-zm_{1}^{2}-(1-z) m_{2}^{2}  \vert}{4\pi\mu^{2}}))\Big)} \label{eq:57}.
\end{align}

\subsection{One loop renormalization}
\label{sec:one-loop-renorm-1}

\subsubsection{Lagrangian density with counterterms and new Feynman
  rules}
\label{sec:lagr-dens-with}

The first step in the renormalization program is to rewrite the
original Lagrangian density~\eqref{eq:1} by splitting it into two
parts with identical structure
\begin{equation}
  \label{eq:58}
  \begin{aligned}
    \mathcal{L}&=\frac{1}{2}\partial_{\mu}\phi_{1}\partial^{\mu}\phi_{1} + \frac{1}{2}\partial_{\mu}\phi_{2}\partial^{\mu}\phi_{2}\\ &\phantom{AAA}-\frac{1}{2}m_{1}^{2}\phi_{1}^{2}  -\frac{1}{2}m_{2}^{2}\phi_{2}^{2} -\frac{\lambda_{1}}{4!}\phi_{1}^{4} -\frac{\lambda_{2}}{4!}\phi_{2}^{4} -\frac{\lambda_{3}}{4}\phi_{1}^{2}\phi_{2}^{2}\\
    &=\frac{1}{2}\partial_{\mu}\varphi_{1}\partial^{\mu}\varphi_{1}+ \frac{1}{2}\partial_{\mu}\varphi_{2}\partial^{\mu}\varphi_{2}
    +\frac{\delta_{Z_{1}}}{2}\partial_{\mu}\varphi_{1}\partial^{\mu}\varphi_{1}+ \frac{\delta_{Z_{2}}}{2}\partial_{\mu}\varphi_{2}\partial^{\mu}\varphi_{2}\\ &\phantom{AAA}-\frac{1}{2}M_{1}^{2}\varphi_{1}^{2} -\frac{1}{2}M_{2}^{2}\varphi_{2}^{2} -\frac{\Lambda_{1}}{4!}\varphi_{1}^{4} -\frac{\Lambda_{2}}{4!}\varphi_{2}^{4} -\frac{\Lambda_{3}}{4}\varphi_{1}^{2}\varphi_{2}^{2}\\
    &\phantom{AAA}-\frac{\delta_{M_{1}}}{2}\varphi_{1}^{2}
    -\frac{\delta_{M_{2}}}{2}\varphi_{2}^{2}
    -\frac{\delta_{\Lambda_{1}}}{4!}\varphi_{1}^{4}
    -\frac{\delta_{\Lambda_{2}}}{4!}\varphi_{2}^{4}
    -\frac{\delta_{\Lambda_{3}}}{4}\varphi_{1}^{2}\varphi_{2}^{2}.  \end{aligned}
\end{equation}
Here $M_{1}$, $M_{2}$, $\Lambda_{1}$, $\Lambda_{2}$ and $\Lambda_{3}$
are new parameters of the Lagrangian density and the terms containing
$\delta_{Z_{1}}$, $\delta_{Z_{2}}$, $\delta{m_{1}}$, $\delta{m_{2}}$,
$\delta{\lambda_{1}}$, $\delta{\lambda_{2}}$ and $\delta{\lambda_{3}}$
are called \textit{counterterms}. The fields $\varphi_{1}$ and
$\varphi_{2}$ have different normalization than the fields $\phi_{1}$
and $\phi_{2}$. The parameters $m_{1}$, $m_{2}$, $\lambda_{1}$,
$\lambda_{2}$ and $\lambda_{3}$ of the original Lagrangian
density~\eqref{eq:1} are called \textit{bare} parameters and the
following relations hold
\begin{equation}
  \label{eq:59}
  \begin{gathered}
    \phi_{1}=\sqrt{Z_{\varphi_{1}}}\varphi_{1},\quad \phi_{2}=\sqrt{Z_{\varphi_{2}}}\varphi_{2},\quad Z_{\phi_{1}}=1+\delta_{Z_{1}}, \quad Z_{\phi_{2}}=1+\delta_{Z_{2}},\\
    m_{1}^{2}=\frac{M_{1}^{2}+\delta_{M_{1}}}{Z_{\varphi_{1}}},\quad m_{2}^{2}=\frac{M_{2}^{2}+\delta_{M_{2}}}{Z_{\varphi_{2}}},\\
    \lambda_{1}=\frac{\Lambda_{1}+\delta_{\Lambda_{1}}}{Z_{\varphi_{1}}^{2}},\quad
    \lambda_{2}=\frac{\Lambda_{2}+\delta_{\Lambda_{2}}}{Z_{\varphi_{2}}^{2}},\quad
    \lambda_{3}=\frac{\Lambda_{3}+\delta_{\Lambda_{3}}}{Z_{\varphi_{1}}Z_{\varphi_{2}}}.
  \end{gathered}
\end{equation}

Splitting of the Lagrangian density, Eq.~\eqref{eq:58} introduces new
rules for the Feynman diagrams, which are shown in Table~\ref{tab:7},
which have to be used in calculation of the diagrams.

\begin{table}[!ht]\centering
  \caption{\label{tab:7}Analytical expressions corresponding to the
    elements of the Feynman diagrams for the Lagrangian density in
    Eq.~\eqref{eq:58}.}\vspace*{10pt}
  \begin{tabular}{c|rclrcl}
    \multirow{2}{*}{\rot{\parbox[c]{2.4cm}{\centering Propagators}}}&
                                                                      \begin{minipage}[c]{0.2\linewidth}
                                                                        \includegraphics[width=\linewidth,keepaspectratio]{fig02}
                                                                      \end{minipage}&
                                                                      $\longrightarrow$&$\dfrac{i}{p^{2}-M_{1}^{2}+i\epsilon}$
                                                                      &\begin{minipage}[c]{0.2\linewidth}
                                                                        \includegraphics[width=\linewidth,keepaspectratio]{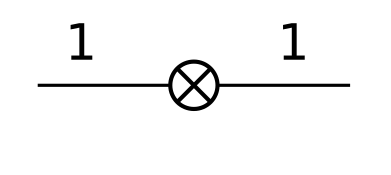}
                                                                      \end{minipage}&
                                                                                      $\longrightarrow$&$-i\delta_{M_{1}}$\\
                                                                    &\begin{minipage}[c][20mm][c]{0.2\linewidth}
                                                                       \includegraphics[width=\linewidth,keepaspectratio]{fig03}
                                                                     \end{minipage}
                                                                                    &$\longrightarrow$&$\dfrac{i}{p^{2}-M_{2}^{2}+i\epsilon}$
                                                                      &\begin{minipage}[c][20mm][c]{0.2\linewidth}
                                                                        \includegraphics[width=\linewidth,keepaspectratio]{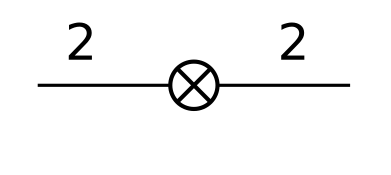}
                                                                      \end{minipage}&
                                                                                      $\longrightarrow$&$-i\delta_{M_{2}}$\\[6pt]
    \hline \multirow{3}{*}{\rot{\parbox[c]{6cm}{\centering Vertices}}}
                                                                    &\begin{minipage}[c][30mm][c]{0.2\linewidth}
                                                                       \includegraphics[width=\linewidth,keepaspectratio]{fig04}
                                                                     \end{minipage}&
                                                                     $\longrightarrow$&$-i\Lambda_{1}$
                                                                     &\begin{minipage}[c][30mm][c]{0.2\linewidth}
                                                                       \includegraphics[width=\linewidth,keepaspectratio]{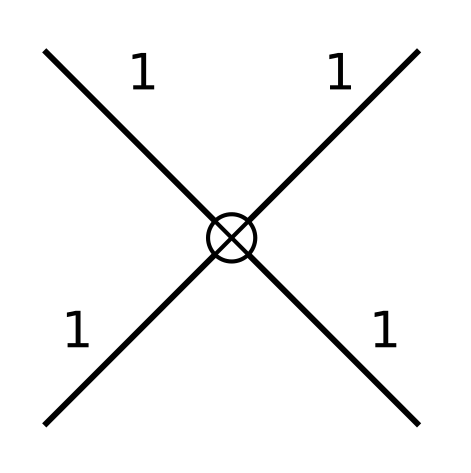}
                                                                     \end{minipage}&
                                                                                     $\longrightarrow$&$-i\delta_{\Lambda_{1}}$\\
                                                                    &\begin{minipage}[c]{0.2\linewidth}
                                                                       \includegraphics[width=\linewidth,keepaspectratio]{fig05}
                                                                     \end{minipage}&
                                                                     $\longrightarrow$&$-i\Lambda_{2}$,
                                                                     &\begin{minipage}[c]{0.2\linewidth}
                                                                       \includegraphics[width=\linewidth,keepaspectratio]{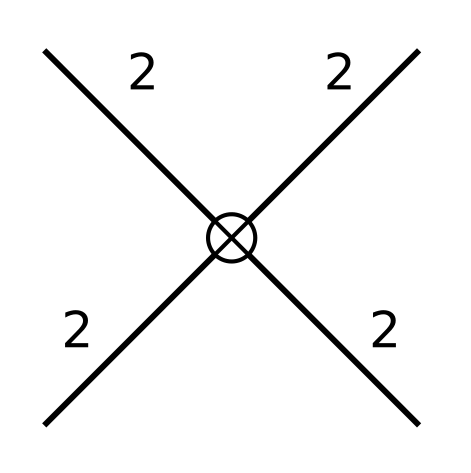}
                                                                     \end{minipage}
                                                                                    &$\longrightarrow$&$-i\delta_{\Lambda_{2}}$,\\
                                                                    &\begin{minipage}[c]{0.2\linewidth}
                                                                      \includegraphics[width=\linewidth,keepaspectratio]{fig06}
                                                                    \end{minipage}&
                                                                    $\longrightarrow$&$-i\Lambda_{3}$
                                                                    &\begin{minipage}[c]{0.2\linewidth}
                                                                      \includegraphics[width=\linewidth,keepaspectratio]{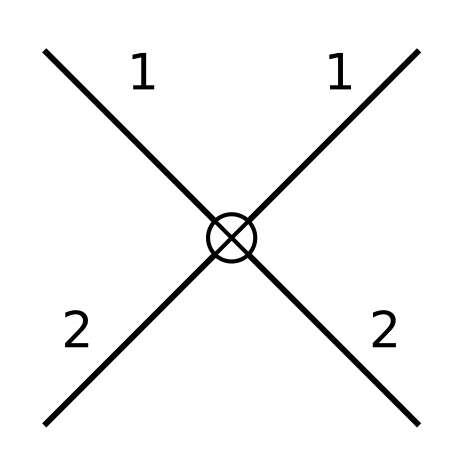}
                                                                    \end{minipage}&
                                                                    $\longrightarrow$&$-i\delta_{\Lambda_{3}}.$
  \end{tabular}
\end{table}
Before proceeding we have to specify, what we mean by the
\textit{physical} coupling constant. From
Subsection~\ref{sec:regul-two-point}, we know that at the order of one
loop the coupling constants do depend on the Mandelstam variables $s$,
$t$ and $u$, so the definition of the \textit{physical} coupling
constant has to be taken at certain values of these variables. In a
determination of the \textit{physical} coupling constant we must use the
Feynman rules from Table~\ref{tab:7}. A possible choice of the
Mandelstam variables is
\begin{equation}\label{eq:51}
  \begin{aligned}
    &\text{for $\Lambda_{1}$:}\quad s=4M_{1}^{2},\quad &&t=0,\quad u=0,\\
    &\text{for $\Lambda_{2}$:}\quad s=4M_{2}^{2},\quad &&t=0,\quad u=0,\\
    &\text{for $\Lambda_{3}$:}\quad s=4M_{1}M_{2},\quad &&t=0,\quad u=0.
  \end{aligned}
\end{equation}
We will consider other choices when we discuss the renormalization
group equations.

\subsubsection{Renormalization of the coupling constants}
\label{sec:renorm-coupl-const}

Using Eq.~\eqref{eq:53}, prescription~\eqref{eq:51} and the Feynman
rules from Table~\ref{tab:7} we get for the $\Lambda_{1}$ four point
Green's function, which should be finite at the one loop level
\begin{multline}\label{eq:60}
  \displaystyle\Lambda_{1}\mu^{4-d} + \frac{\Lambda_{1}^{2}\mu^{4-d}}{16\pi^{2}}\Big(\frac{3}{d-4} +\frac{3}{2}\gamma+ \frac{1}{2}\big(F_{\Sigma}(4M_{1}^{2},0,0,M_{1}^{2},\mu^{2})\big)\Big)\\
  +\frac{\Lambda_{3}^{2}\mu^{4-d}}{16\pi^{2}}\Big(\frac{3}{d-4}
  +\frac{3}{2}\gamma+
  \frac{1}{2}\big(F_{\Sigma}(4M_{1}^{2},0,0,M_{2}^{2},\mu^{2})\big)\Big)
  +\delta_{\Lambda_{1}},
\end{multline}
so in order to cancel the infinite terms in Eq.~\eqref{eq:60} we choose $\delta_{\Lambda_{1}}$ to be equal
\begin{multline}\label{eq:61}
  \delta_{\Lambda_{1}}=
  -\frac{\Lambda_{1}^{2}\mu^{4-d}}{16\pi^{2}}\Big(\frac{3}{d-4} +\frac{3}{2}\gamma+ \frac{1}{2}\big(F_{\Sigma}(4M_{1}^{2},0,0,M_{1}^{2},\mu^{2})\big)\Big)\\
  -\frac{\Lambda_{3}^{2}\mu^{4-d}}{16\pi^{2}}\Big(\frac{3}{d-4}
  +\frac{3}{2}\gamma+
  \frac{1}{2}\big(F_{\Sigma}(4M_{1}^{2},0,0,M_{2}^{2},\mu^{2})\big)\Big).
\end{multline}
Analogously we obtain
\begin{multline}\label{eq:77}
  \delta_{\Lambda_{2}}=
  -\frac{\Lambda_{2}^{2}\mu^{4-d}}{16\pi^{2}}\Big(\frac{3}{d-4} +\frac{3}{2}\gamma+ \frac{1}{2}\big(F_{\Sigma}(4M_{2}^{2},0,0,M_{2}^{2},\mu^{2})\big)\Big)\\
  -\frac{\Lambda_{3}^{2}\mu^{4-d}}{16\pi^{2}}\Big(\frac{3}{d-4}
  +\frac{3}{2}\gamma+
  \frac{1}{2}\big(F_{\Sigma}(4M_{2}^{2},0,0,M_{1}^{2},\mu^{2})\big)\Big).
\end{multline}
For $\Lambda_{3}$ four point Green's function we have
\begin{multline}
  \label{eq:90}
\Lambda_{3}\mu^{4-d} + \frac{\Lambda_{1}\Lambda_{3} \mu^{4-d}}{16\pi^{2}}\Big(\frac{1}{d-4} +\frac{1}{2}(\gamma+ F(4M_{1}M_{2},M_{1}^{2},\mu^{2}))\Big)\\ + \frac{\Lambda_{2}\Lambda_{3} \mu^{4-d}}{16\pi^{2}}\Big(\frac{1}{d-4} +\frac{1}{2}(\gamma+ F(4M_{1}M_{2},M_{2}^{2},\mu^{2}))\Big)\\ +\frac{\Lambda_{3}^{2}\mu^{4-d}}{8\pi^{2}}\Big(\frac{1}{d-4} +\frac{1}{2}(\gamma+ \int_{0}^{1}\dd{z}\ln(\frac{\vert z(1-z)4M_{1}M_{2}-zM_{1}^{2}-(1-z) M_{2}^{2}  \vert}{4\pi\mu^{2}}))\Big)\\ +\frac{\Lambda_{3}^{2}\mu^{4-d}}{8\pi^{2}}\Big(\frac{1}{d-4} +\frac{1}{2}(\gamma+ \int_{0}^{1}\dd{z}\ln(\frac{\vert zM_{1}^{2}+(1-z) M_{2}^{2}  \vert}{4\pi\mu^{2}}))\Big)
\end{multline}
and $\delta_{\Lambda_{3}}$ is equal
\begin{multline}\label{eq:78}
  \delta_{\Lambda_{3}}=
-\frac{\Lambda_{1}\Lambda_{3} \mu^{4-d}}{16\pi^{2}}\Big(\frac{1}{d-4} +\frac{1}{2}(\gamma+ F(4M_{1}M_{2},M_{1}^{2},\mu^{2}))\Big)\\ - \frac{\Lambda_{2}\Lambda_{3} \mu^{4-d}}{16\pi^{2}}\Big(\frac{1}{d-4} +\frac{1}{2}(\gamma+ F(4M_{1}M_{2},M_{2}^{2},\mu^{2}))\Big)\\ -\frac{\Lambda_{3}^{2}\mu^{4-d}}{8\pi^{2}}\Big(\frac{1}{d-4} +\frac{1}{2}(\gamma+ \int_{0}^{1}\dd{z}\ln(\frac{\vert z(1-z)4M_{1}M_{2}-zM_{1}^{2}-(1-z) M_{2}^{2}  \vert}{4\pi\mu^{2}}))\Big)\\ -\frac{\Lambda_{3}^{2}\mu^{4-d}}{8\pi^{2}}\Big(\frac{1}{d-4} +\frac{1}{2}(\gamma+ \int_{0}^{1}\dd{z}\ln(\frac{\vert zM_{1}^{2}+(1-z) M_{2}^{2}  \vert}{4\pi\mu^{2}}))\Big).
\end{multline}

Putting all previous results together and taking the limit
$d\rightarrow 4$ we obtain the following result for the one loop
amplitude of the process of elastic scattering of two particles of the
type~1
\begin{multline}
  \label{eq:62}
  \displaystyle\Lambda_{1} + \frac{\Lambda_{1}^{2}}{32\pi^{2}} \big(F_{\Sigma}(s,t,u,M_{1}^{2},\mu^{2})-F_{\Sigma} (4M_{1}^{2},0,0,M_{1}^{2},\mu^{2})\big)\\
  +\frac{\Lambda_{3}^{2}}{32\pi^{2}}
  \big(F_{\Sigma}(s,t,u,M_{2}^{2},\mu^{2})
  -F_{\Sigma}(4M_{1}^{2},0,0,M_{2}^{2},\mu^{2})\big).
\end{multline}

The coupling constants $\Lambda_{1}$, $\Lambda_{2}$ and $\Lambda_{3}$
are chosen to be finite. From Eqs.~\eqref{eq:59} it then follows that the
bare coupling constants $\lambda_{1}$, $\lambda_{2}$ and $\lambda_{3}$
must have a pole at~$d=4$.

The one loop amplitude for the process of elastic scattering of two
particles of the type 2 is equal
\begin{multline}
  \label{eq:63}
  \displaystyle\Lambda_{2} + \frac{\Lambda_{2}^{2}}{32\pi^{2}} \big(F_{\Sigma}(s,t,u,M_{2}^{2},\mu^{2})-F_{\Sigma} (4M_{2}^{2},0,0,M_{2}^{2},\mu^{2})\big)\\
  +\frac{\Lambda_{3}^{2}}{32\pi^{2}}
  \big(F_{\Sigma}(s,t,u,M_{1}^{2},\mu^{2})
  -F_{\Sigma}(4M_{2}^{2},0,0,M_{1}^{2},\mu^{2})\big)
\end{multline}
and the one loop amplitude for the process of elastic scattering of
two particles, one of type 1 and the other of type 2 is equal
\begin{multline}
  \label{eq:64}
  \Lambda_{3}+ \frac{\Lambda_{1}\Lambda_{3}}{16\pi^{2}}(F(s,M_{1}^{2},\mu^{2}) -F(4M_{1}M_{2},M_{1}^{2},\mu^{2}))\\ + \frac{\Lambda_{2}\Lambda_{3}}{16\pi^{2}}(F(s,M_{2}^{2},\mu^{2}) -F(4M_{1}M_{2},M_{2}^{2},\mu^{2}))\\
+  \frac{\Lambda_{3}^{2}}{16\pi^{2}}\Big( \int_{0}^{1}\dd{z}\ln(\frac{\vert z(1-z)s-zM_{1}^{2}-(1-z) M_{2}^{2}  \vert} {\vert z(1-z)4M_{1}M_{2}-zM_{1}^{2}-(1-z) M_{2}^{2}  \vert})\Big)\\
  +\frac{\Lambda_{3}^{2}}{16\pi^{2}}\Big(
  \int_{0}^{1}\dd{z}\ln(\frac{\vert z(1-z)u-zM_{1}^{2}-(1-z) M_{2}^{2}
    \vert} {\vert zM_{1}^{2}+(1-z) M_{2}^{2} \vert})\Big).
\end{multline}

\subsubsection{Renormalization of the masses}
\label{sec:renorm-mass}

The propagator of particle~1 is equal (at one loop $\delta_{Z_{1}}=0$ and $\delta_{Z_{2}}=0$)
\begin{equation}
  \label{eq:65}
  \frac{i}{p^{2}-M_{1}^{2}-\delta_{M_{1}}-\Sigma_{1}-\Sigma_{3}^{1}}.
\end{equation}
Here we were using the Feynman rules given in Table~\ref{tab:7} and
$\Sigma_{1}$ an $\Sigma_{3}^{1}$ are equal
\begin{subequations}\label{eq:66}
  \begin{align}
    &\label{eq:66a} \Sigma_{1} =\frac{\Lambda_{1}M_{1}^{2}}{16\pi^{2}}\Big(
      \frac{1}{d-4}+ \frac{1}{2} \big(\gamma-1 +\ln(\frac{M_{1}^{2}}{4\pi\mu^{2}}
      )\big) \Big),\\
    &\label{eq:66b} \Sigma_{3}^{1} =\frac{\Lambda_{3} M_{2}^{2}}{16\pi^{2}}\Big(
      \frac{1}{d-4}+ \frac{1}{2} \big(\gamma-1 +\ln(\frac{M_{2}^{2}}{4\pi\mu^{2}}
      )\big) \Big).
  \end{align}
\end{subequations}
The propagator has a pole at the value of the physical mass, which is
equal to $M_{1}$, so from Eq.~\eqref{eq:65} we obtain the value of
$\delta_{M_{1}}$
\begin{equation}
  \label{eq:68}
  \delta_{M_{1}}=-\Sigma_{1}-\Sigma_{3}^{1}
\end{equation}
and also for $\delta_{M_{2}}$, after the interchange of the indices
$1\leftrightarrow2$
\begin{equation}
  \label{eq:67}
  \delta_{M_{2}}=-\Sigma_{2}-\Sigma_{3}^{2},
\end{equation}
where $\Sigma_{2}$ and $\Sigma_{3}^{2}$ are equal
\begin{subequations}\label{eq:69}
  \begin{align}
    &\label{eq:69a} \Sigma_{2} =\frac{\Lambda_{2} M_{2}^{2}}{16\pi^{2}}\Big(
      \frac{1}{d-4}+ \frac{1}{2} \big(\gamma-1 +\ln(\frac{M_{2}^{2}}{4\pi\mu^{2}}
      )\big) \Big),\\
    &\label{eq:69b} \Sigma_{3}^{2} =\frac{\Lambda_{3} M_{1}^{2}}{16\pi^{2}}\Big(
      \frac{1}{d-4}+ \frac{1}{2} \big(\gamma-1 +\ln(\frac{M_{1}^{2}}{4\pi\mu^{2}}
      )\big) \Big).    
  \end{align}
\end{subequations}
The corrections $\delta_{M_{1}}$ and $\delta_{M_{2}}$ have poles at
$d=4$, so from Eq.~\eqref{eq:59} it follows that the bare masses
$m_{1}^{2}$ and $m_{2}^{2}$ must also have poles at $d=4$.

\begin{figure}[!hb]
  \centering
  \includegraphics[width=0.45\linewidth]{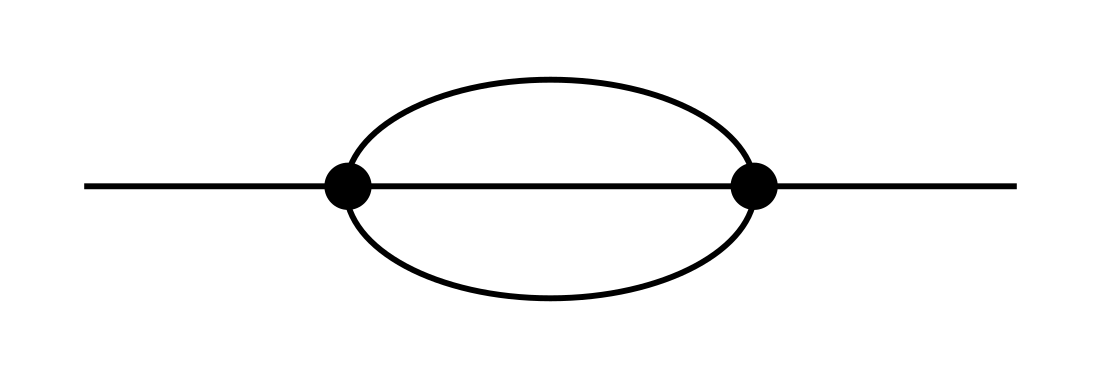}
  \caption{\label{fig:12} The \textit{setting sun} diagram in the
    $\phi^{4}$ theory, which is divergent and requires the
    introduction of a correction, related with the normalization of
    the scalar field.}
\end{figure}

One should mention here that at the next order of the perturbation
theory and also at two loops, there is an additional type of the
divergent diagram of the type shown in Fig.~\ref{fig:12}. This
divergence is removed by the parameters $\delta_{Z_{1}}\neq0$ and
$\delta_{Z_{2}}\neq0$ and the normalization of the fields $\phi_{1}$
and $\phi_{2}$ is modified. It will be discussed in
Sec.~\ref{sec:renorm-at-two}.

\section{One loop renormalization group equations}
\label{sec:renorm-group-at}

\subsection{Derivation of equations}
\label{sec:derivation-equations}

Renormalization group equations (RGE) in field
theory~\cite{Stueckelberg, PhysRev.95.1300} provide a method for the
study of the asymptotic behavior of the Green's functions. The first
step in the derivation of the RGEs is the demonstration that the
Green's functions do not depend on the cut-off parameter $\mu$ and the
determination of the equations for the Green's functions. From the
split form of the Lagrangian density~\eqref{eq:58} it follows that we
have two sets of the Green's functions: one set for the original
Lagrangian density, dependent on the parameters $m_{1}$, $m_{2}$,
$\lambda_{1}$, $\lambda_{2}$, $\lambda_{3}$ and the other set of
Green's functions obtained from the Lagrangian density after
splitting, dependent on $M_{1}$, $M_{2}$, $\Lambda_{1}$,
$\Lambda_{2}$, $\Lambda_{3}$ and $\mu$. These two sets of Green's
functions describe the same theory, so they have to be equal and we
have the following relations between these Green's functions in the
space-time dimension $d$
\begin{multline}
  \label{eq:70}
  G^{(n)}_{i_{1}\ldots i_{n}}(p_{1},\ldots p_{n},\lambda_{1}, \lambda_{2}, \lambda_{3},
  m_{1},m_{2}, d)\\ = Z_{\phi_{1}}^{-n_{1}/2} Z_{\phi_{2}}^{-n_{2}/2} \mathbf{G}^{(n)}_{i_{1}\ldots i_{n}}(p_{1},\ldots, p_{n},\Lambda_{1}, \Lambda_{2}, \Lambda_{3},
  M_{1},M_{2}, d,\mu).
\end{multline}
Here the function $G^{(n)}_{i_{1}\ldots i_{n}}$ is the Green's
function before the splitting of the Lagrangian density and
$\mathbf{G}^{(n)}_{i_{1}\ldots i_{n}}$ is the Green's function after
the splitting, which is finite. The $n_{1}$ and $n_{2}$ denote the
number of fields in the Green's function of type~1 and type~2,
respectively, ($n_{1}+n_{2}=n$).

The left-hand side of Eq.~\eqref{eq:70} does not depend on $\mu$, so
the right-hand side cannot depend on $\mu$ either. If we differentiate
Eq.~\eqref{eq:70} with respect to $\mu$ then we obtain
\begin{multline}
  \label{eq:71}
  \Bigg[\mu\frac{\partial}{\partial\mu} +\mu\frac{\partial\Lambda_{1}} {\partial\mu}\frac{\partial}{\partial\Lambda_{1}} +\mu\frac{\partial\Lambda_{2}} {\partial\mu}\frac{\partial}{\partial\Lambda_{2}} +\mu\frac{\partial\Lambda_{3}} {\partial\mu}\frac{\partial}{\partial\Lambda_{3}}\\ +\mu\frac{\partial M_{1}} {\partial\mu}\frac{\partial}{\partial M_{1}} +\mu\frac{\partial M_{2}} {\partial\mu}\frac{\partial}{\partial M_{2}} -\mu\frac{n_{1}}{2} \frac{\partial\ln Z_{\varphi_{1}}}{\partial\mu}\\ -\mu\frac{n_{2}}{2} \frac{\partial\ln Z_{\varphi_{2}}}{\partial\mu}\Bigg]\mathbf{G}^{(n)}_{i_{1}\ldots i_{n}}(p_{1},\ldots, p_{n},\Lambda_{1}, \Lambda_{2}, \Lambda_{3},
  M_{1},M_{2}, d,\mu) =0.
\end{multline}
Eq.~\eqref{eq:71} is the condition that the Green's functions have to
fulfill that the physical predictions of the theory do not depend on
the choice of the renormalization point $\mu$. Conventionally one
introduces the notation
\begin{equation}
  \label{eq:72}
  \begin{aligned}
    \beta_{\Lambda_{i}}(\Lambda_{1}, \Lambda_{2}, \Lambda_{3}, \frac{M_{1}}{\mu}, \frac{M_{2}}{\mu},d)&=\mu\frac{\partial\Lambda_{i}}{\partial\mu}, \quad  &&i=1,2,3,\\
    \gamma_{d_{i}}(\Lambda_{1}, \Lambda_{2}, \Lambda_{3}, \frac{M_{1}}{\mu}, \frac{M_{2}}{\mu},d)&= \frac{\mu}{2}\frac{\partial\ln Z_{\varphi_{i}}}{\partial\mu}, \quad &&i=1,2,\\
    \gamma_{M_{i}}(\Lambda_{1}, \Lambda_{2}, \Lambda_{3}, \frac{M_{1}}{\mu}, \frac{M_{2}}{\mu},d)&= \frac{\mu}{2}\frac{\partial \ln M^{2}_{i}} {\partial\mu}, \quad &&i=1,2.
  \end{aligned}
\end{equation}

Next, one considers the Green's function with the momenta scaled by the
factor~$t$:
$\mathbf{G}^{(n)}_{i_{1}\ldots i_{n}}(tp_{1},\ldots,
tp_{n},\Lambda_{1}, \Lambda_{2}, \Lambda_{3}, M_{1},M_{2}, d,\mu)$,
which has the dimension $4-n+\frac{(4-d)(n-2)}{2}$ and fulfills the
following scaling equation
  \begin{multline}
    \label{eq:73}
    \Bigg[ \mu\frac{\partial}{\partial\mu}  +t\frac{\partial}{\partial t} +M_{1}\frac{\partial}{\partial M_{1}} +M_{2}\frac{\partial}{\partial M_{2}}   -\Big(4-n+\frac{(4-d)(n-2)}{2}\Big) \Bigg]\\
    \mathbf{G}^{(n)}_{i_{1}\ldots i_{n}}(t p_{1},\ldots, t p_{n},\Lambda_{1}, \Lambda_{2}, \Lambda_{3},
  M_{1},M_{2}, d,\mu)=0.
  \end{multline}
  After subtracting Eq.~\eqref{eq:73} from Eq.~\eqref{eq:71} and taking the limit $d\rightarrow4$ we obtain an equation that describes the scaling properties of the Green's functions
  \begin{multline}
    \label{eq:74}
    \Bigg[ -t\frac{\partial}{\partial t} +\beta_{\Lambda_{1}} \frac{\partial}{\partial\Lambda_{1}} +\beta_{\Lambda_{2}} \frac{\partial}{\partial\Lambda_{2}} +\beta_{\Lambda_{3}} \frac{\partial}{\partial\Lambda_{3}}  +\Big(\gamma_{M_{1}}-1\Big)M_{1} \frac{\partial}{\partial M_{1}}\\ +\Big(\gamma_{M_{2}}-1\Big)M_{2} \frac{\partial}{\partial M_{2}} -n_{1}\gamma_{d_{1}} -n_{2}\gamma_{d_{2}} +4-n\Bigg]\\
    \mathbf{G}^{(n)}_{i_{1}\ldots i_{n}}(t p_{1},\ldots, t
    p_{n},\Lambda_{1}, \Lambda_{2}, \Lambda_{3}, M_{1},M_{2},
    4,\mu)=0.
  \end{multline}
The solution of Eq.~\eqref{eq:74} can be expressed in the following way
\begin{multline}
  \label{eq:75}
  \mathbf{G}^{(n)}_{i_{1}\ldots i_{n}}(t p_{1},\ldots, t
    p_{n},\Lambda_{1}, \Lambda_{2}, \Lambda_{3}, M_{1},M_{2},
    4,\mu)\\ = t^{4-n}\exp\Big[-n_{1}\int_{1}^{t}\frac{\gamma_{M_{1}}(\tau) \dd \tau}{\tau}\Big] \exp\Big[-n_{2}\int_{1}^{t}\frac{\gamma_{M_{2}}(\tau)\dd \tau} {\tau}\Big]\\ \times\mathbf{G}^{(n)}_{i_{1}\ldots i_{n}}(p_{1},\ldots, p_{n},\Lambda_{1}(t), \Lambda_{2}(t), \Lambda_{3}(t), M_{1}(t),M_{2}(t),
    4,\mu).
\end{multline}
Here $\Lambda_{1}(t)$, $\Lambda_{2}(t)$, $\Lambda_{3}(t)$, $M_{1}(t)$,
$M_{2}(t)$ are the \textit{running} coupling constants and masses that
fulfill the following equations and initial conditions
\begin{subequations}
  \label{eq:76}
  \begin{align}
    \label{eq:76a}
    \Scale[0.95]{ t\frac{\dd \Lambda_{i}(t)}{\dd t}}&
\Scale[0.95]{= \beta_{\Lambda_{i}}(\Lambda_{1}(t), \Lambda_{2}(t), \Lambda_{3}(t), M_{1}(t), M_{2}(t)), \quad i=1,2,3,}\\
    \label{eq:76b}
    \Scale[0.98]{ t\frac{\dd M_{i}(t)}{\dd t}} &\Scale[0.98]{ = M_{i}(t)(\gamma_{M_{i}}(\Lambda_{1}(t), \Lambda_{2}(t), \Lambda_{3}(t), M_{1}(t), M_{2}(t))-1), \;  i=1,2.}\\
\Scale[0.98]{  \Lambda_{1}(0)} &\Scale[0.98]{ =\Lambda_{1},\; \Lambda_{2}(0)=\Lambda_{2},\; \Lambda_{3}(0) =\Lambda_{3},\; M_{1}(0)=M_{1},\; M_{2}(0)=M_{2}.}
  \end{align}
\end{subequations}
Eq.~\eqref{eq:75} is the key relation of the renormalization group
method: it relates the Green's functions at scaled momenta and
unscaled coupling constants with Green's functions at unscaled momenta
and scaled coupling constants.  Eqs.~\eqref{eq:76} are called the
\textit{renormalization group} equations for the \textit{running}
coupling constants and masses. The solutions of these equations, with
the help of Eq.~\eqref{eq:75}, give the information about the
asymptotic behavior of the theory. The right-hand side of
Eqs.~\eqref{eq:76}, the functions
$\beta_{\Lambda_{i}}(\Lambda_{1}(t), \Lambda_{2}(t), \Lambda_{3}(t),
M_{1}(t), M_{2}(t))$ and
$\gamma_{M_{i}}(\Lambda_{1}(t), \Lambda_{2}(t), \Lambda_{3}(t),
M_{1}(t), M_{2}(t))$, can only be determined perturbatively and we
will find now their form at order of one loop.

\subsection{Calculation of the $\beta$ functions}
\label{sec:deriv-beta-funct}

The definition of the $\beta$ and $\gamma$ functions is given in
Eqs.~\eqref{eq:72}, which show that in general they depend on the
coupling constants $\Lambda_{i}$ and the masses $M_{i}$. The
$\beta_{\Lambda_{i}}$ functions contain the derivatives of
$\Lambda_{i}$ with respect to $\mu$, the mass parameter in the
dimensional renormalization. The calculation of this derivative is not
a straightforward matter, because we do not know the dependence of
$\Lambda_{i}$ on $\mu$, but we know the dependence of the
\textit{bare} coupling constants $\lambda_{i}$ as functions of the
renormalized coupling constants $\Lambda_{i}$.

Let us start with the determination of the $\beta$ functions, which
are analytic at $d=4$. It means that the expansion around point $d=4$
has the form
\begin{gather}
  \label{eq:79}
  \beta_{\Lambda_{i}}(\Lambda_{1},\Lambda_{2},\Lambda_{3},d,\mu) =\sum_{l=0}^{\infty} b_{l}^{\Lambda_{i}}(\Lambda_{1},\Lambda_{2},\Lambda_{3},\mu)(d-4)^{l}\\
  \intertext{and at $d=4$}
  \mu\frac{\partial\Lambda_{i}}{\partial\mu}\Bigg\vert_{d=4} =\beta_{\Lambda_{i}}(\Lambda_{1},\Lambda_{2},\Lambda_{3},4,\mu) = b_{0}^{\Lambda_{i}},
\end{gather}
so we have to calculate $b_{0}^{\Lambda_{i}}$.

The unrenormalized coupling constant $\lambda_{1}$
which is independent of the cut off $\mu$ has a Laurent expansion
around point $d=4$
\begin{equation}\label{eq:80}
  \lambda_{1}=\frac{\Lambda_{1}+\delta_{\lambda_{1}}}{Z_{\varphi_{1}}^{2}} =
  \mu^{4-d}\Bigg[\Lambda_{1}+\sum_{l=1}^{\infty}\frac{a_{l}^{\Lambda_{1}} (\Lambda_{1},\Lambda_{2},\Lambda_{3})}{(d-4)^{l}}\Bigg].
\end{equation}
The right hand side of Eq.~\eqref{eq:80} depends on $\mu$ explicitly through the factor $\mu^{4-d}$ and implicitly through the coupling constants $\Lambda_{i}$. If we differentiate Eq.~\eqref{eq:80} with respect to $\mu$ we obtain
\begin{multline}
  \label{eq:81}
  0=\mu\frac{\partial \lambda_{1}}{\partial \mu} =(4-d) \mu^{4-d}\Bigg[\Lambda_{1}+\sum_{l=1}^{\infty}\frac{a_{l}^{\Lambda_{1}}} {(d-4)^{l}}\Bigg]\\ +  \mu^{4-d}\mu\frac{\partial \Lambda_{1}} {\partial\mu}\Bigg[1 +\sum_{l=1}^{\infty}\frac{\partial a_{l}^{\Lambda_{1}}} {\partial\Lambda_{1}}\frac{1}{(d-4)^{l}}\Bigg] +  \mu^{4-d}\mu\frac{\partial \Lambda_{2}} {\partial\mu}\sum_{l=1}^{\infty}\frac{\partial a_{l}^{\Lambda_{1}}} {\partial\Lambda_{2}}\frac{1}{(d-4)^{l}}\\
  +  \mu^{4-d}\mu\frac{\partial \Lambda_{3}} {\partial\mu}\sum_{l=1}^{\infty}\frac{\partial a_{l}^{\Lambda_{1}}} {\partial\Lambda_{3}}\frac{1}{(d-4)^{l}} =(4-d) \mu^{4-d}\Bigg[\Lambda_{1}+\sum_{l=1}^{\infty}\frac{a_{l}^{\Lambda_{1}}} {(d-4)^{l}}\Bigg]\\ +  \mu^{4-d}\beta_{\Lambda_{1}}\Bigg[1 +\sum_{l=1}^{\infty}\frac{\partial a_{l}^{\Lambda_{1}}} {\partial\Lambda_{1}}\frac{1}{(d-4)^{l}}\Bigg] +  \mu^{4-d}\beta_{\Lambda_{2}}\sum_{l=1}^{\infty}\frac{\partial a_{l}^{\Lambda_{1}}} {\partial\Lambda_{2}}\frac{1}{(d-4)^{l}}\\
  +  \mu^{4-d}\beta_{\Lambda_{3}}\sum_{l=1}^{\infty}\frac{\partial a_{l}^{\Lambda_{1}}} {\partial\Lambda_{3}}\frac{1}{(d-4)^{l}}.
\end{multline}
The coefficients at various powers of $(d-4)$ in Eq.~\eqref{eq:81} are equal
\begin{subequations}  \label{eq:82}
\begin{align}
  \label{eq:82a}
&    \Scale[0.91]{ -a_{1}^{\Lambda_{1}}+b_{0}^{\Lambda_{1}}+\sum_{l=1}^{\infty} \Big(b_{l}^{\Lambda_{1}} \frac{\partial a_{l}^{\Lambda_{1}}}{\partial\Lambda_{1}} + b_{l}^{\Lambda_{2}} \frac{\partial a_{l}^{\Lambda_{1}}}{\partial\Lambda_{2}} +b_{l}^{\Lambda_{3}} \frac{\partial a_{l}^{\Lambda_{1}}}{\partial\Lambda_{3}}\Big)=0\quad \text{at }(d-4)^{0}}\\
  \label{eq:82b}
&      \Scale[0.91]{ -\Lambda_{1}+b_{1}^{\Lambda_{1}}+\sum_{l=1}^{\infty} \Big( b_{l+1}^{\Lambda_{1}} \frac{\partial a_{l}^{\Lambda_{1}}}{\partial\Lambda_{1}} + b_{l+1}^{\Lambda_{2}} \frac{\partial a_{l}^{\Lambda_{1}}}{\partial\Lambda_{2}} +b_{l+1}^{\Lambda_{3}} \frac{\partial a_{l}^{\Lambda_{1}}}{\partial\Lambda_{3}}\Big)=0\quad \text{at }(d-4)^{1}}\\
  \label{eq:82c}
&      \Scale[0.91]{ b_{k}^{\Lambda_{1}}+\sum_{l=1}^{\infty} \Big(b_{l+k}^{\Lambda_{1}} \frac{\partial a_{l}^{\Lambda_{1}}}{\partial\Lambda_{1}} + b_{l+k}^{\Lambda_{2}} \frac{\partial a_{l}^{\Lambda_{1}}}{\partial\Lambda_{2}} +b_{l+k}^{\Lambda_{3}} \frac{\partial a_{l}^{\Lambda_{1}}}{\partial\Lambda_{3}}\Big)=0\quad \text{at }(d-4)^{k},\; k\geq2}\\
  \label{eq:82d}
&   \Scale[0.91]{ -a_{k+1} +\sum_{l=1}^{\infty} \Big(b_{l}^{\Lambda_{1}} \frac{\partial a_{l+k}^{\Lambda_{1}}}{\partial\Lambda_{1}} + b_{l}^{\Lambda_{2}} \frac{\partial a_{l+k}^{\Lambda_{1}}}{\partial\Lambda_{2}} +b_{l}^{\Lambda_{3}} \frac{\partial a_{l+k}^{\Lambda_{1}}}{\partial\Lambda_{3}}\Big)=0\quad \text{at }(d-4)^{-k},\;k\geq1.}
\end{align}
\end{subequations}
Similar equations hold for
$\mu\frac{\partial\lambda_{2}}{\partial\mu}$ and
$\mu\frac{\partial\lambda_{3}}{\partial\mu}$. Eqs.~\eqref{eq:82} are
the recursive equations for the coefficients $b_{k}^{\Lambda_{1}}$ of
expansion of the beta functions in terms of the coefficients
$a_{k}^{\Lambda_{i}}$ that are calculated in the renormalization process. From Eqs.~\eqref{eq:82c}, which are linear homogeneous equations it follows that
\begin{equation}
  \label{eq:83}
  b_{k}^{\Lambda_{i}}=0\quad \text{for }k\geq2\text{ and }i=1,2,3.
\end{equation}
It means that we have the following equations for the coefficients $b_{0}^{\Lambda_{i}}$ and $b_{1}^{\Lambda_{i}}$
\begin{equation}\left.
  \label{eq:84}
  \begin{aligned}
    -a_{1}^{\Lambda_{i}}+b_{0}^{\Lambda_{i}}+ \Big(b_{1}^{\Lambda_{1}} \frac{\partial a_{1}^{\Lambda_{i}}}{\partial\Lambda_{1}} + b_{1}^{\Lambda_{2}} \frac{\partial a_{1}^{\Lambda_{i}}}{\partial\Lambda_{2}} +b_{1}^{\Lambda_{3}} \frac{\partial a_{1}^{\Lambda_{i}}}{\partial\Lambda_{3}}\Big)=0\\
    -\Lambda_{i}+b_{1}^{\Lambda_{i}}=0
  \end{aligned}\right\}\quad i=1,2,3,
\end{equation}
which immediately give
\begin{equation}
  \label{eq:85}
 \beta_{\Lambda_{i}} = b_{0}^{\Lambda_{i}}= a_{1}^{\Lambda_{i}} -\Big(\Lambda_{1} \frac{\partial a_{1}^{\Lambda_{i}}}{\partial\Lambda_{1}} + \Lambda_{2} \frac{\partial a_{1}^{\Lambda_{i}}}{\partial\Lambda_{2}} +\Lambda_{3} \frac{\partial a_{1}^{\Lambda_{i}}}{\partial\Lambda_{3}}\Big).
\end{equation}

Let us calculate now the functions $\beta_{\Lambda_{i}}$ at one loop. From Eqs.~\eqref{eq:59} and~\eqref{eq:61} we have ($Z_{\phi_{i}}=1$ at one loop)
\begin{multline}\label{eq:86}
    \lambda_{1}=\Lambda_{1}+\delta_{\lambda_{1}}
  =\Lambda_{1}
  -\frac{\Lambda_{1}^{2}\mu^{4-d}}{16\pi^{2}}\Big(\frac{3}{d-4} +\frac{3}{2}\gamma+ \frac{1}{2}\big(F_{\Sigma}(4M_{1}^{2},0,0,M_{1}^{2},\mu^{2})\big)\Big)\\
  -\frac{\Lambda_{3}^{2}\mu^{4-d}}{16\pi^{2}}\Big(\frac{3}{d-4}
  +\frac{3}{2}\gamma+
  \frac{1}{2}\big(F_{\Sigma}(4M_{1}^{2},0,0,M_{2}^{2},\mu^{2})\big)\Big)
\end{multline}
and we see that $a_{1}^{\Lambda_{1}}$ is equal
\begin{equation}
  \label{eq:87}
  a_{1}^{\Lambda_{1}}= -\frac{3(\Lambda_{1}^{2}+\Lambda_{3}^{2})}{16\pi^{2}},
\end{equation}
so $\beta_{\Lambda_{1}}$ is equal
\begin{equation}
  \label{eq:88}
    \beta_{\Lambda_{1}}= \frac{3(\Lambda_{1}^{2}+\Lambda_{3}^{2})}{16\pi^{2}}.
\end{equation}
$\beta_{\Lambda_{2}}$ is obtained by the exchange of the indices
\begin{equation}
  \label{eq:89}
      \beta_{\Lambda_{2}}= \frac{3(\Lambda_{2}^{2}+\Lambda_{3}^{2})}{16\pi^{2}}.
\end{equation}
From Eq.~\eqref{eq:78} we obtain the value of $a_{1}^{\Lambda_{3}}$
\begin{equation}
  \label{eq:91}
  a_{1}^{\Lambda_{3}}=-\frac{(\Lambda_{1}+\Lambda_{2})\Lambda_{3}}{16\pi^{2}} -\frac{\Lambda_{3}^{2}}{4\pi^{2}}
\end{equation}
and $\beta_{\Lambda_{3}}$ 
\begin{equation}
  \label{eq:92}
  \beta_{\Lambda_{3}}=\frac{(\Lambda_{1}+\Lambda_{2})\Lambda_{3}}{16\pi^{2}} +\frac{\Lambda_{3}^{2}}{4\pi^{2}}.
\end{equation}

The one loop renormalization group equations have thus the form
\begin{subequations}
  \label{eq:93}
  \begin{align}
    \label{eq:93a}
    t\frac{\partial\Lambda_{1}}{\partial t} &= \frac{3(\Lambda_{1}^{2}+\Lambda_{3}^{2})}{16\pi^{2}},\\
    t\frac{\partial\Lambda_{2}}{\partial t} &= \frac{3(\Lambda_{2}^{2}+\Lambda_{3}^{2})}{16\pi^{2}},\\
    t\frac{\partial\Lambda_{3}}{\partial t} &= \frac{(\Lambda_{1}+\Lambda_{2})\Lambda_{3}}{16\pi^{2}} +\frac{\Lambda_{3}^2}{4\pi^{2}}.
  \end{align}
\end{subequations}

We will now find the renormalization group equations for the masses~$M_{i}$. The procedure is similar to the derivation of the renormalization group equations for the coupling constant. The functions $\gamma_{M_{i}}$ from Eq.~\eqref{eq:72} are analytic at $d=4$ and have the expansion
\begin{gather}
  \label{eq:94}
  \gamma_{M_{i}}(\Lambda_{1},\Lambda_{2},\Lambda_{3},d,\mu) =\sum_{l=0}^{\infty} g_{l}^{M_{i}}(\Lambda_{1},\Lambda_{2},\Lambda_{3},\mu)(d-4)^{l}\\
  \shortintertext{and at $d=4$}
\label{eq:95}  \frac{\mu}{2}\frac{\partial M_{i}^{2}}{\partial\mu}\Bigg\vert_{d=4} =\gamma_{M_{i}}(\Lambda_{1},\Lambda_{2},\Lambda_{3},4,\mu) = g_{0}^{M_{i}}.  
\end{gather}
We also have the relation between $m_{i}^{2}$ and $M_{i}^{2}$
\begin{gather}
  \label{eq:96}
  m_{i}^{2}=M_{1}^{2}\sum_{l=0}^{\infty}\frac{b_{l}^{M_{i}} (\Lambda_{1},\Lambda_{2},\Lambda_{3})}{(d-4)^{l}}  + M_{2}^{2}  \sum_{l=0}^{\infty}\frac{c_{l}^{M_{i}} (\Lambda_{1},\Lambda_{2},\Lambda_{3})}{(d-4)^{l}},\\
  b_{0}^{M_{1}}=1,\quad b_{0}^{M_{2}}=0,\quad c_{0}^{M_{1}}=0,\quad c_{0}^{M_{2}}=1. \nonumber
\end{gather}
At one loop Eq.~\eqref{eq:96} for $m_{1}^{2}$ reads
\begin{equation}
  \label{eq:97}
  m_{1}^{2}=M_{1}^{2}=\frac{M_{1}^{2}+\delta_{m_{1}}}{Z_{\phi_{1}}} =
  M_{1}^{2}- \frac{\Lambda_{1}M_{1}^{2} + \Lambda_{3}M_{2}^{2} }{16\pi^{2}(d-4)}.
\end{equation}
The bare masses $m_{i}$ do not depend on $\mu$, so differentiating Eq.~\eqref{eq:96} with respect to $\mu$ we obtain
\begin{equation}
  \label{eq:98}
0=  \mu\frac{\partial m_{1}^{2}}{\partial\mu} = \mu\frac{\partial M_{1}^{2}}{\partial\mu}  - \frac{1}{16\pi^{2}(d-4)} \Big( M_{1}^{2} \mu \frac{\partial\Lambda_{1}} {\partial\mu}  + M_{2}^{2} \mu \frac{\partial\Lambda_{3}} {\partial\mu}\Big).
\end{equation}
The derivatives $\mu(\partial\Lambda_{i})/(\partial\mu)$ at order $(d-4)$ are equal
\begin{equation}
  \label{eq:99}
  \mu\frac{\partial\Lambda_{i}}{\partial\mu}=(d-4)\Lambda_{i},
\end{equation}
so we obtain
\begin{equation}
  \label{eq:100}
  \mu\frac{\partial M_{1}^{2}}{\partial\mu}=
  \frac{\Lambda_{1}M_{1}^{2} + \Lambda_{3}M_{2}^{2} }{16\pi^{2}}.
\end{equation}
The functions $\gamma_{M_{i}}$ are thus equal
\begin{equation}
  \label{eq:101}
\gamma_{M_{1}}=  \frac{\Lambda_{1}M_{1}^{2} + \Lambda_{3}M_{2}^{2} } {32\pi^{2}M_{1}^{2}},\quad \gamma_{M_{2}}=  \frac{\Lambda_{2}M_{2}^{2} + \Lambda_{3}M_{1}^{2} } {32\pi^{2}M_{2}^{2}}
\end{equation}
and the one loop renormalization group equations for the masses are equal
\begin{subequations}\label{eq:102}
  \begin{align}
    \label{eq:102a}
    t\frac{\partial M_{1}^{2}}{\partial t}=  \frac{\Lambda_{1}M_{1}^{2} + \Lambda_{3}M_{2}^{2} } {32\pi^{2}}\\
    \label{eq:102b}
    t\frac{\partial M_{2}^{2}}{\partial t}=  \frac{\Lambda_{2}M_{2}^{2} + \Lambda_{3}M_{1}^{2} } {32\pi^{2}}.
  \end{align}
\end{subequations}

\section{Renormalization at two loops}
\label{sec:renorm-at-two}

\begin{figure}[!ht]
  \includegraphics[width=0.99\linewidth]{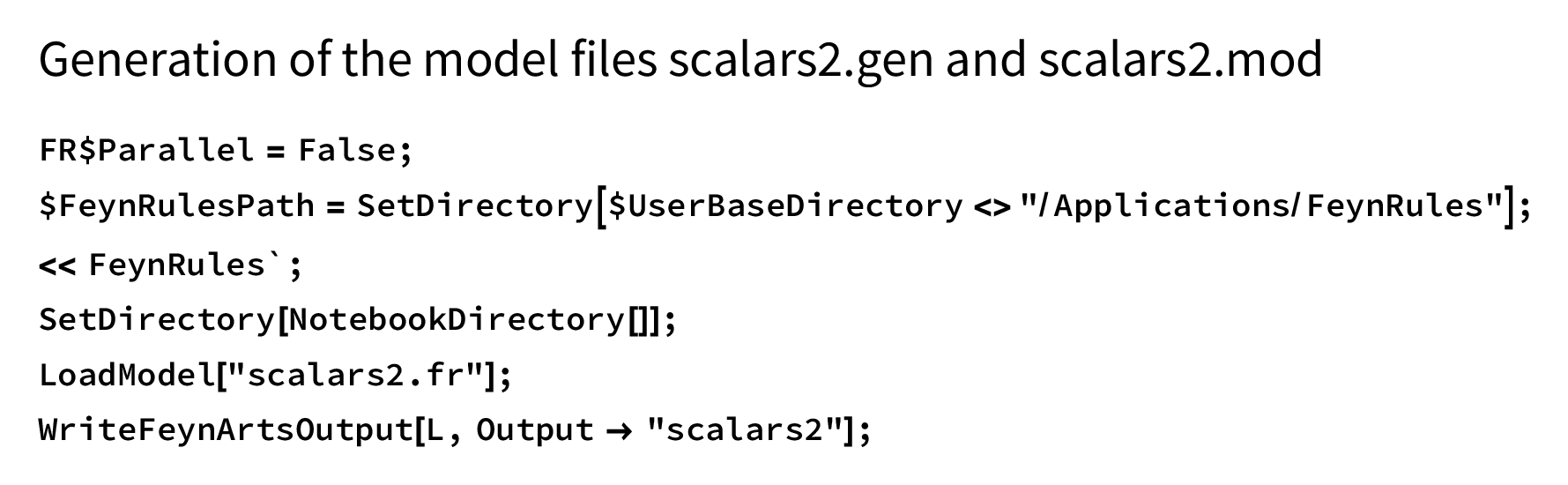}
  \caption{\label{fig:15}Listing of the \textit{Mathematica} program
    for generation of the \textit{driver} files for the
    \textit{FeynArts} package.}
\end{figure}

The complexity of calculations in field theory is rapidly increasing
with the order of the perturbation theory and the number of loops. The
reason is that the number of the Feynman diagrams grows very fast at
each order and its structure becomes more involved. To overcome these
difficulties there have been developed various methods to generate the
diagrams and the corresponding analytical expressions together with
the explicit calculation of the integrals. In this paper we will
consider the \textit{FeynArts}~\cite{FeynArts,HAHN2001418}, which is
the \textit{Mathematica} package for the determination of the Feynman
diagrams of the \textup{QED} and \textit{Standard Model} and some of
its extensions. For each of the considered models there is a special
\textit{driver} file (with the extension ``.mod''), which contains the
information about the Feynman rules. The package also contains a
\textit{generic} file, with the information about the kinematic
(Lorentz) structure of the fields and of the vertices. The model of
two interacting scalar fields is not included in the
\textit{FeynArts}, so we have prepared the corresponding
\textit{driver} file, using the \textit{FeynRules}~\cite{FeynRules,
  ask2012lagrangians, FeynRules_paper}, which is another
\textit{Mathematica} package dedicated to the generation of such
\textit{driver} files directly from the Lagrangian density. The
prepared \textit{FeynRules} file, called \verb|scalar2.fr| is given in
Appendix~\ref{sec:list-feynrules}. The listing of the
\textit{Mathematica} program to generate the files \verb|scalars.gen|
and \verb|scalars.mod| is given in Fig.~\ref{fig:15}
and the listings of the files \verb|scalars2.gen| and
\verb|scalars2.mod| thus obtained are given in
Appendix~\ref{sec:list-FeynArts-files}.

\begin{figure}[!ht]
  \includegraphics[width=0.99\linewidth]{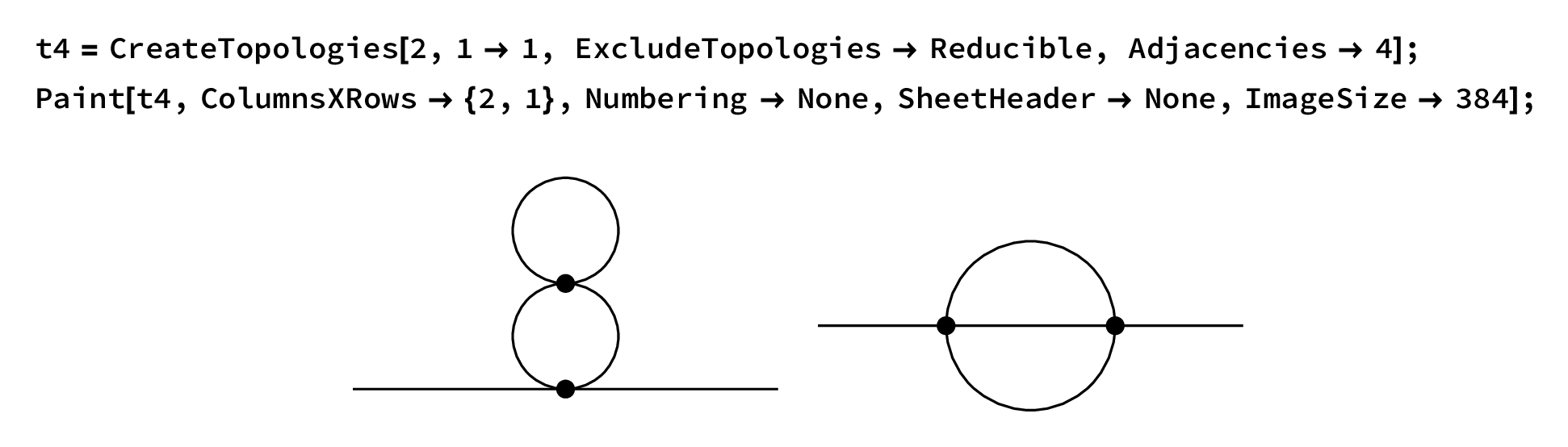}
  \caption{\label{fig:16}Command to generate and draw the two loop
    topology diagrams, using the \textit{FeynArts} package of
    \textit{Mathematica}. The full program is given in
    Appendix~\ref{sec:list-math-progr}.}
\end{figure}

\begin{figure}[!hb]
  \includegraphics[width=0.99\linewidth]{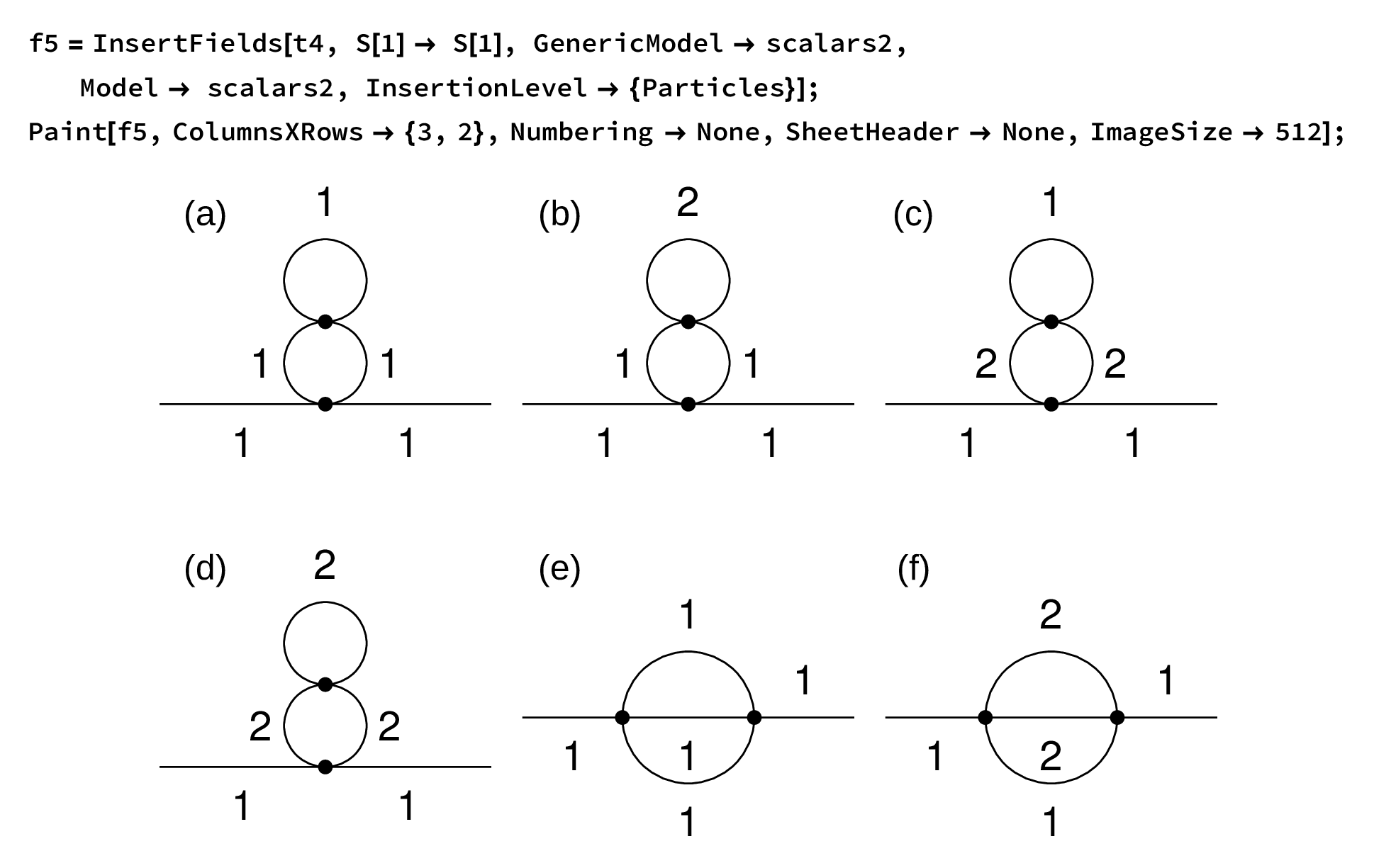}
  \caption{Feynman diagrams at two loops for the propagator of the
    particle~1. The full program is given in
    Appendix~\ref{sec:list-math-progr}. The labels (a)--(f) were added
    after evaluation by \textit{Mathematica}.\label{fig:13}}
\end{figure}

With the help of the \textit{driver} files for our model we can
generate the Feynman diagrams. For this purpose we use the
\textit{Mathematica} program which is given in
Appendix~\ref{sec:list-math-progr}. First, we execute the command,
which generates the topology diagrams for two loop propagator,
which are shown in the listing shown in Fig.~\ref{fig:16}.
The next step is to attach particles and vertices to the topology
diagrams. This is shown in Fig.~\ref{fig:13}.
We see that at two loops we have 6 diagrams for the propagator of
particle~1.

The analytical expressions for the two loop diagrams (a)--(f) are
\begin{subequations}\label{eq:103}
\begin{align}
  \label{eq:103a}
  &i(-i\Lambda_{1})^{2}\frac{1}{4}\int\frac{\dd[4]k_{1}}{(2\pi)^{4}} \frac{i}{k_{1}^{2}-M_{1}^{2}+i\epsilon} \int\frac{\dd[4]k_{2}}{(2\pi)^{4}} \frac{(i)^{2}}{(k_{2}^{2}-M_{1}^{2}+i\epsilon)^{2}}, \\
  \label{eq:103b}
  &i(-i\Lambda_{1})(-i\Lambda_{3})\frac{1}{4}\int\frac{\dd[4]k_{1}}{(2\pi)^{4}} \frac{i}{k_{1}^{2}-M_{2}^{2}+i\epsilon} \int\frac{\dd[4]k_{2}}{(2\pi)^{4}} \frac{(i)^{2}}{(k_{2}^{2}-M_{1}^{2}+i\epsilon)^{2}}, \\
  \label{eq:103c}
  &i(-i\Lambda_{3})^{2}\frac{1}{4}\int\frac{\dd[4]k_{1}}{(2\pi)^{4}} \frac{i}{k_{1}^{2}-M_{1}^{2}+i\epsilon} \int\frac{\dd[4]k_{2}}{(2\pi)^{4}} \frac{(i)^{2}}{(k_{2}^{2}-M_{2}^{2}+i\epsilon)^{2}}, \\
  \label{eq:103d}
  &i(-i\Lambda_{2})(-i\Lambda_{3})\frac{1}{4}\int\frac{\dd[4]k_{1}}{(2\pi)^{4}} \frac{i}{k_{1}^{2}-M_{2}^{2}+i\epsilon} \int\frac{\dd[4]k_{2}}{(2\pi)^{4}} \frac{(i)^{2}}{(k_{2}^{2}-M_{2}^{2}+i\epsilon)^{2}},
\end{align}
\begin{align}
&i(-i\Lambda_{1})^{2}\frac{1}{6}\int\frac{\dd[4]k_{1}}{(2\pi)^{4}} \int\frac{\dd[4]k_{2}}{(2\pi)^{4}} \frac{i}{k_{1}^{2}-M_{1}^{2}+i\epsilon}  \frac{i}{k_{2}^{2}-M_{1}^{2}+i\epsilon}\nonumber\\
  \label{eq:103e} &\hspace*{0.45\linewidth}\times\frac{i}{(p-k_{1}+k_{2})^{2}-M_{1}^{2}+i\epsilon},
  \\
  &i(-i\Lambda_{3})^{2}\frac{1}{4}\int\frac{\dd[4]k_{1}}{(2\pi)^{4}} \int\frac{\dd[4]k_{2}}{(2\pi)^{4}} \frac{i}{k_{1}^{2}-M_{2}^{2}+i\epsilon}  \frac{i}{k_{2}^{2}-M_{2}^{2}+i\epsilon}\nonumber\\
 \label{eq:103f}  &\hspace*{0.45\linewidth}\times\frac{i}{(p-k_{1}+k_{2})^{2}-M_{1}^{2}+i\epsilon}. 
\end{align}
\end{subequations}
The integrals in Eqs.~\eqref{eq:103} are divergent, so their singularities have to be removed by regularization and renormalization. 

\begin{figure}[!ht]
  \includegraphics[width=0.99\linewidth]{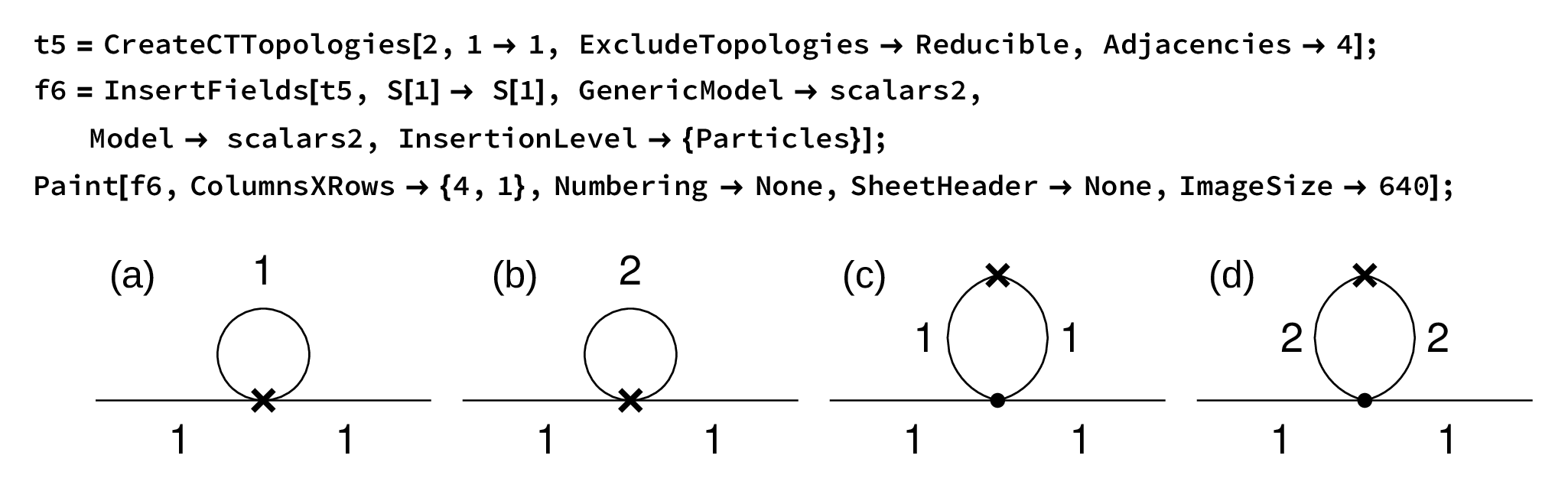}
  \caption{Counter term Feynman diagrams at two loops for the
    propagator of the particle~1. The labels (a)--(d) were added after
    evaluation by \textit{Mathematica}. The full program is given in
    Appendix~\ref{sec:list-math-progr}.\label{fig:14}}
\end{figure}

For a complete renormalization program we also need the Feynman diagrams with counter terms, which are shown in Fig.~\ref{fig:14}.
The analytical expressions corresponding to the diagrams in Fig.~\ref{fig:14} are the following
\begin{subequations}\label{eq:104}
  \begin{align}
    \label{eq:104a}
    i(-i\delta_{\Lambda_{1}})\frac{1}{2}&\int\frac{\dd[4]k}{(2\pi)^{4}} \frac{i}{k^{2}-M_{1}^{2}+i\epsilon},\\
    \label{eq:104b}
    i(-i\delta_{\Lambda_{3}})\frac{1}{2}&\int\frac{\dd[4]k}{(2\pi)^{4}} \frac{i}{k^{2}-M_{2}^{2}+i\epsilon},\\
    \label{eq:104c}
    i(-i\Lambda_{1})(-i\delta_{M_{1}})\frac{1}{2} &\int\frac{\dd[4]k_{2}}{(2\pi)^{4}} \frac{(i)^{2}}{(k_{2}^{2}-M_{1}^{2}+i\epsilon)^{2}},\\
    \label{eq:104d}
    i(-i\Lambda_{3})(-i\delta_{M_{2}})\frac{1}{2} &\int\frac{\dd[4]k}{(2\pi)^{4}} \frac{(i)^{2}}{(k^{2}-M_{2}^{2}+i\epsilon)^{2}}.
  \end{align}
\end{subequations}

In our analysis we will limit ourselves only to the discussion of the divergent terms. The evaluation of the diagrams displayed in Fig.~\ref{fig:13} (a)-(d) with the corresponding formulas given in Eqs.~\eqref{eq:103}~(a)-(d) is simple and the result is the following
\begin{subequations}\label{eq:106}
  \begin{align}
    \label{eq:106a}
    \Scale[0.95]{ i(-i\Lambda_{1}\mu^{4-d})^{2}\frac{1}{4}i(-i)i\frac{\Gamma(1-\frac{d} {2})}{(4\pi)^{\frac{d}{2}}\Gamma(1)}\frac{1}{(M_{1}^{2})^{1-\frac{d}{2}}}
    \frac{\Gamma(2-\frac{d} {2})}{(4\pi)^{\frac{d}{2}}\Gamma(2)}\frac{1}{(M_{1}^{2})^{2-\frac{d}{2}}},}&\\
    \label{eq:106b}
    \Scale[0.95]{ i(-i\Lambda_{1}\mu^{4-d}) (-i\Lambda_{3}\mu^{4-d}) \frac{1}{4}i(-i)i\frac{\Gamma(1-\frac{d} {2})}{(4\pi)^{\frac{d}{2}}\Gamma(1)}\frac{1}{(M_{2}^{2})^{1-\frac{d}{2}}}
\frac{\Gamma(1-\frac{d} {2})}{(4\pi)^{\frac{d}{2}}\Gamma(1)}\frac{1}{(M_{1}^{2})^{1-\frac{d}{2}}},}&\\
    \label{eq:106c}
    \Scale[0.95]{ i(-i\Lambda_{3}\mu^{4-d})^{2}\frac{1}{4}i(-i)i\frac{\Gamma(1-\frac{d} {2})}{(4\pi)^{\frac{d}{2}}\Gamma(1)}\frac{1}{(M_{1}^{2})^{1-\frac{d}{2}}}
    \frac{\Gamma(2-\frac{d} {2})}{(4\pi)^{\frac{d}{2}}\Gamma(2)}\frac{1}{(M_{2}^{2})^{2-\frac{d}{2}}},}&\\
    \label{eq:106d}
    \Scale[0.95]{ i(-i\Lambda_{2}\mu^{4-d}) (-i\Lambda_{3}\mu^{4-d}) \frac{1}{4}i(-i)i\frac{\Gamma(1-\frac{d} {2})}{(4\pi)^{\frac{d}{2}}\Gamma(1)}\frac{1}{(M_{2}^{2})^{1-\frac{d}{2}}}
\frac{\Gamma(1-\frac{d} {2})}{(4\pi)^{\frac{d}{2}}\Gamma(1)}\frac{1}{(M_{2}^{2})^{1-\frac{d}{2}}}.}&
  \end{align}
\end{subequations}

Next we calculate the diagrams corresponding to the couterterms given
in Fig.~\ref{fig:14}~(a)-(d). Using the formulas from
Eqs.~\eqref{eq:104} we get
\begin{subequations}\label{eq:107}
  \begin{align}
    \label{eq:107a}
    i(-\delta_{\Lambda_{1}})i(-i)\frac{1}{2}\frac{\Gamma(1-\frac{d}{2})} {(4\pi)^{\frac{d}{2}}\Gamma(1)}\frac{1}{(M_{1}^{2})^{1-\frac{d}{2}}},&\\
    \label{eq:107b}
    i(-\delta_{\Lambda_{2}})i(-i)\frac{1}{2}\frac{\Gamma(1-\frac{d}{2})} {(4\pi)^{\frac{d}{2}}\Gamma(1)}\frac{1}{(M_{2}^{2})^{1-\frac{d}{2}}},&\\
    \label{eq:107c}
i(-i\Lambda_{1})(-i\delta_{M_{1}}) \frac{\Gamma(2-\frac{d}{2})} {(4\pi)^{\frac{d}{2}}\Gamma(2)}\frac{1}{(M_{1}^{2})^{2-\frac{d}{2}}},&\\
    \label{eq:107d}
i(-i\Lambda_{3})(-i\delta_{M_{2}}) \frac{\Gamma(2-\frac{d}{2})} {(4\pi)^{\frac{d}{2}}\Gamma(2)}\frac{1}{(M_{2}^{2})^{2-\frac{d}{2}}}.&
  \end{align}
\end{subequations}

The following step is to find the sum of all diagrams from
Figs.~\ref{fig:13}~(a)-(d) and~\ref{fig:14} and calculate the
divergent contributions for the spacetime dimension $d=4$. It turns
out that all divergent contributions cancel at this order~\footnote{It
  should be noticed that such a cancellation does not occur for the
  case of only one scalar field}.

We will now consider the remaining two \textit{setting sun} diagrams
in Figs.~\ref{fig:13}~(e) and~(f). These diagrams are an example of
the \textit{overlapping divergencies} and such diagrams are more
difficult to calculate~\cite{ramond1997field, doi:10.1142/4733}. From
Eqs.~\eqref{eq:103e} and~\eqref{eq:103f} one can see that they are
quadratically divergent. The analytic structure of both equations is
identical, but Eq.~\eqref{eq:103f} is more general, because it
contains two different masses. Such a diagram is not present for the
case of one scalar field. Another important property of these diagrams
is their dependence on the external momentum~$p$ of the particle. The
integral in Eq.~\eqref{eq:103f} converted to the Euclidean space
in~$d$ dimensions has the form
\begin{equation}
  \label{eq:105}
  I(p^{2})=\int\frac{\dd[d]k_{1}}{(2\pi)^{d}} \int\frac{\dd[d]k_{2}}{(2\pi)^{d}} \frac{1}{k_{1}^{2}+M_{2}^{2}}  \frac{1}{k_{2}^{2}+M_{2}^{2}}  \frac{1}{(p-k_{1}+k_{2})^{2}+M_{1}^{2}}
\end{equation}
and it is the function of~$p^{2}$, so its Taylor expansion in powers of~$p^{2}$ is
\begin{equation}
  \label{eq:108}
  I(p^{2})=I(0)+p^{2}\Big(\frac{p^{\mu}}{p^{2}}\frac{\partial I(p^{2})}{\partial p^{\mu}}\Big)\Big\vert_{p^{2}=0} + \frac{1}{2!}\, p^{4} \Big(\frac{p^{\mu}p^{\nu}}{p^{4}}\frac{\partial^{2} I(p^{2})}{\partial p^{\mu} \partial p^{\nu}}\Big)\vert_{p^{2}=0} +\cdots
\end{equation}
The first term $I(0)$ is quadratically divergent and the next term, linear in~$p^{2}$ is logarithmically divergent. The third term and all higher terms are convergent. The fact that the second term is divergent is important, because it has to be included in the renormalization procedure and it gives contribution to the term~$\delta_{Z_{1}}$ in Eq.~\eqref{eq:58}.

The integral for the first term in Eq.~\eqref{eq:108} is equal
\begin{equation}
  \label{eq:109}
  I(0)=\int\frac{\dd[d]k_{1}}{(2\pi)^{d}} \int\frac{\dd[d]k_{2}}{(2\pi)^{d}} \frac{1}{k_{1}^{2}+M_{2}^{2}}  \frac{1}{k_{2}^{2}+M_{2}^{2}}  \frac{1}{(k_{1}-k_{2})^{2}+M_{1}^{2}}.
\end{equation}
The first step in the calculation of this integral is to apply the
scheme of t'Hooft and Veltman~\cite{tHooft_Veltman} which consists in
inserting the following expression into the integral
\begin{equation*}
  1=\frac{1}{2d}\Big(\frac{\partial (k_{1})_{\mu}}{\partial(k_{1})_{\mu}} +\frac{\partial (k_{2})_{\mu}}{\partial(k_{2})_{\mu}}\Big)
\end{equation*}
and then integrating by parts. After such an operation one obtains
\begin{multline}
  \label{eq:110}
  I(0)=-\frac{1}{2d}\Big(-6I(0)\\ +2M_{1}^{2} \int\frac{\dd[d]k_{1}}{(2\pi)^{d}} \int\frac{\dd[d]k_{2}}{(2\pi)^{d}} \frac{1}{k_{1}^{2}+M_{2}^{2}}  \frac{1}{k_{2}^{2}+M_{2}^{2}}  \frac{1}{((k_{1}-k_{2})^{2}+M_{1}^{2})^{2}}\\ +
  4M_{2}^{2} \int\frac{\dd[d]k_{1}}{(2\pi)^{d}} \int\frac{\dd[d]k_{2}}{(2\pi)^{d}} \frac{1}{(k_{1}^{2}+M_{2}^{2})^{2}}  \frac{1}{k_{2}^{2}+M_{2}^{2}}  \frac{1}{(k_{1}-k_{2})^{2}+M_{1}^{2}}\Big)
\end{multline}
so $I(0)$ becomes
\begin{multline}
  \label{eq:111}
  I(0)=-\frac{1}{d-3}\\ \times\Big(M_{1}^{2} \int\frac{\dd[d]k_{1}}{(2\pi)^{d}} \int\frac{\dd[d]k_{2}}{(2\pi)^{d}} \frac{1}{k_{1}^{2}+M_{2}^{2}}  \frac{1}{k_{2}^{2}+M_{2}^{2}}  \frac{1}{((k_{1}-k_{2})^{2}+M_{1}^{2})^{2}}\\ +
  2M_{2}^{2} \int\frac{\dd[d]k_{1}}{(2\pi)^{d}} \int\frac{\dd[d]k_{2}}{(2\pi)^{d}} \frac{1}{(k_{1}^{2}+M_{2}^{2})^{2}}  \frac{1}{k_{2}^{2}+M_{2}^{2}}  \frac{1}{(k_{1}-k_{2})^{2}+M_{1}^{2}}\Big).
\end{multline}
\renewcommand*{\sk}{0.79}
The second integral in Eq.~\eqref{eq:111} is calculated as follows
\begin{subequations}\label{eq:112}
\begingroup
\allowdisplaybreaks
  \begin{align}
  \label{eq:112a}
    \Scale[\sk]{ \int}&\Scale[\sk]{\frac{\dd[d]k_{1}}{(2\pi)^{d}} \int\frac{\dd[d]k_{2}}{(2\pi)^{d}} \frac{1}{(k_{1}^{2}+M_{2}^{2})^{2}}  \frac{1}{k_{2}^{2}+M_{2}^{2}}  \frac{1}{(k_{1}-k_{2})^{2}+M_{1}^{2}}}\\
  \label{eq:112b}
  &\Scale[\sk]{=\int\frac{\dd[d]k_{1}}{(2\pi)^{d}} \int\frac{\dd[d]k_{2}}{(2\pi)^{d}} \int_{0}^{1} \dd x \frac{1}{(k_{1}^{2}+M_{2}^{2})^{2}}  \frac{1}{(x(k_{2}^{2}+M_{2}^{2})+(1-x)((k_{1}-k_{2})^{2}+M_{1}^{2}))^{2}}}\\
    &\Scale[\sk]{=\int\frac{\dd[d]k_{1}}{(2\pi)^{d}} \int\frac{\dd[d]k_{2}}{(2\pi)^{d}} \int_{0}^{1} \dd x \frac{1}{(k_{1}^{2}+M_{2}^{2})^{2}}}\nonumber\\
  \label{eq:112c}
    &\phantom{AAAAAAAA}\Scale[\sk]{  \times\frac{1}{(k_{2}^{2}-2(1-x)k_{1}k_{2} +(1-x)k_{1}^{2} + xM_{2}^{2}+(1-x)M_{1}^{2})^{2}}}\\
  \label{eq:112d}
  &\Scale[\sk]{=\int\frac{\dd[d]k_{1}}{(2\pi)^{d}} \int\frac{\Gamma(2-\frac{d}{2})}{(4\pi)^{\frac{d}{2}}} \int_{0}^{1} \dd x \frac{1}{(k_{1}^{2}+M_{2}^{2})^{2}}  \frac{1}{(x(1-x)k_{1}^{2}+xM_{2}^{2}+(1-x)M_{1}^{2})^{2-\frac{d}{2}}}}\\
  \label{eq:112e}
    &\Scale[\sk]{=\int\frac{\dd[d]k_{1}}{(2\pi)^{d}} \int\frac{\Gamma(2-\frac{d}{2})}{(4\pi)^{\frac{d}{2}}} \int_{0}^{1} \dd x \frac{1}{(k_{1}^{2}+M_{2}^{2})^{2}}  \frac{(x(1-x))^{\frac{d}{2}-2}}{\big(k_{1}^{2} +\frac{xM_{2}^{2}+(1-x)M_{1}^{2}}{x(1-x)}\big)^{2-\frac{d}{2}}}}\\
    \label{eq:112f}
  &\Scale[\sk]{=\int\frac{\dd[d]k_{1}}{(2\pi)^{d}} \frac{\Gamma(4-\frac{2}{d})}{(4\pi)^{\frac{d}{2}}} \int_{0}^{1} \dd x \int_{0}^{1} \dd y   \frac{(x(1-x))^{\frac{d}{2}-2}y^{1-\frac{d}{2}}(1-y)} {\big((1-y)(k_{1}^{2}+M_{2}^{2}) +y\big(k_{1}^{2} +\frac{xM_{2}^{2}+(1-x)M_{1}^{2}}{x(1-x)}\big)\big)^{4-\frac{d}{2}}}}\\
    \label{eq:112g}
  &\Scale[\sk]{= \frac{\Gamma(4-d)}{(4\pi)^{d}} \int_{0}^{1} \dd x \int_{0}^{1} \dd y   \frac{(x(1-x))^{\frac{d}{2}-2}y^{1-\frac{d}{2}}(1-y)} {\big((1-y)M_{2}^{2} +y\big( \frac{xM_{2}^{2}+(1-x)M_{1}^{2}}{x(1-x)}\big)\big)^{4-d}}}\\
    &\begin{multlined}[b][0.85\linewidth]
      \Scale[\sk]{= -\frac{\Gamma(4-d)}{(4\pi)^{d}(2-\frac{d}{2})} \int_{0}^{1} \dd x \int_{0}^{1} \dd y  (x(1-x))^{\frac{d}{2}-2}y^{2-\frac{d}{2}}}\\
    \label{eq:112h}
    \Scale[\sk]{
      \times\frac{\dd}{\dd y} \left(\frac{(1-y)} {\big((1-y)M_{2}^{2} +y\big( \frac{xM_{2}^{2}+(1-x)M_{1}^{2}}{x(1-x)}\big)\big)^{4-d}}\right)}.
  \end{multlined}
  \end{align}
\end{subequations}
\endgroup
Let us explain each step in Eqs.~\eqref{eq:112}
\begin{description}
\item[Eq.~\eqref{eq:112a}$\rightarrow$~\eqref{eq:112b}] Feynman
  prescription.
\item[Eq.~\eqref{eq:112b}$\rightarrow$~\eqref{eq:112c}] Transformation of the denominator under the integral.
\item[Eq.~\eqref{eq:112c}$\rightarrow$~\eqref{eq:112d}] Integration over $k_{2}$.
\item[Eq.~\eqref{eq:112d}$\rightarrow$~\eqref{eq:112e}] Transformation of the fraction under the integral.
\item[Eq.~\eqref{eq:112e}$\rightarrow$~\eqref{eq:112f}] Feynman prescription.
\item[Eq.~\eqref{eq:112f}$\rightarrow$~\eqref{eq:112g}] Integration over $k_{1}$.
\item[Eq.~\eqref{eq:112g}$\rightarrow$~\eqref{eq:112h}] Integration by parts, using the identity
\begin{equation*}
y^{1-\frac{d}{2}}= \frac{1}{2-\frac{d}{2}} \frac{\dd y^{2-\frac{d}{2}}}{\dd y}.
\end{equation*}
\end{description}
The integrated function in Eq.~\eqref{eq:112h} has no singularity at $d=4$ and it can be expanded in the Taylor series at $d=4$. The first two terms of this expansion can be explicitly integrated and the final result for the integral~\eqref{eq:112h} at this order is
\begin{multline}
  \label{eq:113}
  \int\frac{\dd[d]k_{1}}{(2\pi)^{d}} \int\frac{\dd[d]k_{2}}{(2\pi)^{d}} \frac{1}{(k_{1}^{2}+M_{2}^{2})^{2}}  \frac{1}{k_{2}^{2}+M_{2}^{2}}  \frac{1}{(k_{1}-k_{2})^{2}+M_{1}^{2}}\\ =  \frac{\Gamma(4-d)}{(4\pi)^{d}(2-\frac{d}{2})}\big(1+(d-4)\big(-\frac{1}{2}+\ln M_{2}^{2}\big)\big).
\end{multline}
In analogy with the first integral in Eq.~\eqref{eq:111} we obtain
\begin{multline}
\label{eq:114}
  \int\frac{\dd[d]k_{1}}{(2\pi)^{d}} \int\frac{\dd[d]k_{2}}{(2\pi)^{d}} \frac{1}{k_{1}^{2}+M_{2}^{2}}  \frac{1}{k_{2}^{2}+M_{2}^{2}}  \frac{1}{((k_{1}-k_{2})^{2}+M_{1}^{2})^{2}}\\ =  \frac{\Gamma(4-d)}{(4\pi)^{d}(2-\frac{d}{2})}\big(1+(d-4)\big(-\frac{1}{2}+\ln M_{1}^{2}\big)\big).
\end{multline}

We will now calculate the second term in Eq.~\eqref{eq:108}. It is equal
\begin{equation}
  \label{eq:115}
    \frac{p^{\mu}}{p^{2}}\frac{\partial}{\partial p^{\mu}}\int\frac{\dd[d]k_{1}}{(2\pi)^{d}} \int\frac{\dd[d]k_{2}}{(2\pi)^{d}} \frac{1}{k_{1}^{2}+M_{2}^{2}}  \frac{1}{k_{2}^{2}+M_{2}^{2}}  \frac{1}{(p-k_{1}+k_{2})^{2}+M_{1}^{2}}\Big\vert_{p^{2}=0}.
\end{equation}
Let us calculate the integral in Eq.~\eqref{eq:115}. After making the same transformations as in the previous integral we obtain
\begin{subequations}\label{eq:116}
\renewcommand*{\sk}{0.97}
\begingroup
\allowdisplaybreaks
\begin{align}
  & \begin{multlined}[b]
    \Scale[\sk]{-\frac{p^{\mu}}{2p^{2}} \frac{\Gamma(3-d)}{(4\pi)^{d}} \frac{\partial}
      {\partial p^{\mu}}\int_{0}^{1}\dd (x(1-x))^{\frac{d}{2}-2}} \\
  \label{eq:116a}
  \Scale[\sk]{\times \int_{0}^{1}\dd y y^{1-\frac{d}{2}} \left.\frac{1}{\big(p^{2}y(1-y) + (1-y)M_{2}^{2} +y\big( \frac{xM_{2}^{2}+(1-x)M_{1}^{2}} {x(1-x)}\big)\big)^{3-d}}\right\vert_{p^{2}=0}}
\end{multlined}\\
&\begin{multlined}[b][0.87\linewidth]
  \Scale[\sk]{=\frac{(3-d)\Gamma(3-d)}{(4\pi)^{d}} \int_{0}^{1}\dd (x(1-x))^{\frac{d}{2}-2} \int_{0}^{1}\dd y y^{2-\frac{d}{2}}(1-y)}\\
  \label{eq:116b}
  \Scale[\sk]{\times  \left.\frac{1}{\big(p^{2}y(1-y) + (1-y)M_{2}^{2} +y\big( \frac{xM_{2}^{2}+(1-x)M_{1}^{2}}{x(1-x)}\big)\big)^{4-d}}\right\vert_{p^{2}=0}}
\end{multlined}\\
&\begin{multlined}[b][0.87\linewidth]
  \Scale[\sk]{=\frac{\Gamma(4-d)}{(4\pi)^{d}} \int_{0}^{1}\dd (x(1-x))^{\frac{d}{2}-2} \int_{0}^{1}\dd y y^{2-\frac{d}{2}}(1-y)}\\
  \label{eq:116c}
  \Scale[\sk]{\times \frac{1}{\big((1-y)M_{2}^{2} +y\big( \frac{xM_{2}^{2}+(1-x)M_{1}^{2}}{x(1-x)}\big)\big)^{4-d}}}.
\end{multlined}
\end{align}
\end{subequations}
\endgroup
The singular term at $d=4$ of the integral in Eq.~\eqref{eq:116c} is equal
\begin{equation}
  \label{eq:117}
  \frac{\Gamma(4-d)}{2(4\pi)^{d}}.
\end{equation}
The contribution of the same type of the integral in Fig.~\ref{fig:13}~(f) is the same as for the diagram in Fig.~\ref{fig:13}~(e) and the sum of the all divergent contributions to the diagrams in Figs.~\ref{fig:13} and~\ref{fig:14} is equal
\renewcommand*{\sk}{0.88}
\begin{multline}
  \label{eq:118} \Scale[\sk]{\frac{3M_{1}^{2}(\Lambda_{1}\mu^{4-d})^{2}}{6(d-3)}\frac{\Gamma(4-d)}{(4\pi)^{d}(2-\frac{d}{2})}\big(1+(d-4) \big(-\frac{1}{2}+\ln M_{1}^{2}\big)\big)}\\
\Scale[\sk]{+ \frac{(\Lambda_{3}\mu^{4-d})^{2}}{6(d-3)}\frac{\Gamma(4-d)} {(4\pi)^{d}(2-\frac{d}{2})} \times\Big( M_{1}^{2} \big(1+(d-4)\big(-\frac{1}{2}+\ln M_{1}^{2}\big)\big)}\\
\Scale[\sk]{+2M_{2}^{2} \big(1+(d-4)\big(-\frac{1}{2}+\ln M_{2}^{2}\big)\big)
  \Big)
  -p^{2}\frac{\Gamma(4-d)}{2(4\pi)^{d}}\Big(\frac{(\Lambda_{1}\mu^{4-d})^{2}}{6} +\frac{(\Lambda_{3}\mu^{4-d})^{2}}{4}\Big).}
\end{multline}
To obtain the counterterms one has to expand the formula in Eq.~\eqref{eq:118} around the point $d=4$ and keep only the divergent terms. The result of this procedure is
\begin{multline}
  \label{eq:119}
  \frac{1}{2(d-4)^{2}}(2M_{1}^{2}\hat{\Lambda}_{1}^{2}  +(M_{1}^{2}+2M_{2}^{2})\hat{\Lambda}_{3}^{2})\\
  +\frac{1}{4(d-4)}(M_{1}^{2}(2\hat{\Lambda}_{1}^{2} +\hat{\Lambda}_{3}^{2})
  (-3+2\gamma+2\ln(\frac{M_{1}^{2}}{4\pi\mu^{2}})))\\
    +\frac{1}{4(d-4)}(2M_{2}^{2}(2\hat{\Lambda}_{3}^{2} +\hat{\Lambda}_{3}^{2})
    (-3+2\gamma+2\ln(\frac{M_{2}^{2}}{4\pi\mu^{2}})))\\
    +\frac{p^{2}}{24(d-4)}(2\hat{\Lambda}_{1}^{2}+3\hat{\Lambda}_{3}^{2}).
  \end{multline}
  Here we introduced the notation
  \begin{equation*}
    \hat{\Lambda}_{i}=\frac{\Lambda_{i}}{(4\pi)^{2}}.
  \end{equation*}
  Equation~\eqref{eq:119} defines the two loop counterterms for the
  propagator of the particle~1. It consists of two different types of
  terms. The first type does not depend on the external momentum
  $p^{2}$ and it gives the two loop contribution to
  $\delta_{M_{1}}$. The term proportional to $p^{2}$ cannot be
  included in the counterterm $\delta_{M_{1}}$ and it is included in
  the counterterm $\delta_{Z_{1}}$. The results for the propagator of
  the particle~2 are obtained from those for the particle~1 by a
  simple exchange of the indices for the masses and coupling
  constants.

  To complete the renormalization program at two loops one has to consider the renormalization of the coupling constants. The necessary diagrams generated by the \textit{FeynArts} program are given in the figure below.
\centerline{\includegraphics[width=\linewidth]{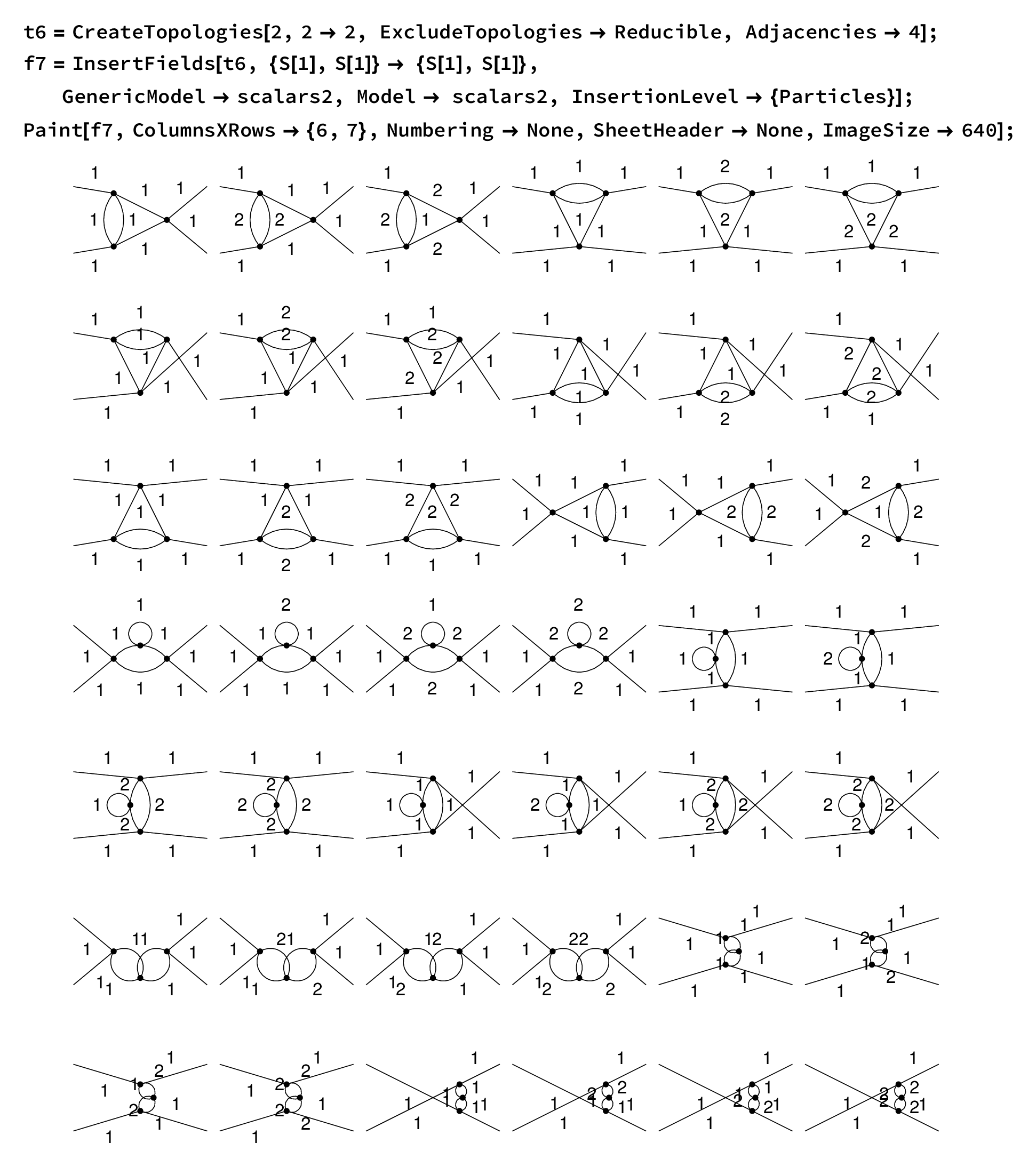}}
Technically, the diagrams shown in this figure do not present any new difficulties and they can be calculated using the methods described above. For this reason we will not calculate these diagrams here.

\section{Conclusions}
\label{sec:conclusions}

We have presented here a detailed description of the renormalization
of the quantum field theory of two interacting scalar fields with an
interaction described by the 4-th order homogeneous polynomial. Such a
theory is an extension of the quantum field theory of one scalar field
with a potential equal to $\phi^{4}$. The theory with one scalar field
is discussed in many textbooks (see for
example~\cite{ryder1996quantum, ramond1997field,
  peskin2018introduction, cheng1984gauge, veltman1994diagrammatica}),
but the most complete discussion can be found in the book by Hagen
Kleinert and Verena Schulte-Frohlinde~\cite{doi:10.1142/4733}. The
theory with two scalar fields shows many similarities with one scalar
field model, but there are some differences. The most important one is
the cancellation of the divergencies in Figs.~\ref{fig:13}~(a)-(d)
and~\ref{fig:14}. Such cancellation does not occur for one scalar
field. Another important feature of the theory with two scalar fields
is the fact that it is a step forward in the path to study possible
symmetries under the finite symmetry group transformations of systems
of several scalar fields. Such symmetries may introduce some new and
unexpected properties of these systems.

\subsection*{Acknowledgments}

This paper was supported in part by Proyecto SIP: 20211112, Secretar\'{\i}a de
Investigaci\'{o}n y Posgrado, Beca EDI y Comisi\'{o}n de Operaci\'{o}n
y Fomento de Actividades Acad\'{e}micas (COFAA) del Instituto
Polit\'{e}cnico Nacional (IPN), M\'{e}xico.

\bibliographystyle{unsrt}
\bibliography{escalar}

\section*{Appendices}
\appendix

\section{Wick's theorem}
\label{sec:wicks-theorem}
\renewcommand{\theequation}{A-\arabic{equation}}
\setcounter{equation}{0}
 
The calculation of the matrix elements of the time ordered products of
the fields is done with the help of the Wick's theorem that transforms the
time ordered product into the normal product of the fields denoted by
$:\cdots:$. Let us consider for simplicity that we have only one
scalar field $\phi(x)$. For the product of $n$ such scalar fields the
Wick's theorem states:
\begin{multline}
  \label{eq:7}
  T(\phi(x_{1})\phi(x_{2})\cdots\phi(x_{n})) =
  :\phi(x_{1})\phi(x_{2})\cdots\phi(x_{n}): +
  :{\contraction{}{\phi}{(x_{1})}{\phi}
    \phi(x_{1})\phi(x_{2})\cdots\phi(x_{n})}:\\ +
  :{\contraction{}{\phi}{(x_{1})\phi(x_{2})}{\phi}
    \phi(x_{1})\phi(x_{2})\phi(x_{3})\cdots\phi(x_{n})}:+\cdots+
  :{\contraction{}{\phi}{(x_{1})\phi(x_{2})\cdots}{\phi}
    \phi(x_{1})\phi(x_{2})\cdots\phi(x_{n})}:\\ +
  :{\contraction{\phi(x_{1})}{\phi}{(x_{2})}{\phi}
    \phi(x_{1})\phi(x_{2})\phi(x_{3})\cdots\phi(x_{n})}:+\cdots+
  :{\contraction{}{\phi}{(x_{1})\phi(x_{2})}{\phi}
    \contraction[2ex]{\phi(x_{1})}{\phi}{(x_{2})\phi(x_{3})}{\phi}
    \phi(x_{1})\phi(x_{2})\phi(x_{3})\phi(x_{4})\cdots\phi(x_{n})}:\\
  +\cdots+:{\contraction{}{\phi}{(x_{1})\phi(x_{2})}{\phi}
    \contraction[2ex]{\phi(x_{1})}{\phi}{(x_{2})\phi(x_{3})\phi(x_{4})\cdots}
    {\phi}
    \phi(x_{1})\phi(x_{2})\phi(x_{3})\phi(x_{4})\cdots\phi(x_{n})}:+\cdots
\end{multline}
Here the symbol of contraction
$\contraction{}{\phantom{\phi}}{(x_{i})}{\phi} \phi(x_{i})\phi(x_{j})$
means that the contracted pair of the fields has to be substituted by
\begin{equation}
  \label{eq:8}
  \contraction{}{\phantom{\phi}}{(x_{i})}{\phi} \phi(x_{i})\phi(x_{j}) \rightarrow \mel{0}{T(\phi(x_{i})\phi(x_{j}))}{0}=i\Delta(x_{i}-x_{j})
\end{equation}
and the contractions have to be taken over all possible pairs of
fields.  If there are more kinds of fields then the contractions have
to be taken only for those pairs of fields that do not commute. The
remaining diagrams obtained from Fig.~\ref{fig:3} are also
logarithmically divergent.

\section{Calculation of $G^{(4)}_{1111}$}
\label{sec:calculation-g4_1111}
\renewcommand{\theequation}{B-\arabic{equation}}
\setcounter{equation}{0}

Let us introduce the shorthand notation
\begin{equation*}
  \phi_{i}^{\text{I}}(x_{k})\rightarrow\phi_{i}^{k}\text{ and }\phi_{i}^{\text{I}}(z)\rightarrow\phi_{i}^{z}.
\end{equation*}
Using this notation and the explicit form of
$\mathcal{V}(\phi_{1}^{\text{I}}(z),\phi_{2}^{\text{I}}(z))$ we have
to calculate three types of the terms
\begin{subequations}\label{eq:11}
  \begin{gather}
    T(\phi_{1}^{1},\phi_{1}^{2}, \phi_{1}^{3},\phi_{1}^{4},\phi_{1}^{z},\phi_{1}^{z}, \phi_{1}^{z},\phi_{1}^{z})\label{eq:11a}\\
    T(\phi_{1}^{1},\phi_{1}^{2}, \phi_{1}^{3},\phi_{1}^{4},\phi_{2}^{z},\phi_{2}^{z}, \phi_{2}^{z},\phi_{2}^{z})\label{eq:11b}\\
    T(\phi_{1}^{1},\phi_{1}^{2},
    \phi_{1}^{3},\phi_{1}^{4},\phi_{1}^{z},\phi_{1}^{z},
    \phi_{2}^{z},\phi_{2}^{z})\label{eq:11c}.
  \end{gather}
\end{subequations}
All of the Wick's contractions for Eq.~\eqref{eq:11a} giving connected
diagrams are shown in Table~\ref{tab:2} and each term in this equation
gives the same contribution which is equal
\begin{equation}
  \label{eq:13}
  T(\phi_{1}^{\text{I}}(x_{1}),\phi_{1}^{\text{I}}(z)) T(\phi_{1}^{\text{I}}(x_{2}),\phi_{1}^{\text{I}}(z)) T(\phi_{1}^{\text{I}}(x_{3}),\phi_{1}^{\text{I}}(z))T(\phi_{1}^{\text{I}}(x_{4}),\phi_{1}^{\text{I}}(z)).
\end{equation}
\begin{table}[!ht]\caption{\label{tab:2}All the Wick's contractions for
    Eq.~\eqref{eq:11a} giving connected diagrams.}\centering
  \begin{tabular}{lll}
    \\
    $\cA\dB\eC\fD\grE$ & $\cA\dB\eD\fC\grE$ &$\cA\dC\eB\fD\grE$\\
    $\cA\dC\eD\fB\grE$ & $\cA\dD\eB\fC\grE$ &$\cA\dD\eC\fB\grE$\\
    $\cB\dA\eC\fD\grE$ & $\cB\dA\eD\fC\grE$ &$\cB\dC\eA\fD\grE$\\
    $\cB\dC\eD\fA\grE$ & $\cB\dD\eA\fC\grE$ &$\cB\dD\eC\fA\grE$\\
    $\cC\dA\eB\fD\grE$ & $\cC\dA\eD\fB\grE$ &$\cC\dB\eA\fD\grE$\\
    $ \cC\dB\eD\fA\grE$ & $\cC\dD\eA\fB\grE$ &$\cC\dD\eB\fA\grE$\\
    $\cD\dA\eB\fC\grE$ & $\cD\dA\eC\fB\grE$ &$\cD\dB\eA\fC\grE$\\
    $\cD\dB\eC\fA\grE$ & $\cD\dC\eA\fB\grE$ &$\cD\dC\eB\fA\grE$
  \end{tabular}
\end{table}
The time ordered products in Eqs.~\eqref{eq:11b} and~\eqref{eq:11c}
lead to the disconnected diagrams and do not contribute.

Concluding, there are~24 equal terms in Eq.~\eqref{eq:11a} so we see
that the factor $1/4!$ in
$\mathcal{V}(\phi_{1}^{\text{I}}(z),\phi_{2}^{\text{I}}(z))$ is
canceled and the final result is
\begin{multline}
  \label{eq:12}
  T(\phi_{1}^{\text{I}}(x_{1}),\phi_{1}^{\text{I}}(x_{2}),
  \phi_{1}^{\text{I}}(x_{3}),\phi_{1}^{\text{I}}(x_{4}),
  \mathcal{V}(\phi_{1}^{\text{I}}(z),\phi_{2}^{\text{I}}(z)))\\ =
  \lambda_{1} T(\phi_{1}^{\text{I}}(x_{1})\phi_{1}^{\text{I}}(z))
  T(\phi_{1}^{\text{I}}(x_{2})\phi_{1}^{\text{I}}(z))
  T(\phi_{1}^{\text{I}}(x_{3})\phi_{1}^{\text{I}}(z))T(\phi_{1}^{\text{I}}(x_{4})\phi_{1}^{\text{I}}(z)).
\end{multline}
Similarly we obtain
\begin{multline}
  \label{eq:14}
  T(\phi_{2}^{\text{I}}(x_{1}),\phi_{2}^{\text{I}}(x_{2}),
  \phi_{2}^{\text{I}}(x_{3}),\phi_{2}^{\text{I}}(x_{4}),
  \mathcal{V}(\phi_{1}^{\text{I}}(z),\phi_{2}^{\text{I}}(z)))\\ =
  \lambda_{2} T(\phi_{2}^{\text{I}}(x_{1}),\phi_{2}^{\text{I}}(z))
  T(\phi_{2}^{\text{I}}(x_{2}),\phi_{2}^{\text{I}}(z))
  T(\phi_{2}^{\text{I}}(x_{3}),\phi_{2}^{\text{I}}(z))T(\phi_{2}^{\text{I}}(x_{4}),\phi_{2}^{\text{I}}(z))
\end{multline}
and this demonstrates that there is no factor $1/4!$ in the Feynman
diagrams with coupling constants $\lambda_{1}$ and $\lambda_{2}$ in
Table~\ref{tab:1}.

\section{Calculation of $G^{(4)}_{1122}$}
\label{sec:calculation-g4_1122}
\renewcommand{\theequation}{C-\arabic{equation}}
\setcounter{equation}{0}

This time we have to calculate three types of terms
\begin{subequations}\label{eq:13x}
  \begin{gather}
    T(\phi_{1}^{1},\phi_{1}^{2}, \phi_{2}^{3},\phi_{2}^{4},\phi_{1}^{z},\phi_{1}^{z}, \phi_{1}^{z},\phi_{1}^{z})\label{eq:13a}\\
    T(\phi_{1}^{1},\phi_{1}^{2}, \phi_{2}^{3},\phi_{2}^{4},\phi_{2}^{z},\phi_{2}^{z}, \phi_{2}^{z},\phi_{2}^{z})\label{eq:13b}\\
    T(\phi_{1}^{1},\phi_{1}^{2},
    \phi_{2}^{3},\phi_{2}^{4},\phi_{1}^{z},\phi_{1}^{z},
    \phi_{2}^{z},\phi_{2}^{z})\label{eq:13c}.
  \end{gather}
\end{subequations}
The time ordered products in Eqs.~\eqref{eq:13a} and~\eqref{eq:13b}
lead to disconnected diagrams so they do not contribute to the
$G^{(4)}_{1122}$.

All the Wick's contractions for Eq.~\eqref{eq:13c} giving connected
diagrams are shown in Table~\ref{tab:3} and each term in this equation
gives the same contribution which is equal
\begin{equation}
  \label{eq:13y}
  T(\phi_{1}^{\text{I}}(x_{1}),\phi_{1}^{\text{I}}(z)) T(\phi_{1}^{\text{I}}(x_{2}),\phi_{1}^{\text{I}}(z)) T(\phi_{1}^{\text{I}}(x_{3}),\phi_{1}^{\text{I}}(z))T(\phi_{1}^{\text{I}}(x_{4}),\phi_{1}^{\text{I}}(z)).
\end{equation}
\begin{table}[!hb]\caption{\label{tab:3}All the Wick's contractions for
    Eq.~\eqref{eq:13c} giving connected diagrams.}\centering
  \begin{tabular}{ll}
    \\
    $\hA\iB\jC\kD\grF$ & $\hA\iB\jD\kC\grF$\\
    $\hB\iA\jC\kD\grF$ & $\hB\iA\jD\kC\grF$
  \end{tabular}
\end{table}

The final result is thus
\begin{multline}
  \label{eq:15}
  T(\phi_{1}^{\text{I}}(x_{1}),\phi_{1}^{\text{I}}(x_{2}),
  \phi_{2}^{\text{I}}(x_{3}),\phi_{2}^{\text{I}}(x_{4}),
  \mathcal{V}(\phi_{1}^{\text{I}}(z),\phi_{2}^{\text{I}}(z)))\\ =
  \lambda_{3}T(\phi_{1}^{\text{I}}(x_{1}),\phi_{1}^{\text{I}}(z))
  T(\phi_{1}^{\text{I}}(x_{2}),\phi_{1}^{\text{I}}(z))
  T(\phi_{2}^{\text{I}}(x_{3}),\phi_{2}^{\text{I}}(z))T(\phi_{2}^{\text{I}}(x_{4}),\phi_{2}^{\text{I}}(z))
\end{multline}
and this demonstrates that there is no factor $1/4$ in the Feynman
diagrams with coupling constants $\lambda_{3}$ in Table~\ref{tab:1}.

\section{The symmetry factors for the diagrams in Fig.~\ref{fig:2}}\label{sec:symm-fact-diagr}
\renewcommand{\theequation}{D-\arabic{equation}}
\setcounter{equation}{0}

The time ordered product in Eq.~\eqref{eq:18a} is
\begin{equation}\label{eq:20}
  \frac{\lambda_{1}}{4!}\mel{0}{T(\phi_{1}^{\text{I}}(x_{1}), \phi_{1}^{\text{I}}(x_{2}), \phi_{1}^{\text{I}}(z_{1}) \phi_{1}^{\text{I}}(z_{1}) \phi_{1}^{\text{I}}(z_{1}) \phi_{1}^{\text{I}}(z_{1}))}{0}.
\end{equation}
The Wick's contractions in Eq.~\eqref{eq:20} are given in
Table~\ref{tab:4}

\begin{table}[!ht]\caption{\label{tab:4}All Wick's contractions for
    Eq.~\eqref{eq:20} giving connected diagrams.}\centering
  \begin{tabular}{lll}
    \\
    $\laA\lbB\leA\grG$
    &$\laA\lbC\ldB\grG$
    &$\laA\lbD\ldA\grG$\\
    $\laB\lbA\leA\grG$
    &$\laB\lbC\lcC\grG$
    &$\laB\lbD\lcB\grG$\\
    $\laC\lbA\ldB\grG$
    &$\laC\lbB\lcC\grG$
    &$\laC\lbD\lcA\grG$\\
    $\laD\lbA\ldA\grG$
    &$\laD\lbB\lcB\grG$
    &$\laD\lbC\lcA\grG$
  \end{tabular}
\end{table}
Every contraction in Table~\ref{tab:4} gives the same contribution and
there are~12 such terms, so $1/4!$ gets multiplied by~12 and there
remains overall factor $1/2$. The symmetry factor is the inverse of
this overall factor and for the diagram (a) in Fig.~\ref{fig:2} it is
equal $S=2$.

The time ordered product in Eq.~\eqref{eq:18b} is
\begin{equation}\label{eq:21}
  \frac{\lambda_{3}}{4}\mel{0}{T(\phi_{1}^{\text{I}}(x_{1}), \phi_{1}^{\text{I}}(x_{2}), \phi_{1}^{\text{I}}(z_{1}) \phi_{1}^{\text{I}}(z_{1}) \phi_{2}^{\text{I}}(z_{1}) \phi_{2}^{\text{I}}(z_{1}))}{0}.
\end{equation}
The Wick's contractions in Eq.~\eqref{eq:21} are given in
Table~\ref{tab:5}.

\begin{table}[!ht]\caption{\label{tab:5}All Wick's contractions for
    Eq.~\eqref{eq:21} giving connected diagrams.}\centering
  \begin{tabular}{ll}
    \\
    $\contraction{}{\phi}{{}_{1}^{1}\phi_{1}^{2}}{\phi}
    \contraction[2ex]{\phi_{1}^{1}}{\phi}{{}_{1}^{2}\phi_{1}^{z}}{\phi}      \contraction[3ex]{\phi_{1}^{1}\phi_{1}^{2}\phi_{1}^{z}\phi_{1}^{z}}{\phi}{{}_{2}^{z}}{\phi}
    \phi_{1}^{1}\phi_{1}^{2}\phi_{1}^{z}\phi_{1}^{z}\phi_{2}^{z}\phi_{2}^{z}$
    &$\contraction{}{\phi}{{}_{1}^{1}\phi_{1}^{2}\phi_{1}^{z}}{\phi}
      \contraction[2ex]{\phi_{1}^{1}}{\phi}{{}_{1}^{2}}{\phi}
      \contraction[3ex]{\phi_{1}^{1}\phi_{1}^{2}\phi_{1}^{z}\phi_{1}^{z}}{\phi}{{}_{2}^{z}}{\phi}
      \phi_{1}^{1}\phi_{1}^{2}\phi_{1}^{z}\phi_{1}^{z}\phi_{2}^{z}\phi_{2}^{z}$
  \end{tabular}
\end{table}
Both contractions in Table~\ref{tab:5} give the same contribution and
there are~2 such terms, so $1/4$ gets multiplied by~2 and there
remains overall factor $1/2$. The symmetry factor is the inverse of
this overall factor and for the diagram (b) in Fig.~\ref{fig:2} we get
$S=2$.

\section{The symmetry factors for the diagrams in Fig.~\ref{fig:3}}
\label{sec:symm-fact-diagr-1}
\renewcommand{\theequation}{E-\arabic{equation}}
\setcounter{equation}{0}

\begin{figure}[ht]
  \includegraphics[width=0.75\linewidth]{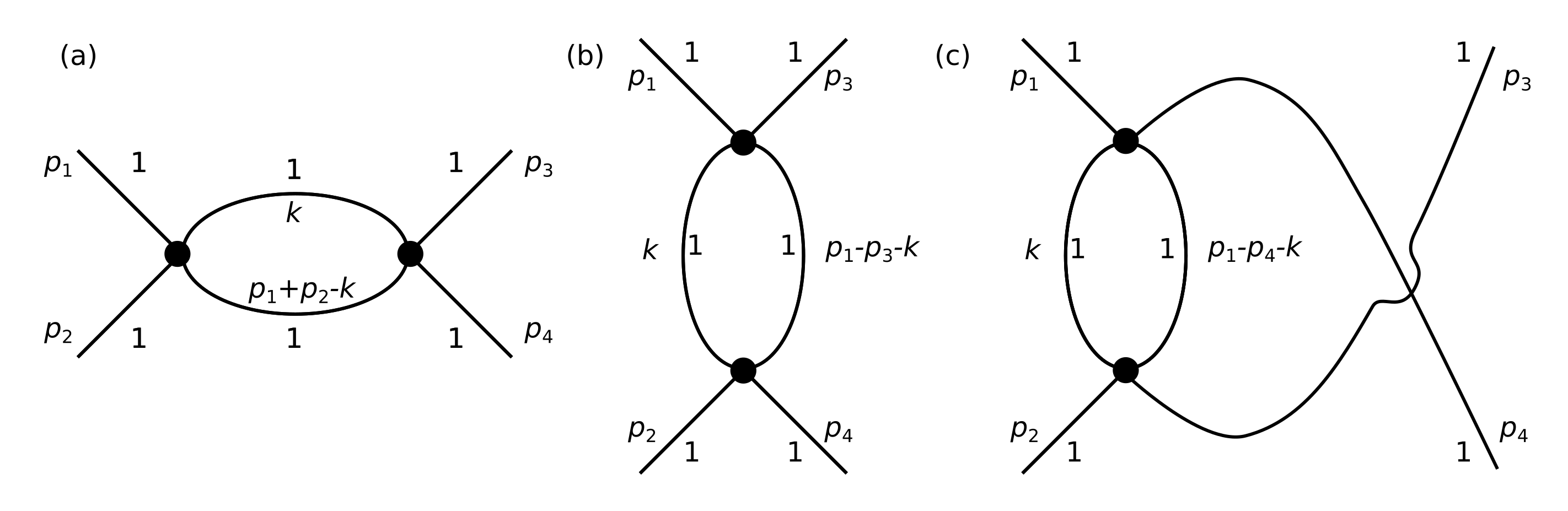}\centering
  \caption{\label{fig:4}Three Feynman diagrams obtained from
    Eq.~\eqref{eq:23a}. $p_{1}, \ldots p_{4}$ are external momenta and
    $k$ is the loop momentum. The momentum is conserved at each
    vertex. All three diagrams have the same symmetry factor.}
\end{figure}

From Eq.~\eqref{eq:23a} we obtain three Feynman diagrams in
Fig.~\ref{fig:4}.
To calculate the symmetry factor of the Feynman diagram in
Fig.~\ref{fig:3}~(a) we have to calculate all contractions in the
following expression
\begin{equation*}
  \phi_{1}^{1}\phi_{1}^{2}\phi_{1}^{3}\phi_{1}^{4} \phi_{1}^{z_{1}}\phi_{1}^{z_{1}}\phi_{1}^{z_{1}}\phi_{1}^{z_{1}} \phi_{1}^{z_{2}}\phi_{1}^{z_{2}}\phi_{1}^{z_{2}}\phi_{1}^{z_{2}}
\end{equation*}
such that there are two contractions between the $\phi_{1}^{1}$,
$\phi_{1}^{2}$ and $\phi_{1}^{z_{1}}$, $\phi_{1}^{z_{1}}$, two
contractions between the $\phi_{1}^{3}$, $\phi_{1}^{4}$ and
$\phi_{1}^{z_{2}}$, $\phi_{1}^{z_{2}}$ and two contractions between
the $\phi_{1}^{z_{1}}$, $\phi_{1}^{z_{1}}$ and $\phi_{1}^{z_{2}}$,
$\phi_{1}^{z_{2}}$ and then it has to be multiplied by~2, because the
vertices can be permuted. This gives
\begin{multline}
  \label{eq:24}
  S^{-1}=\underbrace{\left(\frac{1}{4!}\right)^{2}\frac{1}{2!}}_{\text{factor
      in
      Eq.~\eqref{eq:23a}}}\cdot\underbrace{(4\cdot3)}_{\substack{\text{$\phi_{1}^{1}$,
        $\phi_{1}^{2}$ and $\phi_{1}^{z_{1}}$, $\phi_{1}^{z_{1}}$}\\
      \text{contractions}}}
  \cdot\underbrace{(4\cdot3)}_{\substack{\text{$\phi_{1}^{3}$,
        $\phi_{1}^{4}$ and $\phi_{1}^{z_{2}}$, $\phi_{1}^{z_{2}}$}\\
      \text{contractions}}}\\ \cdot
  \underbrace{2}_{\substack{\text{$\phi_{1}^{z_{1}}$,
        $\phi_{1}^{z_{1}}$ and $\phi_{1}^{z_{2}}$,
        $\phi_{1}^{z_{2}}$}\\ \text{contractions}}}
  \cdot\underbrace{2}_{\text{permutations of vertices}} = \frac{1}{2}.
\end{multline}
The symmetry factor is thus~2 and this means that the factor in front
of this diagram is $1/2$. The symmetry factors for the remaining
patterns in Fig.~\ref{fig:3} are given in Table~\ref{tab:6}.
\begin{table}[ht]
  \caption{\label{tab:6} symmetry factors for the Feynman diagram
    patterns in Fig.~\ref{fig:3}.}\vspace*{10pt}\centering
  \begin{tabular}{l|c}
    Pattern in Fig.~\ref{fig:3}&symmetry factor $S$\\
    \hline
    (a)--(f)&$2$\\
    (g)&$1$
  \end{tabular}
\end{table}

\section{Listing of the \texttt{scalars2.fr} file}
\label{sec:list-feynrules}
\renewcommand{\theequation}{F-\arabic{equation}}
\setcounter{equation}{0}

\begin{lstlisting}
M$ModelName = "scalars2";

M$Information = {
  Authors      -> {"scalars2"},
  Institutions -> {"Cinvestav"},
  Emails       -> {""},
  Date         -> "November 9, 2020"
};

M$Parameters = {

  lambda1 == {
    ParameterType    -> External,
    ParameterName    -> \[Lambda]1,
    Description      -> "self-interaction coupling"
  },
  
  lambda2 == {
    ParameterType    -> External,
    ParameterName    -> \[Lambda]2,
    Description      -> "self-interaction coupling"
  },
  lambda3 == {
    ParameterType    -> External,
    ParameterName    -> \[Lambda]3,
    Description      -> "self-interaction coupling"
  },
  lambda4 == {
    ParameterType    -> External,
    ParameterName    -> \[Lambda]4,
    Description      -> "self-interaction coupling"
  }
};

M$ClassesDescription = {

  S[1] == {
      ClassName     -> phi1,
      ParticleName    -> "1",
      PropagatorLabel -> "1",
      PropagatorType -> Straight,
      SelfConjugate -> True,
      Mass -> mphi1
  },
  S[2] == {
      ClassName     -> phi2,
      ParticleName    -> "2",
      PropagatorLabel -> "2",
      PropagatorType -> Straight,
      SelfConjugate -> True,
      Mass -> mphi2
  }

};

L = 1/2 del[phi1, mu] del[phi1, mu] - 1/(4!) lambda1 * phi1^4
- 1/2 mphi1^2 * phi1^2 + 1/2 del[phi2, mu] del[phi2, mu]
- 1/(4!) lambda2 * phi2^4 - 1/2 mphi2^2 * phi2^2
- 1/4 lambda3 * phi1^2 * phi2^2;
\end{lstlisting}

\section{Listing of the \texttt{scalars2.gen}\\ and \texttt{scalars2.mod} files}
\label{sec:list-FeynArts-files}
\renewcommand{\theequation}{G-\arabic{equation}}
\setcounter{equation}{0}

\noindent \textbf{A. The \texttt{scalars2.gen} file:}
\begin{lstlisting}
(* * * * * * * * * * * * * * * * * * * * * * * * * * * * * * * * * * * * * * * * *)
(*                                                                               *)
(*         This file has been automatically generated by FeynRules.              *)
(*                                                                               *)
(* * * * * * * * * * * * * * * * * * * * * * * * * * * * * * * * * * * * * * * * *)



(*     Kinematic indices    *)

KinematicIndices[ F ] = {};
KinematicIndices[ V ] = {Lorentz};
KinematicIndices[ S ] = {};
KinematicIndices[ SV ] = {Lorentz};
KinematicIndices[ U ] = {};

$FermionLines = True;

(*     Simplification rules    *)

Attributes[ MetricTensor ] = Attributes[ ScalarProduct ] = {Orderless}

FourVector/: -FourVector[ mom_, mu_ ] := FourVector[Expand[-mom], mu]

FourVector[ 0, _ ] = 0

SpinorType[j_Integer, ___] := MajoranaSpinor /; SelfConjugate[F[j]]

SpinorType[_Integer,__] = DiracSpinor

(*     Generic propagators    *)

M$GenericPropagators={

(*general fermion propagator:*)

AnalyticalPropagator[External][s1 F[j1,mom]]==
NonCommutative[SpinorType[j1][-mom,Mass[F[j1]]]],

AnalyticalPropagator[Internal][s1 F[j1,mom]]==
NonCommutative[DiracSlash[-mom]+Mass[F[j1]]]*
I PropagatorDenominator[mom,Mass[F[j1]]],

(*general vector boson propagator:*)

AnalyticalPropagator[External][s1 V[j1,mom,{li2}]]==
PolarizationVector[V[j1],mom,li2],

AnalyticalPropagator[Internal][s1 V[j1,mom,{li1}->{li2}]]==
-I PropagatorDenominator[mom,Mass[V[j1]]]*
(MetricTensor[li1,li2]-(1-GaugeXi[V[j1]])*
FourVector[mom,li1] FourVector[mom,li2]*
PropagatorDenominator[mom,Sqrt[GaugeXi[V[j1]]] Mass[V[j1]]]),

(*general mixing scalar-vector propagator:*)

AnalyticalPropagator[Internal][s1 SV[j1,mom,{li1}->{li2}]]==
I Mass[SV[j1]] PropagatorDenominator[mom,Mass[SV[j1]]]*
FourVector[mom,If[s1==1||s1==-2,li1,li2]],

(*general scalar propagator:*)

AnalyticalPropagator[External][s1 S[j1,mom]]==1,

AnalyticalPropagator[Internal][s1 S[j1,mom]]==
I PropagatorDenominator[mom,Sqrt[GaugeXi[S[j1]]] Mass[S[j1]]],

(*general Fadeev-Popov ghost propagator:*)

AnalyticalPropagator[External][s1 U[j1,mom]]==1,

AnalyticalPropagator[Internal][s1 U[j1,mom]]==
I*PropagatorDenominator[mom,Sqrt[GaugeXi[U[j1]]] Mass[U[j1]]]
}

(*     Generic couplings    *)

M$GenericCouplings = {

	 (* SS *)

AnalyticalCoupling[s1 S[j1, mom1], s2 S[j2, mom2] ] ==
G[+1][s1 S[j1], s2 S[j2]].
{ ScalarProduct[mom1, mom2],
        1 },

	 (* SSSS *)

AnalyticalCoupling[s1 S[j1, mom1], s2 S[j2, mom2], s3 S[j3, mom3], s4 S[j4, mom4] ] ==
G[+1][s1 S[j1], s2 S[j2], s3 S[j3], s4 S[j4]].{1}
}

(* FlippingRules: the flipping rules determines how Dirac
   objects change when the order of fermion fields in the
   coupling is reversed. In other words, it defines how the 
   coupling C[F, -F, ...] is derived from C[-F, F, ...].*)

M$FlippingRules = {
NonCommutative[dm1:_DiracMatrix | _DiracSlash,dm2:_DiracMatrix | _DiracSlash, ChiralityProjector[+1]] ->
 NonCommutative[dm2,dm1, ChiralityProjector[+1]],
NonCommutative[dm1:_DiracMatrix | _DiracSlash,dm2:_DiracMatrix | _DiracSlash, ChiralityProjector[-1]] ->
 NonCommutative[dm2,dm1, ChiralityProjector[-1]],
NonCommutative[dm:_DiracMatrix | _DiracSlash, ChiralityProjector[+1]] ->
-NonCommutative[dm, ChiralityProjector[-1]],
NonCommutative[dm:_DiracMatrix | _DiracSlash, ChiralityProjector[-1]] ->
-NonCommutative[dm, ChiralityProjector[+1]]}

	(* TruncationRules: rule for omitting the wave functions of
	   external Propagators defined in this file. *)

M$TruncationRules = {
  _PolarizationVector -> 1,
  _DiracSpinor -> 1,
  _MajoranaSpinor -> 1 
}
	(* LastGenericRules: the very last rules that are applied to an
	   amplitude before it is returned by CreateFeynAmp. *)

M$LastGenericRules = {
  PolarizationVector[p_, _. mom:FourMomentum[Outgoing, _], li_] :>
    Conjugate[PolarizationVector][p, mom, li]
}
	(* cosmetics: *)

	(*  left spinor in chain + mom incoming -> ar v
	    left spinor in chain + mom outgoing -> ar u
	   right spinor in chain + mom incoming -> u
	   right spinor in chain + mom outgoing -> v *)
Format[
  FermionChain[
    NonCommutative[_[s1_. mom1_, mass1_]],
    r___,
    NonCommutative[_[s2_. mom2_, mass2_]]] ] :=
  Overscript[If[FreeQ[mom1, Incoming], "u", "v"], "_"][mom1, mass1] .
    r . If[FreeQ[mom2, Outgoing], "u", "v"][mom2, mass2]

Format[ DiracSlash ] = "gs"

Format[ DiracMatrix ] = "ga"

Format[ ChiralityProjector[1] ] = SequenceForm["om", Subscript["+"]]

Format[ ChiralityProjector[-1] ] = SequenceForm["om", Subscript["-"]]

Format[ GaugeXi[a_] ] := SequenceForm["xi", Subscript[a]]

Format[ PolarizationVector ] = "ep"

Unprotect[Conjugate];
Format[ Conjugate[a_] ] = SequenceForm[a, Superscript["*"]];
Protect[Conjugate]

Format[ MetricTensor ] = "g"

Format[ ScalarProduct[a__] ] := Dot[a]

Format[ FourVector[a_, b_] ] := a[b]
\end{lstlisting}\vspace*{5pt}

\noindent \textbf{B. The \texttt{scalars2.mod} file:}
\begin{lstlisting}
(* * * * * * * * * * * * * * * * * * * * * * * * * * * * * * * * * * * * * * * * *)
(*                                                                             *)
(*         This file has been automatically generated by FeynRules.            *)
(*                                                                             *)
(* * * * * * * * * * * * * * * * * * * * * * * * * * * * * * * * * * * * * * * * *)


FR$ModelInformation={
  ModelName->"scalars2",
  Authors -> {"scalars2"},
  Institutions -> {"Cinvestav"},
  Emails -> {""},
  Date -> "November 9, 2020"};

FR$ClassesTranslation={};

FR$InteractionOrderPerturbativeExpansion={};

FR$GoldstoneList={};

(*     Declared indices    *)

(*     Declared particles    *)

M$ClassesDescription = {
S[1] == {
    PropagatorLabel -> "1",
    PropagatorType -> Straight,
    SelfConjugate -> True,
    PropagatorArrow -> None,
    Mass -> mphi1,
    Indices -> {} },

S[2] == {
    PropagatorLabel -> "2",
    PropagatorType -> Straight,
    SelfConjugate -> True,
    PropagatorArrow -> None,
    Mass -> mphi2,
    Indices -> {} }
}


(*        Definitions       *)


mphi1[ ___ ] := mphi1;
mphi2[ ___ ] := mphi2;




(*      Couplings (calculated by FeynRules)      *)

M$CouplingMatrices = {

C[ S[1] , S[1] , S[1] , S[1] ] == {{(-I)*\[Lambda]1, (-I)*dL1}},

C[ S[1] , S[1] ] == {{0,(-I)*dM1},{0,(-I)*dM1}},

C[ S[2] , S[2] ] == {{0,(-I)*dM2}i,{0,(-I)*dM2}},

C[ S[1] , S[1] , S[2] , S[2] ] == {{(-I)*\[Lambda]3, (-I)*dL3}},

C[ S[2] , S[2] , S[2] , S[2] ] == {{(-I)*\[Lambda]2, (-I)*dL2}}

}

(* * * * * * * * * * * * * * * * * * * * * * * * * * * * * * * * * * * *)

(* Parameter replacement lists (These lists were created by FeynRules) *)

(* FA Couplings *)

M$FACouplings = {
};
\end{lstlisting}

\section{Listing of the \textit{Mathematica} program for the generation of the Feynman diagrams}
\label{sec:list-math-progr}
\renewcommand{\theequation}{H-\arabic{equation}}
\setcounter{equation}{0}

\begin{lstlisting}
(* QFT of two interacting scalar fields
 Generation of the Feynman diagrams using the  package FeynArts (http \
: // www.feynarts.de/) *)
ClearGlobal[]; Remove["Global`*"];
<< FeynArts`
(* One loop diagrams *)
t1 = CreateTopologies[1, 1 -> 1, ExcludeTopologies -> Reducible, 
   Adjacencies -> 4];
Paint[t1, ColumnsXRows -> {1, 1}, Numbering -> None, 
  SheetHeader -> None, ImageSize -> 192];
f1 = InsertFields[t1, S[1] -> S[1], GenericModel -> scalars2, 
   Model -> scalars2, InsertionLevel -> {Particles}];
Paint[f1, ColumnsXRows -> {2, 1}, Numbering -> None, 
  SheetHeader -> None, ImageSize -> 384];
t2 = CreateCTTopologies[1, 1 -> 1, ExcludeTopologies -> Reducible, 
   Adjacencies -> 4];
Paint[t2, ColumnsXRows -> {1, 1}, Numbering -> None, 
  SheetHeader -> None, ImageSize -> 192];
f2 = InsertFields[t2, S[1] -> S[1], GenericModel -> scalars2, 
   Model -> scalars2, InsertionLevel -> {Particles}];
Paint[f2, ColumnsXRows -> {1, 1}, Numbering -> None, 
  SheetHeader -> None, ImageSize -> 192];
(* Two loop diagrams *)
t3 = CreateTopologies[2, 1 -> 1, ExcludeTopologies -> Reducible, 
   Adjacencies -> 4];
Paint[t3, ColumnsXRows -> {2, 1}, Numbering -> None, 
  SheetHeader -> None, ImageSize -> 384];
f3 = InsertFields[t3, S[1] -> S[1], GenericModel -> scalars2, 
   Model -> scalars2, InsertionLevel -> {Particles}];
Paint[f3, ColumnsXRows -> {3, 2}, Numbering -> None, 
  SheetHeader -> None, ImageSize -> 512];
t4 = CreateCTTopologies[2, 1 -> 1, ExcludeTopologies -> Reducible, 
   Adjacencies -> 4];
Paint[t4, ColumnsXRows -> {3, 1}, Numbering -> None, 
  SheetHeader -> None, ImageSize -> 512];
f4 = InsertFields[t4, S[1] -> S[1], GenericModel -> scalars2, 
   Model -> scalars2, InsertionLevel -> {Particles}];
Paint[f4, ColumnsXRows -> {4, 1}, Numbering -> None, 
  SheetHeader -> None, ImageSize -> 640];
t5 = CreateTopologies[1, 2 -> 2, ExcludeTopologies -> Reducible, 
   Adjacencies -> 4];
Paint[t5, ColumnsXRows -> {3, 1}, Numbering -> None, 
  SheetHeader -> None, ImageSize -> 512];
f5 = InsertFields[t5, {S[1], S[1]} -> {S[1], S[1]}, 
   GenericModel -> scalars2, Model -> scalars2, 
   InsertionLevel -> {Particles}];
Paint[f5, ColumnsXRows -> {3, 2}, Numbering -> None, 
  SheetHeader -> None, ImageSize -> 512];
f6 = InsertFields[t5, {S[1], S[1]} -> {S[2], S[2]}, 
   GenericModel -> scalars2, Model -> scalars2, 
   InsertionLevel -> {Particles}];
Paint[f6, ColumnsXRows -> {4, 1}, Numbering -> None, 
  SheetHeader -> None, ImageSize -> 640];
t6 = CreateTopologies[2, 2 -> 2, ExcludeTopologies -> Reducible, 
   Adjacencies -> 4];
Paint[t6, ColumnsXRows -> {4, 3}, Numbering -> None, 
  SheetHeader -> None, ImageSize -> 512];
f7 = InsertFields[t6, {S[1], S[1]} -> {S[1], S[1]}, 
   GenericModel -> scalars2, Model -> scalars2, 
   InsertionLevel -> {Particles}];
Paint[f7, ColumnsXRows -> {5, 9}, Numbering -> None, 
  SheetHeader -> None, ImageSize -> 640];
f8 = InsertFields[t6, {S[1], S[1]} -> {S[2], S[2]}, 
   GenericModel -> scalars2, Model -> scalars2, 
   InsertionLevel -> {Particles}];
Paint[f8, ColumnsXRows -> {5, 8}, Numbering -> None, 
  SheetHeader -> None, ImageSize -> 640];\end{lstlisting}

\end{document}